\documentclass{iopart}

\pdfoutput=1
\usepackage[top=2.3cm,bottom=2.8cm,left=1.7cm,right=1.6cm,a4paper]{geometry}
\usepackage{epsfig}

\def \Etmiss {{\not}{E_T}}

\def \gevocs {GeV/$c^2$}

\def \D0  {D\O} 
\def \into {\rightarrow}

\newcommand{\met}{\Etmiss}
\newcommand{\ppbar}{p\bar{p}}
\newcommand{\ttbar}{t\bar{t}}
\newcommand{\bbbar}{b\bar{b}}

\newcommand{\plet}{\protect{\it Phys. Lett. }}

\newcommand{\prl}{\PRL}
\newcommand{\prev}{\PR}
\newcommand{\prd}{\PR D}
\newcommand{\nim}{\NIM}
\newcommand{\rpp}{\RPP}

\newcommand{\nphy}{\NP}

\newcommand{\rppbib}[6]{
  \bibitem{#1}
  #2 #3 #4 {\bf #5} #6}

\newcommand{\sub}[1]{\ensuremath{_{\mbox{\scriptsize \,#1}}}}
\newcommand{\supers}[1]{\ensuremath{^{\mbox{\scriptsize #1}}}}

\def\Mt {{\it M}\sub{t}}
\def\mt {{\it M}\sub{t}}
\def\mW {{\it M}\sub{W}}
\def\mH {{\it M}\sub{H}}
\def\Mtbar {{\it M}\ensuremath{_{\,\bar{\mbox{\scriptsize t}}}}}
\def\etal {\protect {\it et al.}}

\newenvironment{mylist} {
    \begin{list} {---} {
    }
} {\end{list}}

\begin{document}

\title[Precision measurements of the top quark mass]{Precision 
measurements of the top quark mass from the Tevatron in the pre-LHC era}
\vspace{1.cm}
\noindent
{Angela Barbaro Galtieri}\\
{\it Ernest O.  Lawrence Berkeley National
    Laboratory, Berkeley, California 94720} \\
{Fabrizio Margaroli} \\
{\it Purdue University, West Lafayette, Indiana 47907}\\
{Igor Volobouev}\\
{\it Texas Tech University, Lubbock, Texas 79409}

%\vspace{1.0cm}
%\centerline{\bf Abstract} 
%\vspace{1.cm}

\begin{abstract}
The top quark is the heaviest of the six quarks of the Standard Model.
Precise knowledge of its mass is important for imposing constraints on
a number of physics processes, including interactions of the as
yet unobserved Higgs boson. The Higgs boson is the only missing particle
of the Standard Model, central to the electroweak symmetry breaking
mechanism and generation of particle masses.
In this Review, experimental measurements of the top quark mass
accomplished at the Tevatron, a~proton-antiproton collider located
at the Fermi National Accelerator Laboratory, are described.
Topologies of top quark events and methods used to separate signal
events from background sources are discussed. Data analysis techniques
used to extract information about the top mass value are reviewed. The
combination of several most precise measurements performed with the
two Tevatron particle detectors, CDF and \D0 ,
yields a value of $\Mt = 173.2 \pm 0.9$~GeV/$c^2$.

\end{abstract}

%\vspace{1.0cm}
\twocolumn

\newpage
% table of Content
\tableofcontents
%
% this is Introduction.tex
\section{Introduction}
\label{sec:intro}

The Standard Model of Particle Physics unifies the weak and
electromagnetic forces into a single quantum field theory.
The addition  of Quantum Chromodynamics (QCD), which describes the
strong interactions that bind quarks into protons and neutrons,
completes the Standard Model (SM).
The elements of this unified theory are six quarks, six leptons and five gauge
bosons. The gauge bosons are the 
$W^{\pm}$ and $Z$ (carriers of the weak force), the photon (carrier of the 
electromagnetic force) and the gluon (carrier of the strong force). 
An additional neutral scalar boson, the Higgs
boson, is necessary to explain electroweak (EWK) symmetry breaking, {\it i.e.},
the observation of non-zero masses of the $W^{\pm}$ 
and $Z$ bosons. It also generates quark and lepton masses through the Yukawa 
interaction. A recent review of EWK symmetry breaking scenarios can be found
in an earlier issue of this journal~\cite{ewksb-rev}.

\begin{figure}[h]
\centerline{
\epsfig{file=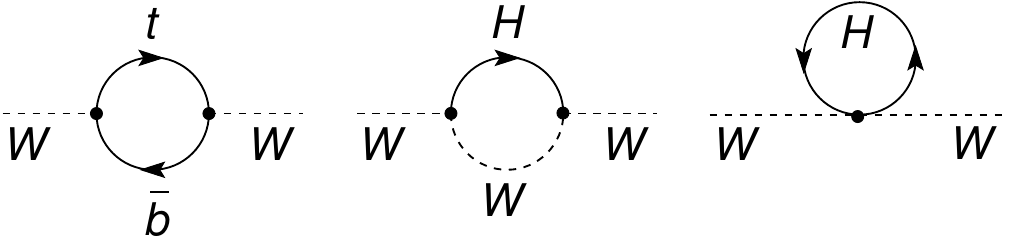,width=.48\textwidth}
}
\caption{Lowest order diagrams that correlate $M\sub{W}$,
$M\sub{t}$, and $M\sub{H}$.
%$\Delta M\sub{W} \propto \mt^2$ in the left diagram, 
%$\Delta M\sub{W} \propto \ln(M\sub{H})$ in the central diagram
%XXX for the third diagram 
}
\label{loop-cor}
\end{figure}

The top quark is the heaviest fundamental fermion.
Prior to its direct observation, its mass was predicted
through a fit to a number of EWK observables sensitive to
virtual top quark effects. This prediction, however, had a very large
uncertainty (for historical details, see a plot of top mass expectations and
measurements versus time in~\cite{quigg}).
The mass of the still unobserved Higgs boson, $M\sub{H}$,
% is not predicted by the theory, but 
is related within the electroweak theory to the $W$ boson mass,
$M\sub{W}$, and the
top quark mass, $M\sub{t}$, through
quantum loop corrections. Some of the lowest order 
diagrams that link $M\sub{W}$, $M\sub{t}$, and $M\sub{H}$ are shown in 
Figure~\ref{loop-cor}. 
% (add the third one at lowest order, but
% use the simpler ones)  

\begin{figure}[htbp]
\centerline{
 \epsfig{file=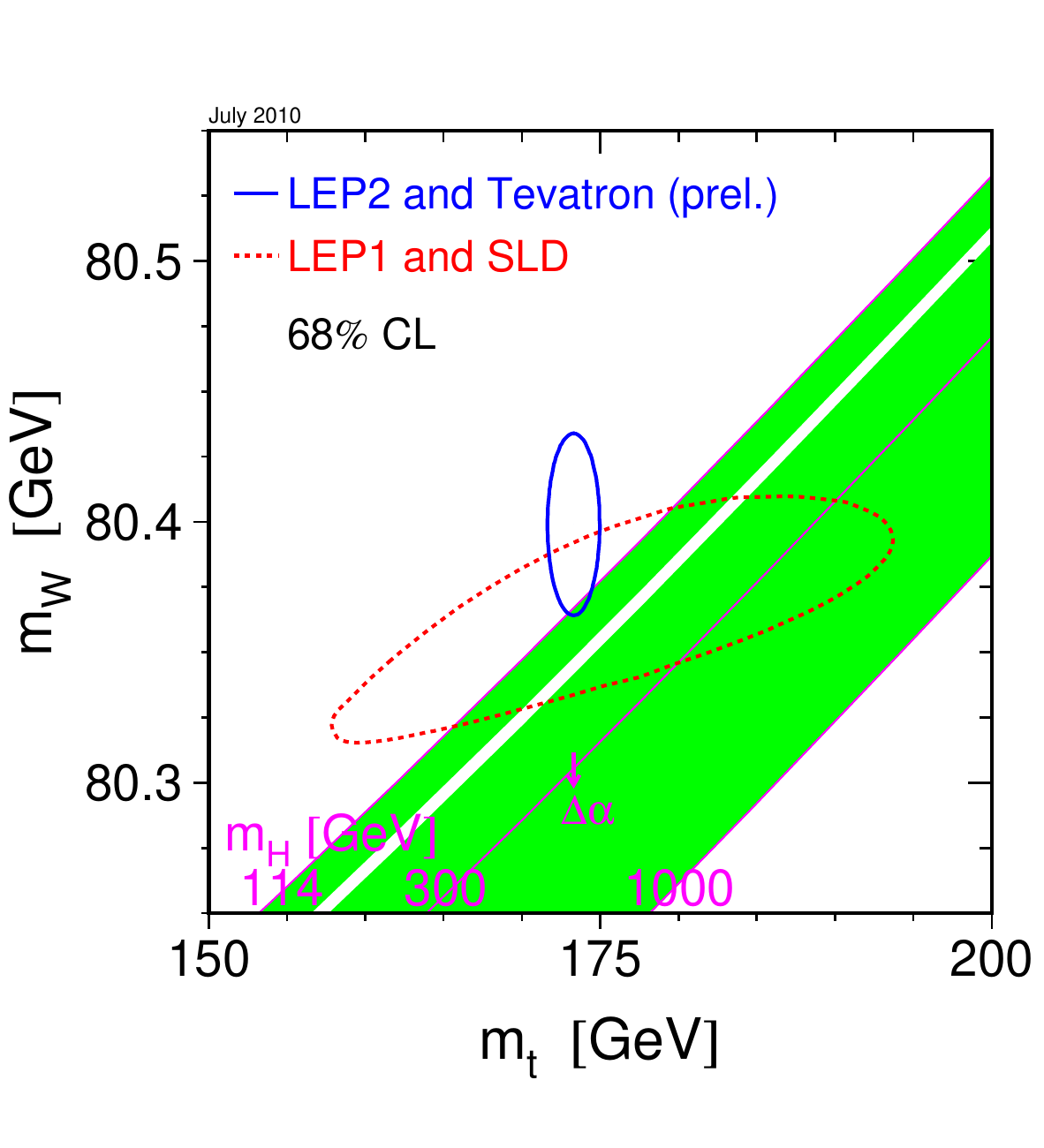,width=.40\textwidth}
}
\caption{
The relationship between \mW, $\mt$, and \mH.
For each value of \mH, the Standard Model constraints possible 
values of $M\sub{W}$ and $\mt$ so that they have to lie along the
corresponding diagonal band.
The dashed contour is the indirect constraint on \mW ~and  $\mt$ from 
measurements of 18 EWK observables; the solid contour is the expectation 
from the  \mW ~and $\mt$ direct measurements. All contours are for the
68\% CL fit result~\cite{lep-ewkwg}.} 
\label{mh-mwmt}
\end{figure}

% Thus, the precision measurement 
Precision measurements of the masses of the $W$ boson and the top quark
are essential to predict the mass of the Higgs boson. An overall fit
of EWK observables including the $W$ and the top masses can put
constraints on the Higgs mass~\cite{lep-ewkwg}.
%The results of the fit to these observable are shown 
%in Figure~\ref{mh-ewkall}.
%It shows all the EWK measurements including  \mW ~and $\mt$,
%the result of the fit and the number of standard deviations between
%the measurement and the fit result.  
Figure~\ref{mh-mwmt} illustrates the relationship between
the three masses, given
current measurements. The Standard Model fit of 18 EWK observables (without
the mass measurements) constrains the Higgs mass to lie inside 
the dashed contour, while the precision with which
the $W$ and top masses are currently known
constrains the Higgs mass to the smaller solid
contour. From the latter we see that a change of 1 \gevocs \ in the
top mass shifts the predicted central value of the Higgs mass 
by $\sim 10$ \gevocs. 
% Clearly the precision measurements of \mW ~and
% $\mt$ ~provide the best information.
%A shift of yy MeV/c$^2$ in \mW shifts \mH by zz \gevocs.   

% two separate figures
%\begin{figure}[htbp]
%\centerline{
% \epsfig{file=s10_show_pull_18.pdf,width=.45\textwidth}}
%\caption{
%Top results of the fit to all EWK measurements. }
%\label{mh-ewkall}
%\end{figure}
%

The discovery of the bottom quark in 1977~\cite{lederman}
set in motion the
search for its partner in the third fermion doublet. 
Experimental lower limits on the top mass 
slowly increased from a few \gevocs \ until the
top quark was observed and its mass was directly measured at the
Tevatron 18 years later~\cite{top_dis-cdf,top_dis-d0}.
% (a~plot in~\cite{quigg} can be consulted for historical detail).
% Based on the fit of measurements
% of electroweak quantities, as explained above,
%predictions on the mass of the top quark were made years before it 
%was discovered, albeit with a large theoretical uncertainty (see plot 
%in reference~\cite{quigg}). 
%The history of the expected top mass values and the 
%actually measured values is shown in Figure~\ref{quigg_hist} ~\cite{quigg}. 
A first hint for the top quark was reported by the CDF collaboration 
in~\cite{CDF94}, together with a mass value of
 $174 \pm 10 \pm13$ \gevocs . Today, the 
measured value of the top quark mass 
is not very far from this very early estimate. Increased statistics, better
understanding of detector performance, and better measurement techniques
have reduced the uncertainty considerably.

%\begin{figure}[htbp]
%\centerline{
%\epsfig{file=quigg_tmass_history.pdf,width=.45\textwidth}
%}
%\caption{History of top mass predictions and measurements (courtesy of
%  Chris Quigg). SHOULD WE DROP THIS FIG?}
%\label{quigg_hist}
%\end{figure}
    
% \noindent

The top quark is much heavier than its partner,
the bottom quark, whose mass is about $5$~\gevocs \ 
(see~\cite{PDG-qmass} for a review on quark masses).
The Yukawa coupling of 
the top quark, $\lambda_t$ = 2$^{3/4}$ G$_F^{1/2} \mt$,
is of order unity. This raises the question if the top quark is
distinct from  the other quarks, {\it i.e.,}
does it have a special role in the electroweak symmetry breaking?
A dynamical breaking of EWK theory by a top quark condensate was proposed 
even before the top quark was discovered~\cite{bardeen}, later extended to
a topcolor model~\cite{top-color}. So far no experimental evidence for  
the validity of such a model has been found.

\subsection{Top mass definition}
\label{sec:tmass-def}
 When referring to quark masses, it is important to define which      
 theoretical framework is used for the given value of the mass.              
 For example, in the overall fit of              
 electroweak measurements
 % mentioned above,
 the top quark mass            
 needs to be expressed in the ${\overline{MS}}$ 
 renormalization scheme. It is not
 completely clear
 how to relate the mass measured in $\ttbar$ production experiments
 with the mass used in the EWK fit. It is normally assumed that what is 
 being measured is the pole mass, $M\sub{pole}$.
 The relation between $M\sub{pole}$ and the 
 mass in the ${\overline{MS}}$ scheme, $M_{\overline{MS}}$,
 can be computed within perturbative QCD. Using approximate 
 next-to-next-to-leading order (NNLO) calculations, the difference 
 is about 10 \gevocs, the ${\overline{MS}}$ mass being 
 smaller~\cite{langenfeld,hoang2}. 

 The top quark mass can be determined from
% One way to determine the top quark
a measurement of the total $\ttbar$ production cross 
section. The cross section dependence on the mass can be 
calculated in any renormalization scheme, and the results for the 
${\overline{MS}}$ scheme are given in~\cite{langenfeld}. The
\D0 ~collaboration has extracted a top mass in the ${\overline{MS}}$
scheme from such a measurement~\cite{d0-sigma-MS}, but the
precision of this method is not comparable with that achieved by direct 
top mass measurements from the top decay products, with a subsequent change 
from $M\sub{pole}$ to $M_{\overline{MS}}$.
   
A number of theoretical questions arise in relating the 
mass measured directly  to the pole mass~\cite{hoang2,hoang1}.
First, the pole mass is sensitive to an infrared renormalon, which implies 
that the value of the pole mass is modified by an amount of the order 
of $\Lambda\sub{QCD}$ as the order in perturbation theory
is changed~\cite{smith-will}.
Second, there are some doubts as to the precise definition of the measured 
mass.
The direct measurements reported here are all calibrated with Monte     
Carlo generators, therefore what is measured is the mass parameter
used in 
the generators, $M\sub{MC}$. 
% Work is being done to answer the question on the
% relation of  $M\sub{MC}$ with $M\sub{pole}$. 
%
The relationship between 
$M\sub{pole}$ and $M\sub{MC}$ can be represented in the general 
form~\cite{hoang2,buckley-all}:
\[
          M\sub{pole} =  M\sub{MC}  + Q_0[ \alpha _s (Q_0)c_1 +...],
\]
\noindent
 where the coefficient $c_1$ is not known (it depends on 
 parton shower implementation details in each particular generator),
 but likely to 
 be of the order unity. The main question is what the 
 appropriate value of the scale $Q_0$ should be.
 It has been argued~\cite{hoang2} that the cutoff on radiation in the parton shower evolution of order 1~GeV employed
 by generators like PYTHIA~\cite{PYTHIA} 
 implies that $Q_0$ is of order 1~GeV 
as well. The difference between the measured mass and the 
 pole mass would then be $\cal{O}$(1~\gevocs).
% plus possible other terms. 
Furthermore, it is not clear if other approximations used in the parton 
shower development alter the resulting mass, as
discussed in Appendix C of~\cite{buckley-all}. 
Theoretical studies of this problem are in 
progress along the lines presented in~\cite{fleming}.
It is expected that
the relation between $M\sub{MC}$ and $M\sub{pole}$ 
will be understood in the not too distant future.

% Apart from this theoretical uncertainty, a precision measurement of the top
% quark mass in $\ttbar$ production data is important, as top events
% share common experimental signatures
% with a number of ``beyond the
% Standard Model'' physics processes. Simulated top event samples, 
% used for evaluation of backgrounds from top production,  are
% generated at the measured  $M\sub{MC}$. 

\subsection{Notation}
\label{sec:notation}

In the subsequent text, we refer to a number of
kinematic and physics quantities.
The most common ones are denoted by the following symbols:

$\eta$ --- Pseudorapidity which characterizes direction
           of a vector ({\it e.g.,} particle momentum)
           with respect to the colliding beam axis (the z axis). It is
           related to the vector polar angle, $\theta$, by
           $\eta = -\ln \left[\tan (\theta / 2)\right]$.

$\phi$ --- Azimuthal angle of a vector.

$\Delta R$ --- Distance in the $\eta$-$\phi$ space: 
$\Delta R = \sqrt{\Delta \eta ^{2}+ \Delta \phi ^{2}}$. 
Equation $\Delta R < c$, with respect to a certain direction
and with some constant $c$, 
defines a circle in the $\eta$-$\phi$ space. In the context
of jet reconstruction algorithms, such circle is usually
referred to as a ``cone''.

$p_T$ --- Transverse momentum: $p_T = p \sin \theta$, where $p$ is the momentum magnitude.

$E_T$ --- Transverse energy, usually defined by $E_T = E \sin \theta$, where
$E$ is the particle energy.

$\met$ --- Missing transverse energy. It approximates the
           transverse momentum carried away by neutrinos
           in the assumption of a fully hermetic particle detector.

$H_T$ --- Scalar sum of the $E_T$ of all charged leptons and jets in the event
          added to the $\met$.

\Mt\ --- Experimentally measured mass of the top quark.
         Unless noted otherwise, $\Mt \equiv M\sub{MC}$.

$m_t$ --- An estimate of the top quark mass obtained in a single
          collider event.

$\int \mathcal{L} dt$ --- Integrated luminosity accumulated by an experiment. $\mathcal{L}$ is the instantaneous luminosity.

\section{Top production and decay}
\label{sec:top_pro_dec}

A precision measurement of the top quark mass relies heavily on the SM
predictions for top production and decay processes.
It is therefore essential to 
validate underlying physics and detector response models by comparing
experimental measurements with theoretical expectations. Any deviations
found will have to be understood and properly represented in the models.
Effects not taken into account explicitly are treated as sources of systematics
uncertainties (as discussed in Section~\ref{sec:systematics}).

\subsection{Cross section for $\ttbar$ production}
\label{sec:top-sig}
\begin{figure}[htbp]
\centerline{
\epsfig{file=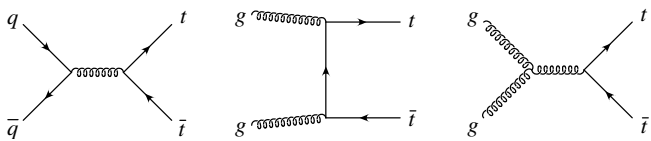,width=.48\textwidth}
}
\caption{Some of the Feynman diagrams that contribute to $\ttbar$ production.}
\label{top-pro}
\end{figure}

\begin{figure}[htbp]
\centerline{
\epsfig{file=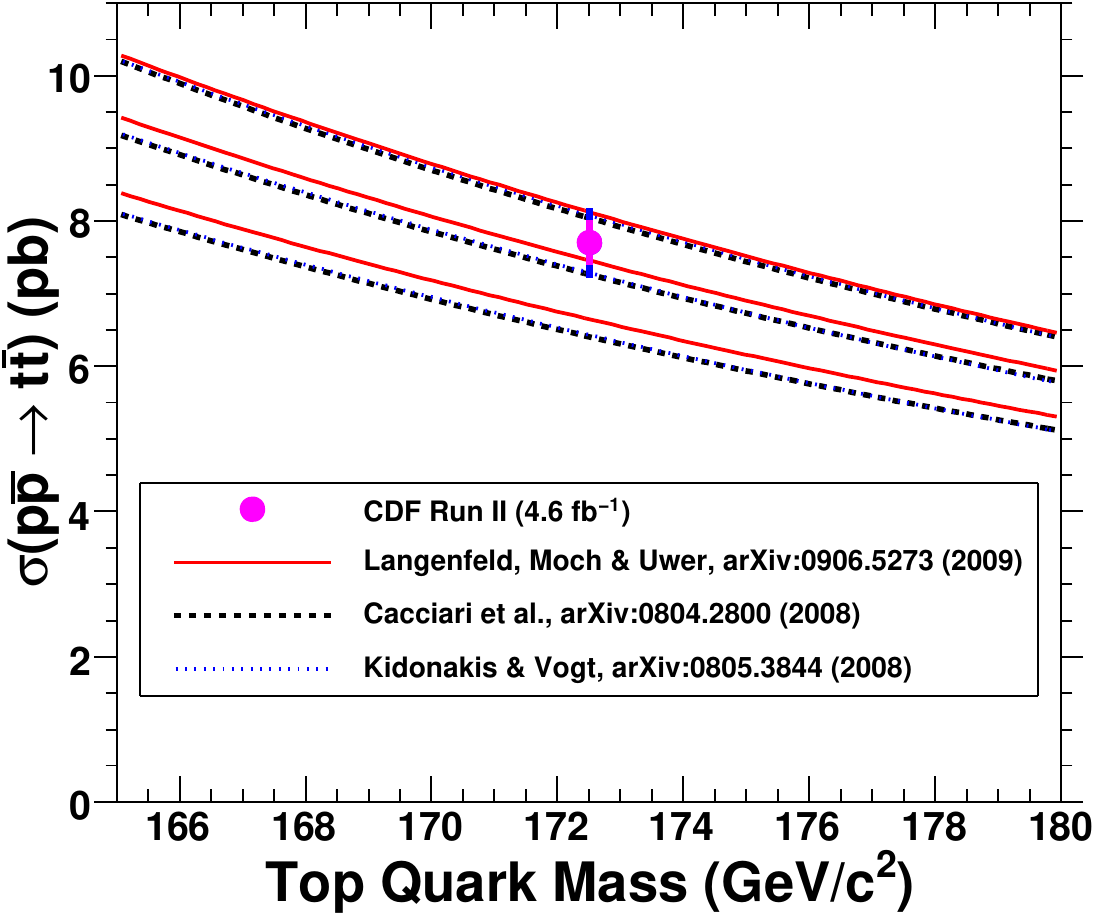,width=.50\textwidth}
}
\caption{Top cross section vs. top quark mass as calculated by different 
  authors~\cite{langenfeld,2008-cacciari,kido-2008}. For each computation, the middle curve shows the central value for the theoretical prediction, while the upper and lower curves show the $\pm 1\sigma$ computations respectively; the curves from the
  last two authors are very similar.  The data point~\cite{cdf-best-sigma}
  is the average of two CDF measurements which use the 
  $Z$ boson production cross section to eliminate the luminosity 
  systematics~\cite{Aaltonen:2010ic}.}
\label{sig-mass}
\end{figure}

The production of top-antitop pairs is a process that can be calculated in 
perturbative QCD. Figure~\ref{top-pro} 
illustrates some of the leading order (LO) QCD diagrams that contribute
to $\ttbar$ production.
At the Tevatron, the process is dominated by the quark annihilation diagram 
(leftmost in the figure), whereas the gluon fusion
diagrams contribute only 15\% of
the total cross section~\cite{tt-qqgg,ttgg-other}. In contrast, at 
LHC energies the gluon diagrams are expected to contribute
more  than the quark diagrams. 
% Next-to-leading order (NLO) diagrams, as initial and final state radiation,
% and NNLO contribute as well to the total cross section 
% and the kinematics of top events.
The top cross section measured by CDF at 
the Tevatron ($\sqrt{s}$ = 1.96 \gevocs) as a function
of the top quark mass is shown in 
Figure~\ref{sig-mass}~\cite{cdf-best-sigma}.
The calculations were performed at NLO+NLL (next-to-leading log) order 
(Cacciari \etal)~\cite{2008-cacciari}, at NLO+NNLL 
(next-to-next-to-leading log) at NLO and approximate NNLO by
(Langenfeld \etal)~\cite{langenfeld} and by (Kidonakis
\etal)~\cite{kido-2008}.
The measurement point is placed at $\mt$ = 172.5 GeV/$c^2$ because
this mass value was used to model $t\bar{t}$ production~\cite{Aaltonen:2010ic}.
This measurement, $\sigma_{t\bar{t}} = 7.50 \pm 0.48$~pb,
is the most precise at the time of this writing.
% CDF measurements 
% (with the ``lepton+jets'' channel)
% that uses the $Z$ boson production 
% cross section to eliminate luminosity 
% systematics. The value is $\sigma = 7.50 \pm 0.48$ pb.
It is in a good agreement with the SM calculations. 

Measured properties of $\ttbar$ events are in agreement with SM
expectations (see recent reviews of Tevatron top-quark physics
results~\cite{tev-tprop,dan-tprop} for more details). Latest
confirmations include the correlation 
between the spins of the top and antitop
quarks~\cite{Aaltonen:2010nz,Abazov:2011qu}, 
the fraction of gluon-gluon 
contribution to the $\ttbar$ production cross section~\cite{pekka-gluon},
and measurements of the $t\bar{t}$ differential cross
sections~\cite{CDFdiff,D0diff} 
(see Section~\ref{sec:mc_models} for a plot). 
One known exception is the forward-backward asymmetry in
top and antitop directions. Recent measurements by both
CDF~\cite{back-for-asy-cdf} and \D0 ~\cite{back-for-asy-d0}
differ by  $\sim3$ standard deviations
from the SM expectation calculated at NLO~\cite{fb-NLO} and approximate 
NNLO~\cite{fb-NNLO}.
This anomalous effect is currently under further investigation. 
\subsection{Top decay modes}
\label{sec:top-decay}

\begin{figure}[htbp]
\includegraphics[width=.3\textwidth]{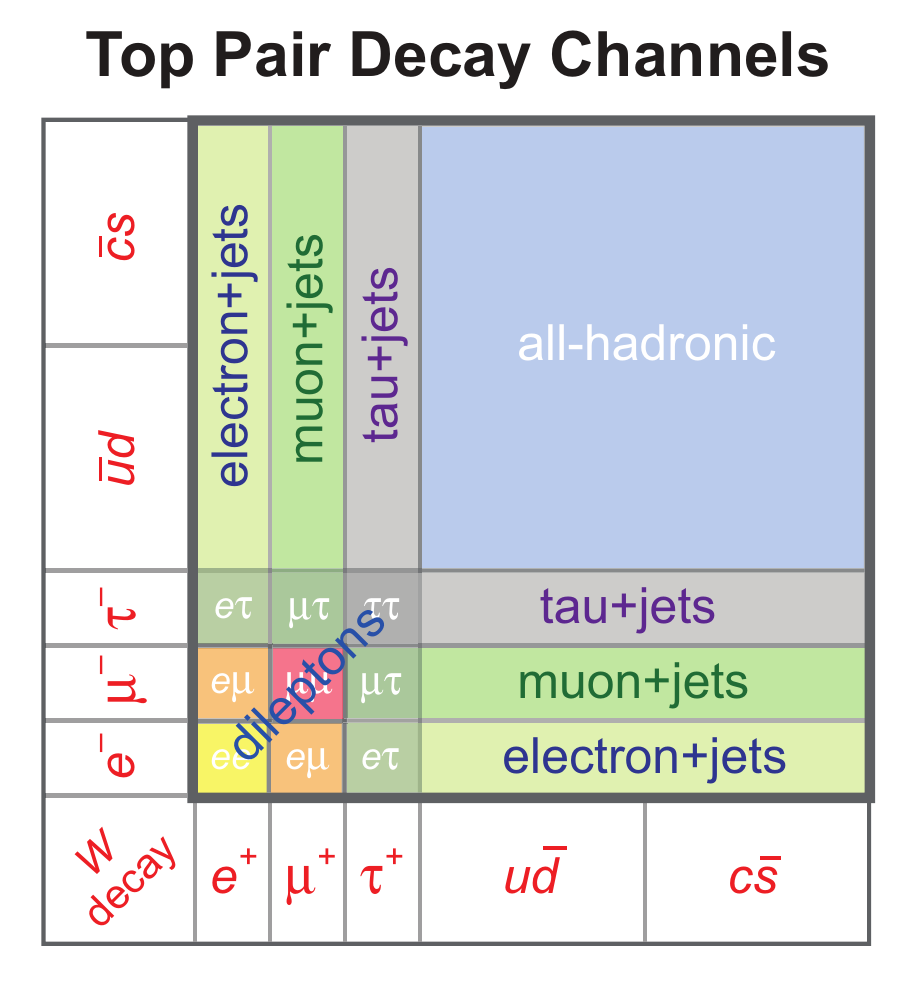}
\caption{Final states of the $t\bar{t}$ system.}
 %Bottom: Fraction of top events in the different topologies 
 %(?taken from Paul's Thesis). } 
\label{ttbar-decays}
\end{figure}

Within the Standard Model, the expected top quark width is
$\sim 1.3$ GeV and its lifetime is about $0.5 \times 10^{-24}$ seconds.
With such a short lifetime ($\ll 1/\Lambda\sub{QCD}$), top quark decays before hadronizing. 
Observation of its decay products allows for a direct measurement
of its mass, a unique feature among quarks~\cite{PDG-qmass}.
The top quark is expected to decay almost exclusively into a
$W$ boson and a bottom quark, so that the intermediate state of a 
$\ttbar$ event is $W^{+} b W^{-} {\bar{b}}$. Each of the two $W$ bosons
decays further into a
charged lepton and a neutrino or into
a quark-antiquark pair (multiple flavor and color assignments are possible),
as illustrated in Figure~\ref{ttbar-decays}.
The quarks produced in $W$ decays are
not observed as such, but they hadronize producing jets of particles
which are subsequently observed and measured in the detectors.
At leading order, 10.6\%  of the $\ttbar$ events will have 
two charged leptons (e, $\mu$, $\tau$), two neutrinos and two jets; 
43.9\% of the final states will have one charged lepton, one neutrino and
four jets, and 45.5\% will have six jets. 
Neutrinos escape direct
detection but their presence can be inferred via
an excess of missing transverse energy in the event.
While $W \rightarrow \ell \nu_{\ell}$ 
decays, with $\ell = e$ or $\mu$, result in
unambiguous experimental
signatures, the case $W \rightarrow \tau \nu_{\tau}$ is more complicated as
tau leptons subsequently decay in a variety of ways.
%
% Additional jets can be due to 
% NLO and NNLO production. 
%
According to the number of electrons and muons produced in the
$W$ decay chains,
the final states (``topologies'') of the $\ttbar$ system
that lead to distinct experimental signatures are called
``dilepton'',``$\ell$+jets'', and ``all-hadronic''.
%Figure~\ref{ttbar-decays} shows all the possible final states for a 
%$\ttbar$ event.
Taking into account $\sim35$\% leptonic decay branching
fraction of the $\tau$,
only 6.4\% of the events end up in the
dilepton category. Similar considerations affect the
$\ell$+jets topology which accounts for 34.1\% of the events
when hadronic $\tau$ decays are excluded.

%To these three major cathegories others
%are added due to detector regions not sufficiently instrumented
%(cracks, low efficiency, etc.). See Sec~\ref{sec:other_ms}.  

Measurements of branching fractions of top quark
decays~\cite{Acosta:2005hr,Abazov:2008yn} have confirmed that
the top quark decays predominantly into $W b$, as predicted by the SM.
A number of decay modes of the top quark have been searched for and
excluded: decay to a charged
Higgs boson (at the level of 10\%) and decays through Flavor Changing Neutral 
Currents (at the level of $10^{-4}$), as summarized in the recent 
reviews~\cite{tev-tprop,dan-tprop}. 
Other properties of top quarks and their decays, such as $W$ boson helicity
fractions~\cite{Aaltonen:2008ei,Abazov:2010jn}, top charge~\cite{Aaltonen:2010js},
and top 
width~\cite{Aaltonen:2010ea,Abazov:2010tm}, have been studied as well,
and no significant deviations from the SM expectations
have been found.

\section{Identifying top events}
\label{sec:top_id}

Stringent requirements are imposed on detectors used for precision
measurements of the top quark mass. High resolution charged particle
tracking together with high granularity, precise energy determination
in calorimeters is necessary for both lepton and jet measurements.
Muon momentum measurements and identification are accomplished, in addition to
good tracking, with dedicated muon systems located outside the calorimeters.
Jet energy measurements require precision calorimetry and good 
segmentation, as jet shapes play an important
role both in the $\met$ measurement 
and in disambiguation between quark and gluon jets 
(see Section~\ref{sec:all-had}). Hermeticity is essential for the $\met$ 
measurement. High resolution tracking is also needed
to identify bottom jets, in particular by reconstructing secondary vertices
in jets. Backgrounds to $\ttbar$ production are considerably
reduced by requiring $b$-jet presence in the event
(see Section~\ref{sec:backgrounds}).

\subsection{Detectors}
\label{sec:detect}

The CDF detector took its first data in 1985,
whereas the \D0 ~detector took its first 
data in 1992. The top quark was discovered in 1995 and the first direct
top mass 
measurements were made with the initial configuration of these detectors 
(Tevatron Run~I measurements). Both the detectors and the Tevatron accelerator
complex went through major 
upgrades during the 2000-2001 shutdown. Precision measurements of
the top mass were performed with the upgraded detectors installed for 
Run II at the Tevatron.
%We will discuss here mostly measurements done with the final
%detector configurations, as they are the most precise.

\begin{figure}[htbp]
\includegraphics[width=.48\textwidth]{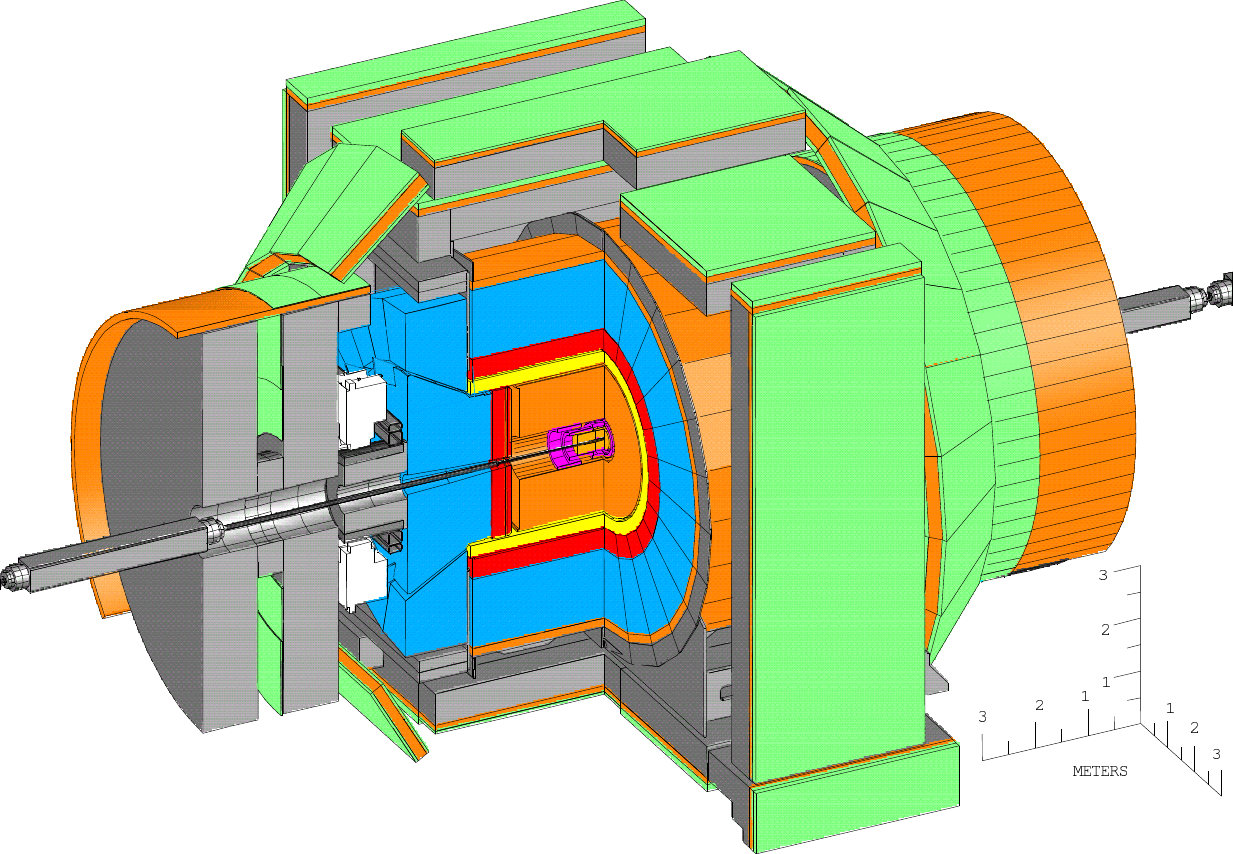}
\caption{The CDF detector componenents from the collision point outward: the silicon system (purple), 
 the central outer tracker (inner part in orange), the solenoid (yellow), the electromagnetic (red) and hadronic (blue)
 calorimeters, and the the muon chambers (green)~\cite{cdf-lep-emu}.} 
% Right: details of the tracking region.}
\label{cdf_detect}
\end{figure}

The Run II CDF detector is shown in  
Figure~\ref{cdf_detect}~\cite{cdf-lep-emu}.   
Precision tracking is achieved with: large number of points along the track, 
high precision space point measurements, long lever arm, and closeness to 
the production vertex for precision impact parameter determination. 
Placement of silicon detectors in close vicinity of the beam pipe has
allowed CDF to achieve a single-particle
transverse momentum determination precision
 $ \frac{\sigma (p_T)}{p_T} = 0.07\% p_T $ 
when the primary vertex is included in the track fit~\cite{cdf-pt-hres}.
The CDF tracking system consists of a silicon microstrip detector
and an open-cell drift chamber immersed in a 1.4 T solenoidal magnetic
field. The silicon detector (SVXII) consists of five double sided cylindrical 
layers of detectors augmented by an inner layer mounted on the beam pipe 
and two more layers on the outside. They  
provide precise charged particle tracking in the radial range 
from 1.5 to 28 cm. Outside this
region, the Central Outer Tracker (COT) extends the tracking system to
137 cm providing 96 additional points for track reconstruction. In 
combination, the COT and the silicon detectors provide
excellent tracking up to a pseudorapidity $|\eta| \sim$ 1.0. Additional
layers of silicon detectors extend tracking coverage 
to $|\eta| \sim$ 2.

% The pseudorapidity is defined 
% as $\eta = -\ln(tan(\theta/2))$, where $\theta$ is the polar angle.

The central electromagnetic (CEM) calorimeter contains $X_0 >$ 18
radiation lengths of lead-scintillator layers. Proportional chambers 
embedded at a depth of $\sim 6 X_0$ (shower maximum) provide shower shape 
information for electron identification. The resolution of the CEM calorimeter 
for electron measurements is  
$ \frac{\sigma (E_T)}{E_T} = \frac{13\%}{\sqrt{E_T}}  \oplus  1.5\%$. 
Projective geometry is used: dense segmentation in the CEM, coarser in
the hadronic calorimeter. The central hadronic calorimeter is composed 
of alternating layers of iron plates and scintillators for a total 
of 4 nuclear interaction lengths. This gives a resolution for charged pions
$\frac{\sigma (E_T)}{E_T} = \frac{50\%}{\sqrt{E_T}} \oplus 3\%$.
In the forward region (1$ < |\eta| <$ 3.6)  a tile calorimeter is used, with
an electron resolution
$\frac{\sigma (E)}{E} = \frac{16\%}{\sqrt{E}} \oplus$ 1.0\% 
for the electromagnetic  component and a pion resolution
$\frac{\sigma (E)}{E} = \frac{80\%} {\sqrt{E}}\oplus 5\%$ 
for the hadronic component. 
% Proportional chambers are placed  
% at shower maximum for electron identification.
An additional hadronic 
calorimeter covers the region between the the central calorimeter and the
plug calorimeter, thus providing hermeticity of the detector. Its
resolution is $\frac{\sigma (E_T)}{E_T} =\frac{75\%}{\sqrt{E}} \oplus
4\%$ for charged pions that do not interact in the CEM~\cite{cdf-jets}.   

The muon chambers coverage extends only to $|\eta| <$ 1.0.
In the center, two muon detectors cover the 
$|\eta| <$ 0.6 region. They consist of four layers of proportional chambers 
each, with the magnet return yoke in between them, thus providing an
additional 60 cm of steel absorber. An additional set of four layers
of drift chambers cover the region 0.6 $<  |\eta| <$ 1.0.

\begin{figure}[htbp]
\epsfig{file=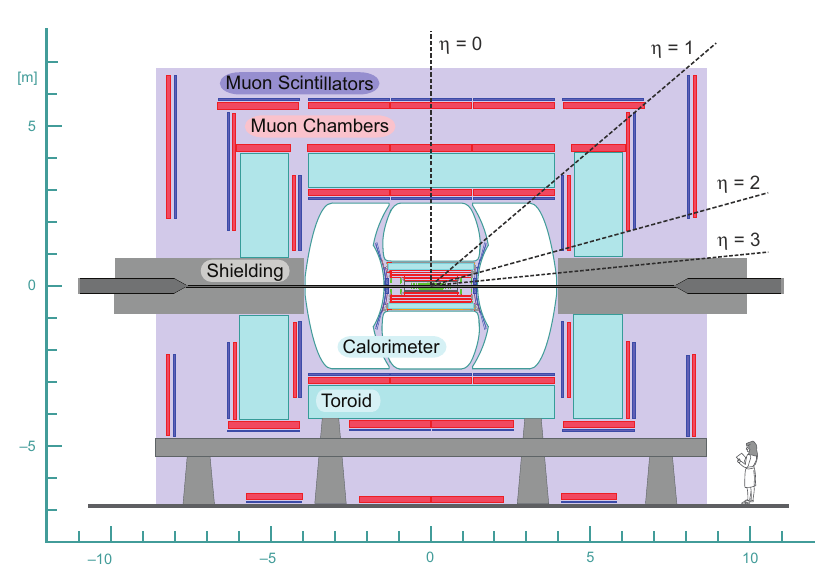,width=.48\textwidth}
\caption{The \D0 ~detector~\cite{d0-public-web}.}
\label{d0_detect}
\end{figure}

The \D0 ~detector is shown schematically in 
Figure~\ref{d0_detect}~\cite{d0-public-web,d0-det}.
The tracking system consists of a silicon microstrip tracker and a
central fiber tracker inside a 2 T solenoid. The silicon detector
includes four
layers of single and double sided detectors in the central region. In 2006
an additional layer of silicon sensors was added on the beam pipe (the data 
before and after this addition are identified as Run~IIa and Run~IIb). 
In the forward direction disks equipped with silicon detectors
complete the tracking system. Outside of the silicon detector system,
16 layers of scintillating fibers (8 axial and 8 stereo),
placed in the radial region 20-52~cm,
provide additional tracking information. 
The combination of the tracking devices provides 
efficient tracking up to pseudorapidity $|\eta| <$ 3. Tracking resolution 
is $ \frac{\sigma (p_T)}{p_T} = 0.2\% p_T  \oplus  1.4\%$.

A preshower detector is placed outside
of the magnet and inside the calorimeter.
% (this is also included on the CDF detector). 
The \D0 ~calorimeter is 
a Uranium-Liquid-Argon system inside a cryostat. It consists of
a central and two endcap components. The EM part has a depth of 
$\sim$ 20 radiation lengths.
% (X$_0$).
The energy resolution for
electrons is  $ \frac{\sigma (E)}{E} = \frac{15\%}{\sqrt{E}}  \oplus 4\%$
in the central calorimeter. In the End Cup calorimeter the resolution
is   $ \frac{\sigma (E)}{E} = \frac{21\%}{\sqrt{E}}  \oplus 4\%$ 
~\cite{d0-ele-res}. The depth of the hadronic section varies from
7.2 nuclear interaction lengths ($\lambda$) at $|\eta|$ = 0 to
$\lambda$ = 10.3 at $|\eta|$ = 1.
% the eta=2 is still to be checked
The calorimeter extends to  $|\eta|$ = 4 with a region of low
efficiency at $1.0 < |\eta|< 1.4$.

The muon system surrounds the calorimeter and consists of
tracking detectors and scintillators covering a pseudorapidity region
up to $|\eta| = 1.0$. A toroidal 1.7~T iron magnet completes the central 
muon system. The forward muon system covers the region
1.0 $<|\eta|<$ 2.0. It consists of mini drift chambers and 
two toroidal magnets with 1.6~T average field.  

Both CDF and \D0 ~use a three-level trigger system to select the events
to analyze. The Tevatron delivers collisions to the detectors every 396 ns, 
which means, at the present luminosity 
($\sim3 \times 10^{32}$~cm$^{-2}$~s$^{-1}$), 
about $2.5 \times 10^6$ bunch crossings per second, with an average of five collisions per crossing.
The first level hardware trigger 
reduces the event rate to
$\sim 10$~kHz/2~kHz (CDF/\D0 ). The second level uses trigger 
processors that reduce the rate to 200~Hz/1000~Hz (CDF/\D0 ).
Finally, the third 
level is based on limited event reconstruction that reduces the rate to
40~Hz/50~Hz (CDF/\D0 ). Digitized detector
readouts for the accepted events are stored on tape for
subsequent offline analysis.    

The two detectors have adequate capabilities to perform many
 precision measurements as well as to explore the vast physics landscape 
 at hand. 
% They are somewhat complementary: CDF has better tracking, 
% while \D0 ~has larger muon coverage and slightly better calorimetry 
% (COMPARISON OK?).

%%%%%%%%%%%%%%%%%%%%%%%%%%%%%%%%%%%%%%%%%%%%%%%

\subsection{Lepton identification} 
\label{sec:leptons}
Electron, muon and neutrino ($\met$) reconstruction utilizes
event information from all detector subsystems. 
Electrons are identified using the tracking system ($p_T$), 
the electromagnetic calorimeter ($E_T$), the electromagnetic shower shape, and
the information from the hadronic calorimeter. CDF obtains the shower shape
information from proportional chambers located at shower
maximum~\cite{cdf-lep-emu}, whereas \D0 ~uses information from finely
segmented layers in the liquid argon calorimeter~\cite{d0-lep-emu}.
Electrons that are likely to come from photon conversions are removed 
using appropriate algorithms. Both experiments have requirements
on the $E/p$ ratio to reduce backgrounds from QCD jets. 
% looked at the D0 reference: tehre are two many details that we cannot include
% also not clear what is the difference of the Et/pt, compared with E/P. 
For electrons coming from $W$ decays, an isolation requirement
is imposed after all other information is processed. This requirement,
which helps in rejecting jets faking electrons,
consists in vetoing significant additional energy in a cone of 
$\Delta R = 0.4$
radius around the electron direction.
For both experiments, there are several levels of electron criteria: ``tight'',
``medium'', and ``loose'', but most of the top mass measurements use
the ``tight'' electron requirements.
Both experiments use $Z \rightarrow e^+ e^-$ decays to calibrate
the electron energy. 

Muons are identified using information from the muon chambers, the tracking 
system, and the calorimeters. Having found a signal in the muon 
chambers,
a ``muon stub'', {\it i.e.}, a track segment,  is reconstructed. Next
step is to match this segment to a track found in the tracking system that 
extrapolates to the muon chambers within a small distance of the segment.
%$|\Delta X|$. 
%of the muon stub. 
This distance is different for the different 
components of the muon system for both CDF~\cite{cdf-lep-emu} 
and \D0 ~\cite{d0-lep-emu}, as it depends on the resolutions of the
chambers and the tracking system
as well as on the distance between them. The track
is also required to originate from the event primary vertex. 
The energy deposited in the EM and hadronic calorimeters by the muon is 
subject to requirements that are different for the two experiments. 
For muons produced in $W$ decays,
a calorimeter isolation as well as a track isolation is required in
both experiments. Muons from cosmic rays are removed. 
% There
% are different levels of requirements to obtain ``tight'', ``medium''
% and ``loose'' categories.
The absolute energy calibration is obtained
with $Z \rightarrow \mu^+ \mu^-$ decays.

The missing transverse energy, $\met$, is determined
from the transverse momentum imbalance
in the event. It is calculated by adding vectorially all the
calorimeter towers, with the direction defined by the vector
connecting the primary vertex to the center of the tower. Corrections
to this sum are made for jets and muons in the event. The muon
correction is obtained by subtracting the energy deposited by the muon in
the calorimeter and by adding its track
momentum to the vectorial sum of the tower
energies. The energy of the jets in the event is corrected by various
coefficients, as described in the next section, and the $\met$
vector is readjusted to take this correction into account.

% SHOULD WE PUT A FORMULA HERE?
  
%%%%%%%%%%%%%%%%%%%%%%%%%%%%%%%%%%%%%%%%
\subsection{Jet reconstruction and calibration}
\label{sec:jet-rec}

Hadronic jets in $t\bar{t}$ events are reconstructed from the energy
flow data collected by the CDF and \D0 \ respective
calorimeters. Initially, all energy depositions are grouped into
projective ``towers'' whose size is consistent with the hadronic
calorimeter granularity. All towers associated with identified
electrons are excluded, electronic noise is subtracted. The
remaining towers are clustered into jets using seeded variations
of the iterative cone algorithm~\cite{blazey-jets} (which is known as the
``mean shift'' algorithm~\cite{ref:meanShift}
in the pattern recognition literature).
Although a number of deficiencies have been discovered in the
cone-based approach to jet reconstruction~\cite{ellis-jets, ref:igvjetnote},
extensive studies of
various particle processes performed with this algorithm
resulted over time in good understanding of jet properties and in consistent
calibrations. Fixed jet size permits a very simple correction
for the presence of underlying event and multiple interactions
occurring in a single bunch crossing. 

\begin{figure}[htbp]
\epsfig{file=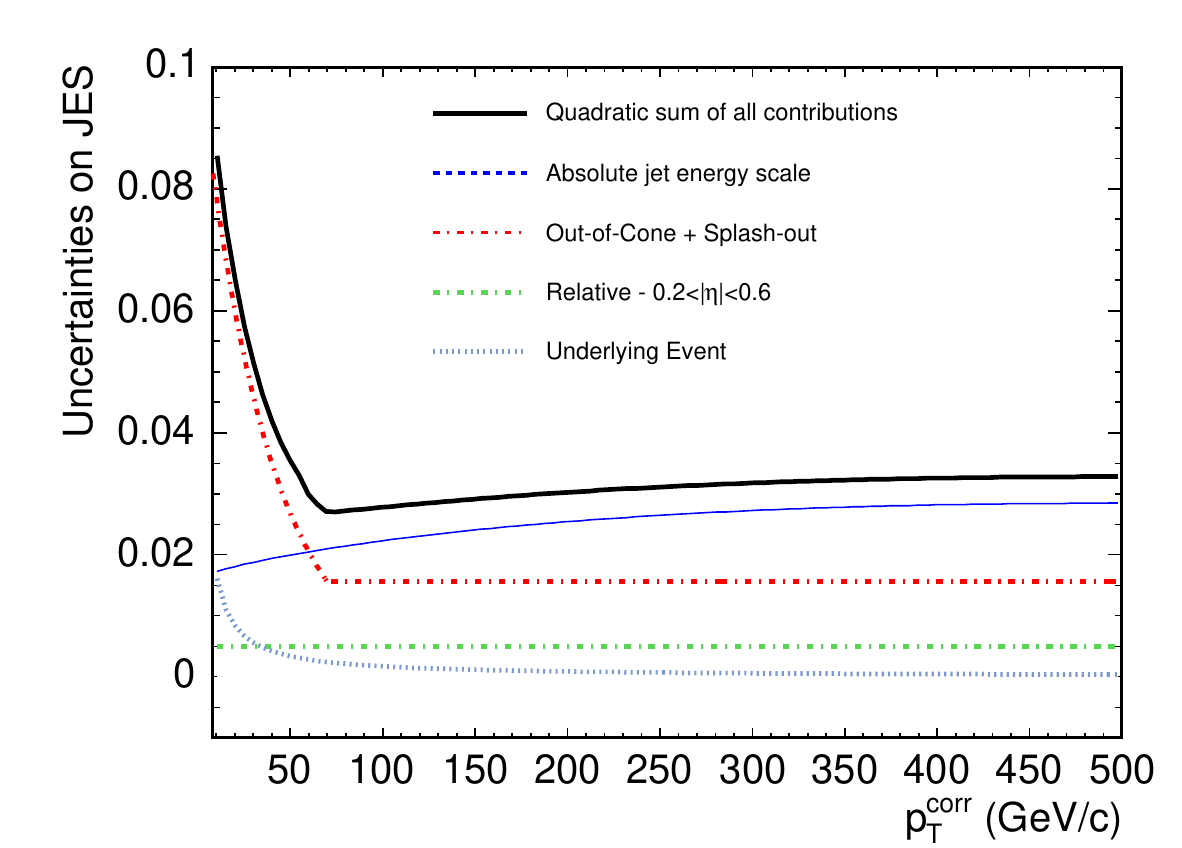,width=.50\textwidth}
\caption{Systematic uncertainties on the jet energy scale
for the CDF experiment~\cite{cdf-jes}.}
\label{jet-sys-cdf}
\end{figure}

The $\eta$-$\phi$ cone radius used by
CDF to reconstruct jets in $t\bar{t}$ events, $\Delta R$ = 0.4,
was chosen to optimize efficiency and to
minimize energy sharing among different jets.
The CDF jet energy resolution is approximately 
$\sigma(E_T)$ = (0.1 ($E_T$/GeV) + 1.0)~GeV
for a 0.7 cone radius~\cite{cdf-jets}. 
The detector response is simulated with the GFLASH 
parametrization~\cite{gflash} interfaced with GEANT3~\cite{GEANT}. 
Test beam data, in addition to pion and electron measurements 
from collider data~\cite{cdf-jes}, are used to tune the GFLASH 
parameters. The $E/p$ distributions at several energies for pions and electrons
are compared to the Monte Carlo expectation to obtain the tuning parameters. 
The jet energy scale is set by the 50~GeV/$c$ test beam data point for pions 
and by the $Z \rightarrow e^+ e^-$ collider data for electrons. 
The jets are corrected for non-linear response of the calorimeter;
poorly instrumented regions; ``multiple interactions'', {\it i.e.}, 
the extra energy from the additional collisions occurring in
the same bunch crossing (pile-up); ``underlying event'', {\it i.e.},
the energy from the 
$p\bar{p}$ remnants; and ``out-of-cone'' energy.
After tuning the simulation to the individual particle response,
an ``absolute'' jet correction is derived from simulation. This 
correction is obtained 
by comparing the jet $p_T$ at the particle level to the jet $p_T$ after 
simulation. The same cone algorithm is used in both cases. The derived 
correction is valid if the number of tracks and the $p_T$ distribution of 
the tracks in data and in simulations are the same. This was verified using 
PYTHIA Monte Carlo (MC) samples with jet $p_T$ up to 600~GeV/$c$.  
The uncertainty of this procedure contributes
1\% to the jet systematic uncertainty, independent of $p_T$.
Calibration, done in the central part of the detector (0.2 $< |\eta| <$ 0.6), 
is extended to all the $\eta$ regions using a dijet imbalance algorithm.
Systematic uncertainties arise 
from uncertainties on the corrections, stability 
of calorimeter response, calorimeter simulation for EM particles, and 
calorimeter simulation for hadrons.  The different components that 
contribute to the jet energy scale (JES) uncertainty are shown in Figure~\ref{jet-sys-cdf}.
The corrections and systematics have been validated using several data samples:
dijet, $\gamma$+jets, $Z$+jets, and with hadronic $W$ boson decays 
in $\ttbar$ events~\cite{cdf-jes}.  

The \D0 \ jet energy resolution is $\sigma(E_{T}$) = 9\% at 100 GeV for a 0.7 
cone radius. The $p_T$ dependence is parameterized as
$\frac{\sigma(p_T)}{p_T} = \sqrt{\frac{N^2}{p_T^2}+\frac{S^2}{p_T}+C}$,
where $N$ determines the magnitude of the noise term, 
$S$ is the stochastic term, and $C$ is a constant.
The three parameters are evaluated at several $\eta$ 
intervals~\cite{voutilainen-thesis}. 
\D0 \ uses the ``midpoint'' version of the
cone algorithm with a radius of 0.5 for top analyses. The multiple interaction 
correction is determined from events taken with a zero bias trigger during 
physics data taking. It accounts for noise and energy pile-up.
 Additional corrections for muons in jets and for
radiation outside of the cone are also made.
The jet response for jets with $\eta < 0.4$  
is derived from a high statistics $\gamma$+jets sample, using 
the $p_T$ imbalance of these events. The photon energy scale is assumed 
to be the same as that of the electrons calibrated with 
$Z \rightarrow e^+ e^-$ 
collider data. The extension beyond  
$\eta = 0.4$ is obtained by $p_T$ imbalance of dijet events, with
one jet in the $\eta < 0.4$ region. 

\begin{figure}[htbp]
\epsfig{file=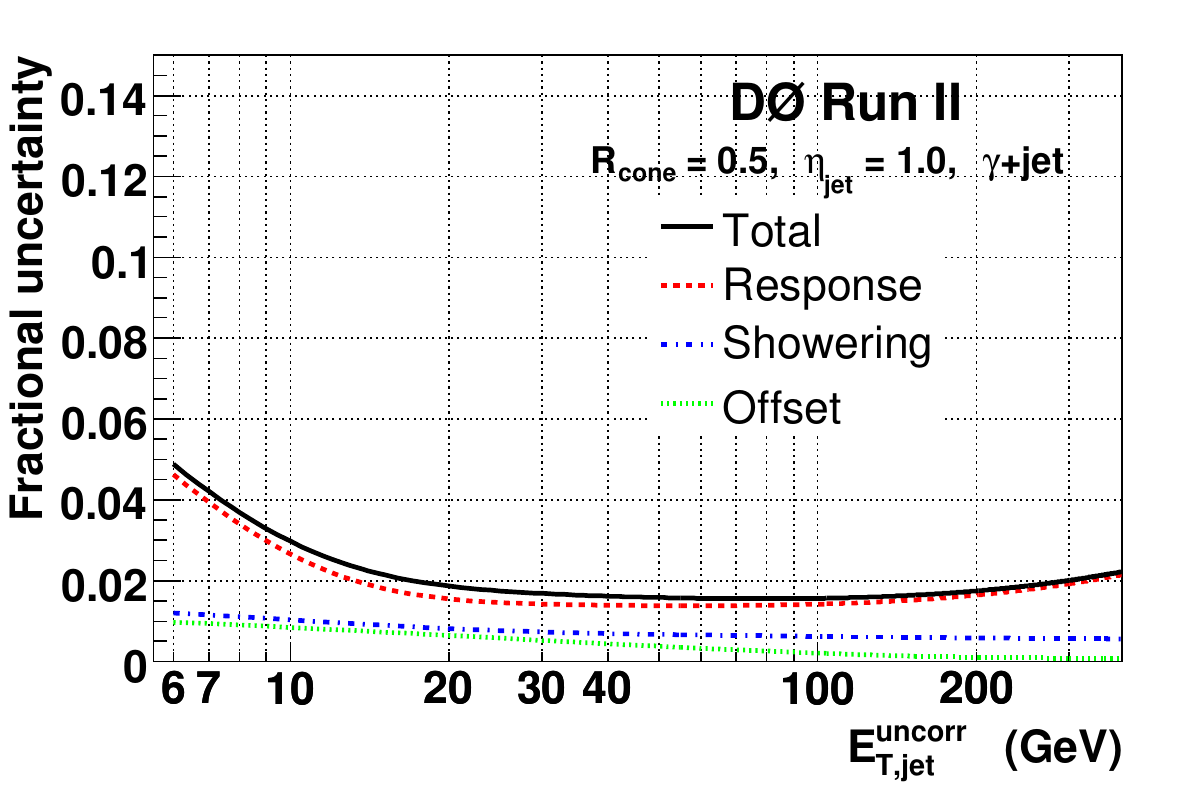,width=.50\textwidth}
\caption{Systematic uncertainties on the jet energy scale for the 
  \D0 ~experiment~\cite{d0-public-web}.}
\label{jet-sys-d0}
\end{figure}

The methods employed for the
top mass measurements use jets of particles obtained with
Monte Carlo generators followed by a GEANT-based
simulation~\cite{GEANT}. The jet
energy scale is evaluated in the data and in the simulation. The
differences are found to be independent from jet $p_T$ and $\eta$ and
are taken into account in the systematic
uncertainties~\cite{d0-jet-jes,ref:d0.ljets.ideogram}. Figure~\ref{jet-sys-d0} 
shows the major components of the systematic uncertainties.

Both CDF and \D0 ~use hadronic $W$ decays
% $W \rightarrow$~jet-jet while performing the top 
% mass measurement  
to determine an overall value of the jet 
energy scale from the $t\bar{t}$ data,
as will be described in more detail in Sections~\ref{sec:systematics} and~\ref{sec:tev_runii}.

%\begin{figure}[htbp]
%\centerline{
%\epsfig{file=jets_cartoon.pdf=.99\textwidth}
%}
%\caption{Schematics of jet reconstruction.}
%\label{jet_cartoon}
%\end{figure}

%%%%%%%%%%%%%%%%%%%%%%%%%%%%%%%%%%%%%%%%%%%%%%5555
\subsection{Tagging $b$ jets}
\label{sec:btag-eff}

One very important ingredient of precision top mass measurements is
identification (tagging)
of heavy flavor jets. It is essential for suppressing  
backgrounds from QCD processes as well as 
for reducing the combinatorics in top events with high jet multiplicity. 

\begin{figure}[htbp]
\centerline{
\epsfig{file=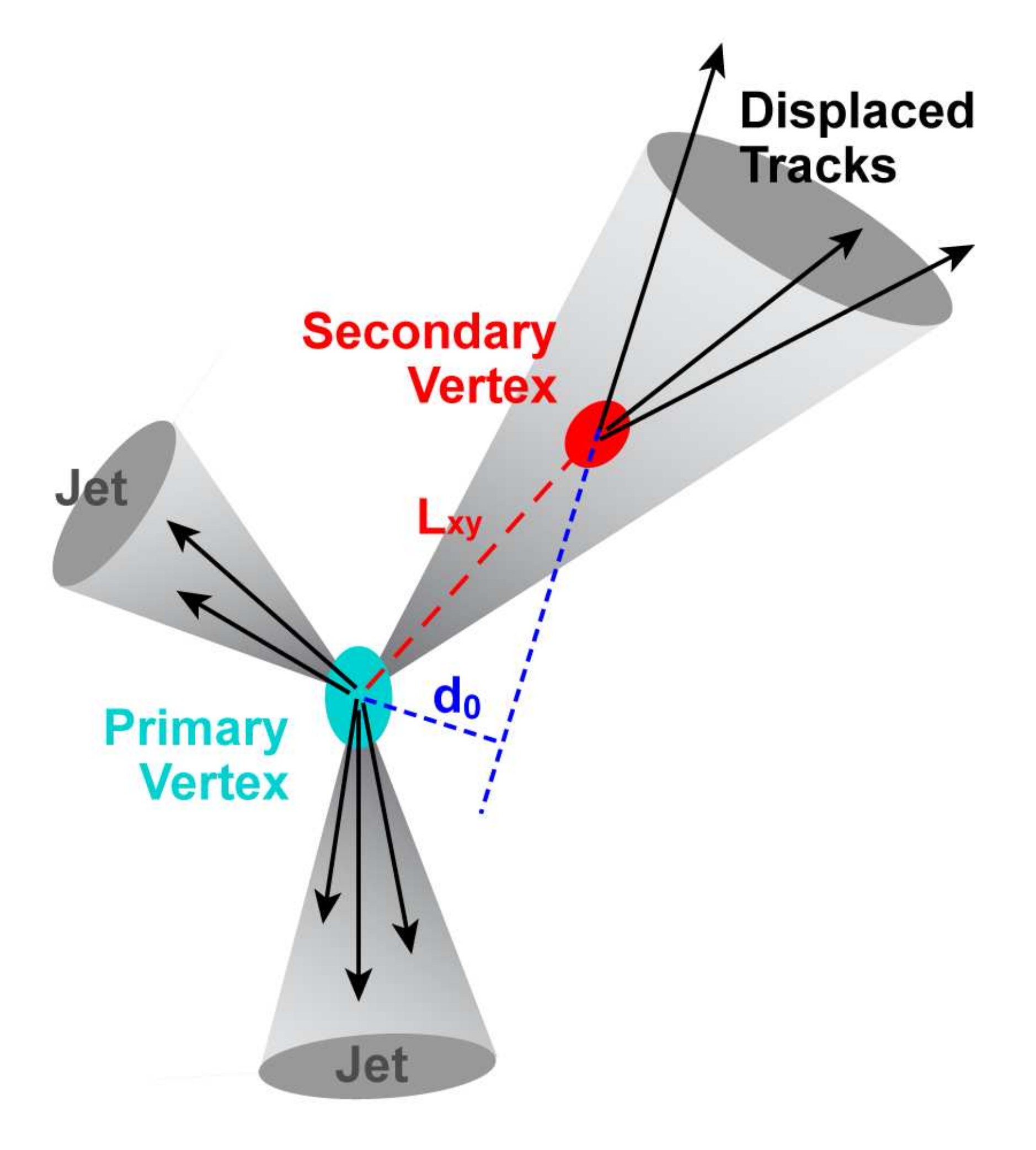,width=.3\textwidth}
}
\caption{Secondary vertex reconstruction. L\sub{xy} is the distance of
  the secondary vertex from the primary vertex in the plane
  orthogonal to the proton beam direction. The
  impact parameter of a track is marked as d$_0$. }
\label{tev-b-detect}
\end{figure}

Tagging of 
heavy flavor jets is done in several ways: (a) by identifying muons or
electrons originating from bottom or charm
hadron decays within the jet;
(b) by reconstructing a secondary vertex in jets, due to the 
decay of a heavy flavor hadrons;
% (c) by evaluating the probability for a jet to be a heavy flavor jet 
(c) by identifying tracks within the jet that originate away from the primary vertex.
% (c) reconstructing a secondary vertex in jets, due to the 
% decay of a heavy flavor hadrons. 
Most of the top mass measurements performed at the 
Tevatron use methods (b) and (c)
for $b$-jet identification, {\it i.e.}, use the L\sub{xy} distance of the
secondary vertex or the impact parameter (d$_0$) of each track,
as illustrated in Figure~\ref{tev-b-detect}.

\begin{figure}[htbp]
\epsfig{file=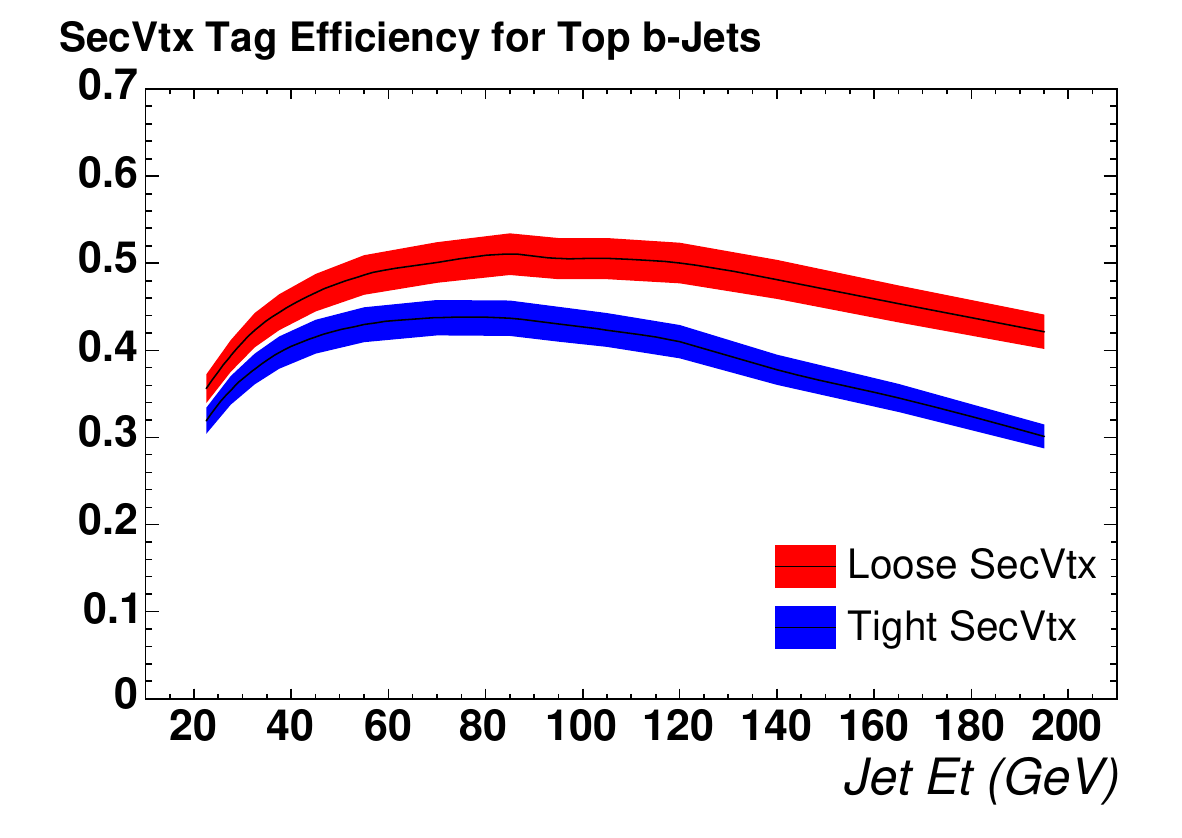,width=.5\textwidth}
\caption{The CDF $b$-tag efficiency for ``tight'' and ``loose'' $b$-tagging 
   requirements~\cite{cdf-btag-eff-mistag}.
   The average $E_T$ of $b$ jets in $\ttbar$ events 
   produced at the Tevatron is $\sim 70$~GeV.
}
\label{cdf-btag-eff}
\end{figure}

The impact parameter resolution of the CDF tracking system is 35 $\mu$m
for tracks with $p_T > 2$~GeV/$c$~\cite{cdf-pt-hres}.
\D0 \ tracking provides impact parameter measurements with respect
to the primary vertex with a precision between 20 and 50 $\mu m$,
depending on the number of hits in the silicon detector~\cite{nim-d0-btag}.
The reconstruction of the secondary vertex follows a similar procedure
for the CDF~\cite{cdf-btag} and \D0  ~\cite{nim-d0-btag} algorithms. A
primary vertex for the event is reconstructed using tracks with a
small impact parameter significance, $d_0/ \sigma_{d_0}<$ 3. The tracks 
in jets within a cone of $\Delta R <$ 0.4 (CDF) or 0.5 (\D0 ) and
with large impact parameter significance $d_0/ \sigma_{d_0}>$ 3.5 
are used to find a secondary vertex. Additional criteria are 
set for the significance of the L\sub{xy} measurement. A jet is tagged as 
a heavy-flavor jet if all criteria, including a good fit of the secondary 
vertex, are met. The probability that the displaced vertex is produced
by a charm jet is
also evaluated, assessing the invariant mass of the tracks that form the
secondary vertex.  Both MC and data samples are used to evaluate the 
efficiency of the tagger in top events and the mistag rate, {\it i.e.}, 
the probability that a light quark jet is tagged as a heavy flavor jet. 
The CDF $b$-tagging efficiency as a function of jet $E_T$ is shown in 
Figure~\ref{cdf-btag-eff} ~\cite{cdf-btag-eff-mistag}. The efficiency 
for a ``tight'' $b$-tagging requirement used for the top mass measurements
is about 40\%. The mistag rate is increasing from
0.4\% at 15 GeV to 2.3\% at 180 GeV.

%\begin{figure}[htbp]
%\epsfig{file=cdf_mistag_eff_ttbar.pdf,width=.5\textwidth}
%\caption{The CDF mistag rate for the  ``tight'' and ``loose'' tagging 
%   requirements~\cite{cdf-btag-eff-mistag}.}
%\label{cdf-mistag}
%\end{figure}

\begin{figure}[htbp]
\centerline{
\epsfig{file=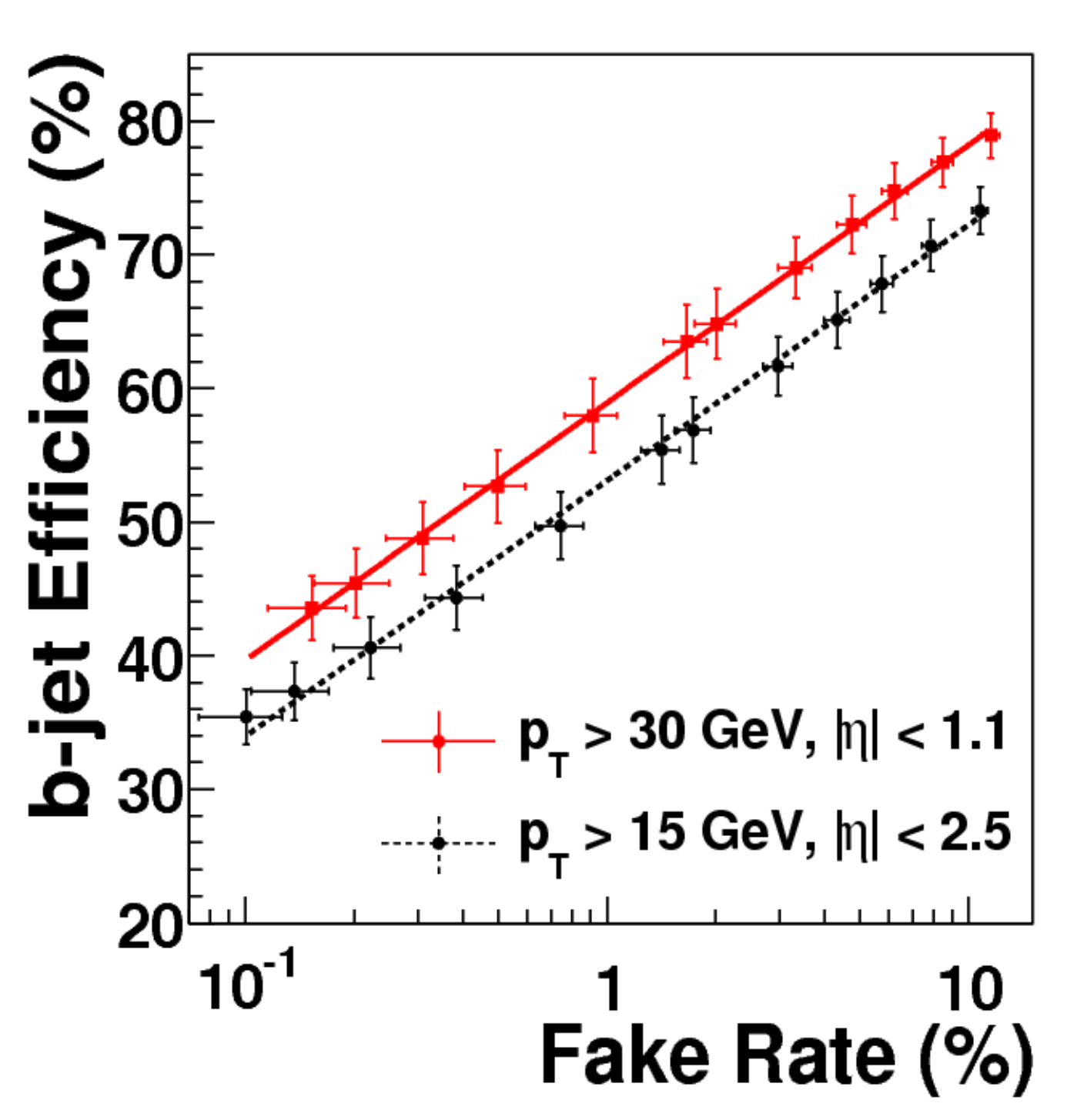,width=.45\textwidth}
}
\caption{\D0 ~$b$-tag efficiency vs. fake rate for two 
         $p_T$ and $\eta$ regions~\cite{nim-d0-btag}.}
\label{d0-btag}
\end{figure}

Both CDF and \D0 ~ have developed more sophisticated algorithms for
$b$-tagging. They use more information than just the secondary vertex
fit result. These neural network (NN) algorithms help reduce the
background from light quark jets that mimic heavy flavor jets. The CDF
algorithm~\cite{keung-cdf} employs other variables in addition to secondary 
vertex information, but has not been used for top mass
measurements. The  \D0 ~NN algorithm~\cite{d0-btag-algo} was 
used for top mass measurements. It takes into account information from impact
parameter measurements, the secondary vertex tagger, and the soft
lepton tagger (SLT). The SLT attempts to identify semileptonic decays
within a jet. Separate efficiencies for $b$ and $c$ quark jets as a function
of $p_T$ and $\eta$ are obtained, and the mistag rate as a function
of the same variables is evaluated~\cite{d0-btag-algo}. The NN
is trained on QCD $\bbbar$
jets and on light quark Monte Carlo samples, and its performance is
measured in data. Figure~\ref{d0-btag} shows the fake (mistag)
rate-vs-efficiency for jets in two different regions of $p_T$ and $\eta$.
   
\subsection{Major background sources}
\label{sec:backgrounds}
There are two substantially different
types of backgrounds: physics processes that have 
topologies similar to $t\bar{t}$ events and instrumental backgrounds
originating from processes that have incompatible final states
but can nevertheless appear top-like due to detector imperfections.
Physics backgrounds are difficult to eliminate completely, but
their rates can be usually reduced by imposing requirements that
exploit the differences between signal and background
kinematic distributions.
Modeling of physics backgrounds
is discussed in Section~\ref{sec:backg}. Instrumental backgrounds are
caused by 
misidentification of charged leptons, neutrinos ($\met$), or $b$ jets.
These backgrounds can be reduced by stringent requirements on particle
identification criteria (charged leptons), topological cuts for the   
$\met$ (neutrinos), and sophisticated algorithms for $b$-tagging.  

Backgrounds are different for the three major topologies
in which the $t\bar{t}$ candidate events are found. A brief
description of the major backgrounds in each of these topologies
is provided below. Additional details will be given
in  Section~\ref{sec:tev_runii}, where the actual top mass measurements
are discussed.

%\begin{figure}[htbp]
%\epsfig{file=sigma-vs-mhiggs.pdf,width=0.40\textwidth} 
%\caption{Cross section for different processes in $\ppbar$
%     collisions at $\sqrt{s} =$ 1.96 TeV. The scale on the right shows all
%     cross sections relative to the $\ttbar$ cross section set at 1. }
%\label{sig-all-mh}
%\end{figure}

The background in the dilepton topology (e and $\mu$) is mostly due to 
Drell-Yan processes ($Z/\gamma ^* \into \ell^{+}\ell^{-}$); diboson production: 
$WW$ and $WZ$ where a lepton from the $Z$ decay is lost and appears as
$\met$ (oppositely charged leptons of the same type
coming from $Z$ decays are rejected with a
$Z$ mass veto);  $W + \ge 2$~jets
(where one jet mimics an electron or a muon); single top production (lepton
+$\ge$~3~jets where one jet mimics an electron). In addition, ``fake'' muons 
can originate from punch-through or from hadrons decaying in flight.
Details of instrumental backgrounds for electron and muon identification
can be found in~\cite{ref:cdf.template.latest,ref:d0.dilep.run2b}.
The $t\bar{t}$ decay chains with a $\tau$ lepton are included
in the $\ell$+jets or dilepton channels only if the $\tau$ decays
leptonically (in $\sim$ 35\% of the cases). Studies of the processes
involving the $\tau$ lepton are discussed in more detail
in Section~\ref{sec:metjets}.
%For the tau lepton channels the major background comes from QCD multijet 
%events where the energy is mismeasured giving rise to large $\met$. 

\begin{figure}[htbp]
\includegraphics[width=.45\textwidth]{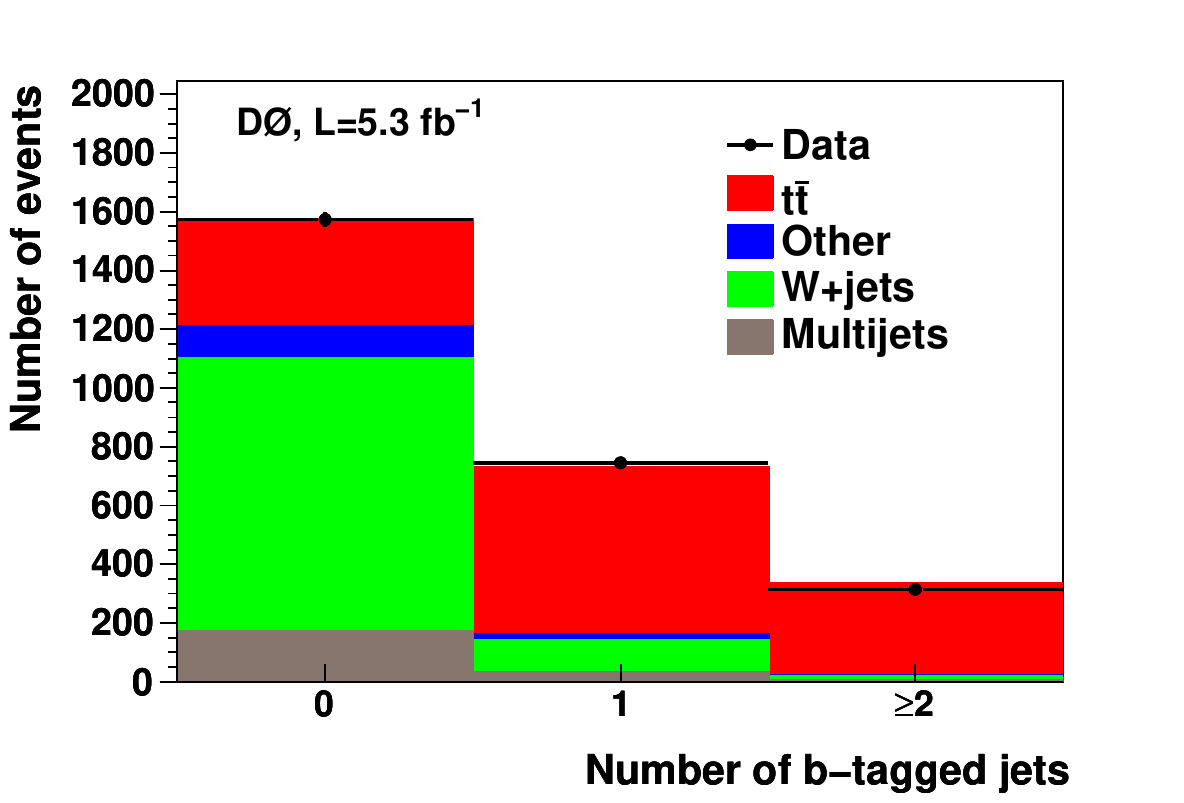}
\caption{Number of candidate $\ttbar$ events vs. the number of tagged jets
in the $\ell$+jets topology for events with 
at least four energetic jets~\cite{Abazov:2011mi}.}
\label{top-njet}
\end{figure}

The major background in the $\ell$+jets topology is $W$+jets
production which has a much larger cross section than $\ttbar$ production.
This background is further classified into
$W$+light flavor and $W$+heavy flavor.
The first one can be reduced considerably by requiring one or more jets
to be tagged as a $b$ jet. 
Another large background (often called non-$W$ QCD) is due to multijet 
production, where a jet fakes an electron or a muon.
Rate estimations for these backgrounds are performed with
data driven methods, developed for top cross 
section measurements. Details can be found in~\cite{cdf-wjets,Aaltonen:2010ic} 
for the CDF analyses and in~\cite{d0-btag-algo} for \D0 \ analyses. 
A good review of these methods can be found in~\cite{tev-tprop}.
Other instrumental backgrounds
are due to mismeasurements of jets which occur in poorly instrumented regions
of the detector and increase the value of $\met$. Requiring a certain
minimum angle between the jet and the $\met$ direction helps to suppress
these backgrounds.
Finally, diboson production, {\it i.e.}, $WW$, $WZ$ and $ZZ$
final states with the same topology as $\ttbar$, and single 
top production also contribute to the background. The dependence of
the background rate on the number of $b$-tagged jets can be seen in
 Figure~\ref{top-njet}.
% taken from ~\cite{Abazov:2011mi}. 
See Section~\ref{sec:lepjets} for the 
contributions of each of these backgrounds to the candidate event
samples after application of all selection requirements.

QCD multijet production is
the major background in the all-hadronic topology.
Without $b$-tagging requirements, this background is 
three orders of magnitude larger than the signal.
%for the simple cuts 
%mentioned below. 
%In addition, the presence of six jets in this topology requires 90 
%permutations giving two jets corresponding to $W$ bosons and three
%jets corresponding to two top quarks. Moved to page 51.  
The sample signal fraction is significantly improved by
demanding that at least two jets are 
tagged as a $b$ jet. Additional kinematic requirements,
{\it e.g.}, a neural network selection which combines a number of 
discriminating variables,
are used to suppress the background further,
as described in Section~\ref{sec:all-had}.

Table~\ref{sig-back} shows the signal-to-background ratio (S/B)
for simple event selection criteria as a rough estimate. 
For the dilepton and $\ell$+jets channels: 
$E_T(leptons) > 20$~GeV, $\met > 20$~GeV, jets with $E_T > 20$~GeV are 
required; for the all-hadronic final state
$E_T > 15$~GeV for the 6-8 jets in the event
is required. 
% The dilepton channel includes tau decays into charged leptons (e, $\mu$). 
The values of S/B for the $\ell$+jets channel can be improved using
additional kinematics requirements, similar to those used
for the all-hadronic topology.
% is added in the form of a likelihood
%~\cite{Abazov:2011mi,cdf-sigma-likely}, 
% as mentioned above for the all-hadronic topology.
See Section~\ref{sec:tev_runii} for the background rates achieved in
various analyses.

\begin{table*}[hbtp]
\caption {S/B for different channels. All samples have a nominal \Mt\ 
  of 175~GeV/$c^2$.}
\begin{center}
\begin{tabular}{c|c|c|c}
\hline \hline
{ Sample} &  Dilepton & $\ell$+jets & All-hadronic \\
       &{  (e,$\mu$)}&{ (e,$\mu$)}   &{ NN selection} \\
%\hline \hline
 { BR}      &{ 6.4\%} &{ 34\%} &{ 46\%} \\
 { Signature} &{ 2$\ell$+2$\nu$+2 jets} &{ 1$\ell$+1$\nu$+4 jets}&{ 6 jets} \\
%\hline 
{ Major Back.} &{ DY, $W/Z$+jets}  &{ QCD, $W/Z$+jets} &{ QCD multijets} \\
\hline
{ 0-$b$-tags S/B} &{  1/1}  &{ 1/4}  &{ 1/20}  \\
{ 1-$b$-tags S/B} &{  4/1}  &{ 4/1}  &{ 1/5}   \\
{ 2-$b$-tags S/B} &{  20/1} &{ 20/1} &{ 1/1}   \\
\hline  
{ Events in 1 fb$^{-1}$} & {25} & 180 & {150 (2 b-tags)} \\
  ( {$\ge$ 1 b-tag)}& & & \\
\hline
\hline
\end{tabular}
\end{center}
\label{sig-back}
\end{table*}[hbtp]

In summary, the $\ell$+jets channel
% {\it a priori}
is the most propitious
for measuring the mass of the top quark at the Tevatron.
With the $b$-tagging requirement, there are 
three signatures for signal identification (a charged lepton, large $\met$,
$b$ hadrons 
in the event). With the appropriate selection, the S/B ratio can be 
reasonably high, and from the branching ratios 34\% of $t\bar{t}$
events are expected 
to manifest themselves as e, $\mu$ + jets. In the dilepton channel there are
fewer events but the S/B ratio is good even without $b$-tagging.
% When the statistic will be high, 
Tagging of $b$-jets can help increase the
S/B ratio even further. Finally, the top mass measurement in the
the all-hadronic channel, with optimal 
choice of variables used for background discrimination, has recently surpassed 
the precision obtained with the dilepton events (see Section~\ref{sec:all-had}).

\section{Monte Carlo modeling of signal and backgrounds}
\label{sec:mc_models}

To perform a precision measurement, it is important   to have the 
most complete understanding of the physics under study as well as 
the best possible description of the
detector response formalized through its computer simulation. The CDF
and \D0 \  collaborations use Monte Carlo
generator programs to model a number
of physics processes relevant to $t\bar{t}$ production and decay:
the hard scattering of the incoming partons,
multiple parton interactions, underlying event arising
from the remnants of the $\ppbar$ system, color reconnection of final
state partons, multiple interactions which happen in the same
bunch-crossing, parton shower, and subsequent hadronization.
Events are subsequently passed through a complete
detector response simulation. The resulting simulated
samples are treated just like the recorded  $p \bar p$
collision data, using the same reconstruction software and particle 
identification algorithms.
% (described in Section\,\ref{sec:top_id}).

A variety of Monte Carlo generators are used to model the signal and
the background and thereby to estimate the event selection efficiencies, 
kinematic distributions, {\it etc.} Among the generators that include all processes 
mentioned earlier are PYTHIA~\cite{PYTHIA} and HERWIG~\cite{ref:herwig}.
These are leading order tree-level generators of $2 \to 2$ parton
processes. The parton shower is modeled using DGLAP 
(Dokshitzer-Gribov-Lipatov-Altarelli-Parisi)
evolution~\cite{Gribov,Altarelli,Dokshitzer},
while the hadronization process is simulated
using a string model in PYTHIA and a cluster model 
in HERWIG. Decays of tau leptons are simulated using
the TAUOLA library~\cite{TAUOLA}. Decays of hadrons containing $c$ and
$b$ quarks are handled with the EVTGEN package~\cite{EVTGEN}. Other
generators, such as ALPGEN~\cite{ALPGEN} and MadEvent~\cite{MADEVENT},
generate only the hard scattering reaction and rely on PYTHIA or HERWIG
for simulation of other processes.

\subsection{Modeling of $t \bar t$ events}
\label{sec:ttbar}
The CDF collaboration employs PYTHIA\,v6.2 as its 
default Monte Carlo generator to model $t\bar{t}$ production and decay
as well as parton fragmentation and hadronization. The \D0 \  collaboration uses instead ALPGEN\,v2~\cite{ALPGEN} in which the probability to radiate additional partons is calculated according to the tree-level
matrix elements of the relevant processes. The subsequent showering and hadronization is performed
with PYTHIA which requires the use of a matching scheme to avoid double-counting of possible final states~\cite{Mangano:2006rw}. Other Monte Carlo packages,
such as HERWIG, MadEvent~\cite{MADEVENT} and MC@NLO~\cite{MCATNLO},
are used  
for studies of systematic uncertainties.
The underlying event is modeled with PYTHIA, with
relevant parameters tuned to reproduce CDF Run~I and Run~II observations. 
This set of parameters is commonly referred to as
Tune~A~\cite{pythia_tune_a_better,pythia_tune_a}.
%Experimental investigations at LEP in $WW$ hadronic decays did not find conclusive evidence of color reconnections
%among final state particles~\cite{Azzurri}. 
PYTHIA\,v6.4 includes color reconnections models tuned to collider
data~\cite{Skands:2009zm} that have been employed in systematic
uncertainty studies.

%The cross section have been scaled to the NLO theoretical cross section predictions~\cite{boh!}.
% The detector response to all particle processes has been modeled with a 
% GEANT-based detector simulation (see Section~\ref{sec:jet-rec}).
 Multiple hadron-hadron interactions within the same bunch-crossing are modeled with 
PYTHIA by the CDF collaboration, while the \D0 \  
collaboration takes extra interactions into account
by overlaying data from random 
$p \bar p$ crossings on top of the MC events. 
CDF normally uses the CTEQ5L~\cite{CTEQ5L} set of
parton distribution functions (PDFs), while \D0 \  
uses the CTEQ6L1 set~\cite{Pumplin:2002vw}.
Other PDFs used are CTEQ6M~\cite{Pumplin:2002vw},
MRST72 and MRST75~\cite{Martin:1998sq}.

\begin{figure} 
\includegraphics[width=.40\textwidth]{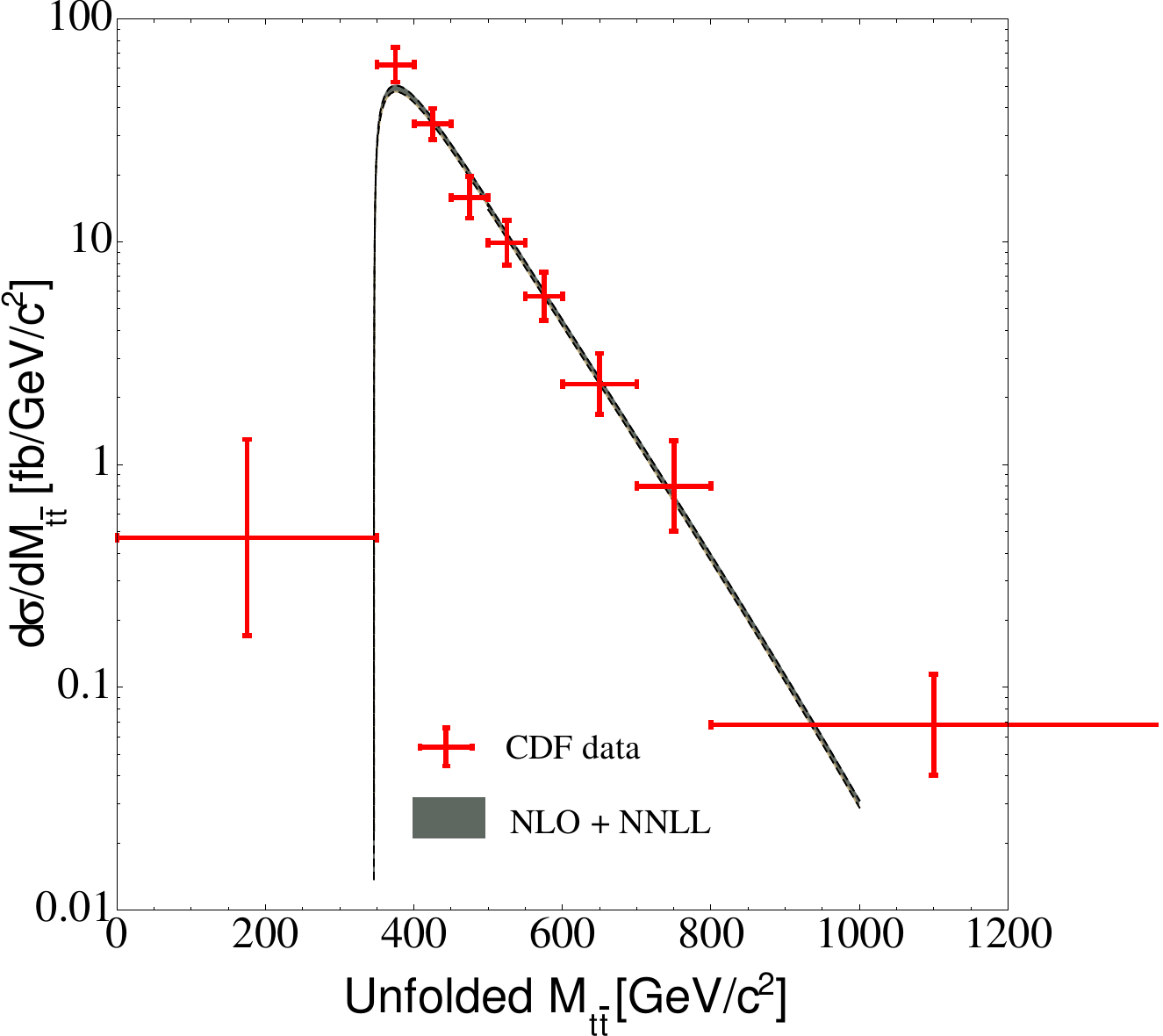} 
\includegraphics[width=.40\textwidth]{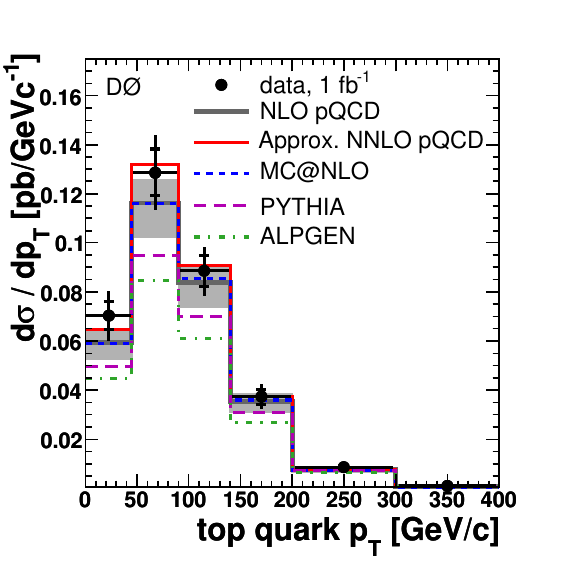} 

\caption{Example tests of the top quark pair production modeling at CDF and \D0 . The left plot shows the differential $\sigma_{t \bar t}$ as a function of the invariant mass of the $t \bar t$ system, together with the approximate NNLO SM predictions~\cite{CDFdiff,Ahrens-NNLL}. The plot on the right shows
the ratio between data and NLO theoretical prediction of the distribution of the top quark $p_T$, together with approximate NNLO predictions and expectations from several Monte Carlo generators~\cite{D0diff}.

%The plot on the left shows the measurement of $d \sigma/ d$.
}
\label{fig:diff} 
\end{figure}

\begin{figure} 
\includegraphics[width=.35\textwidth]{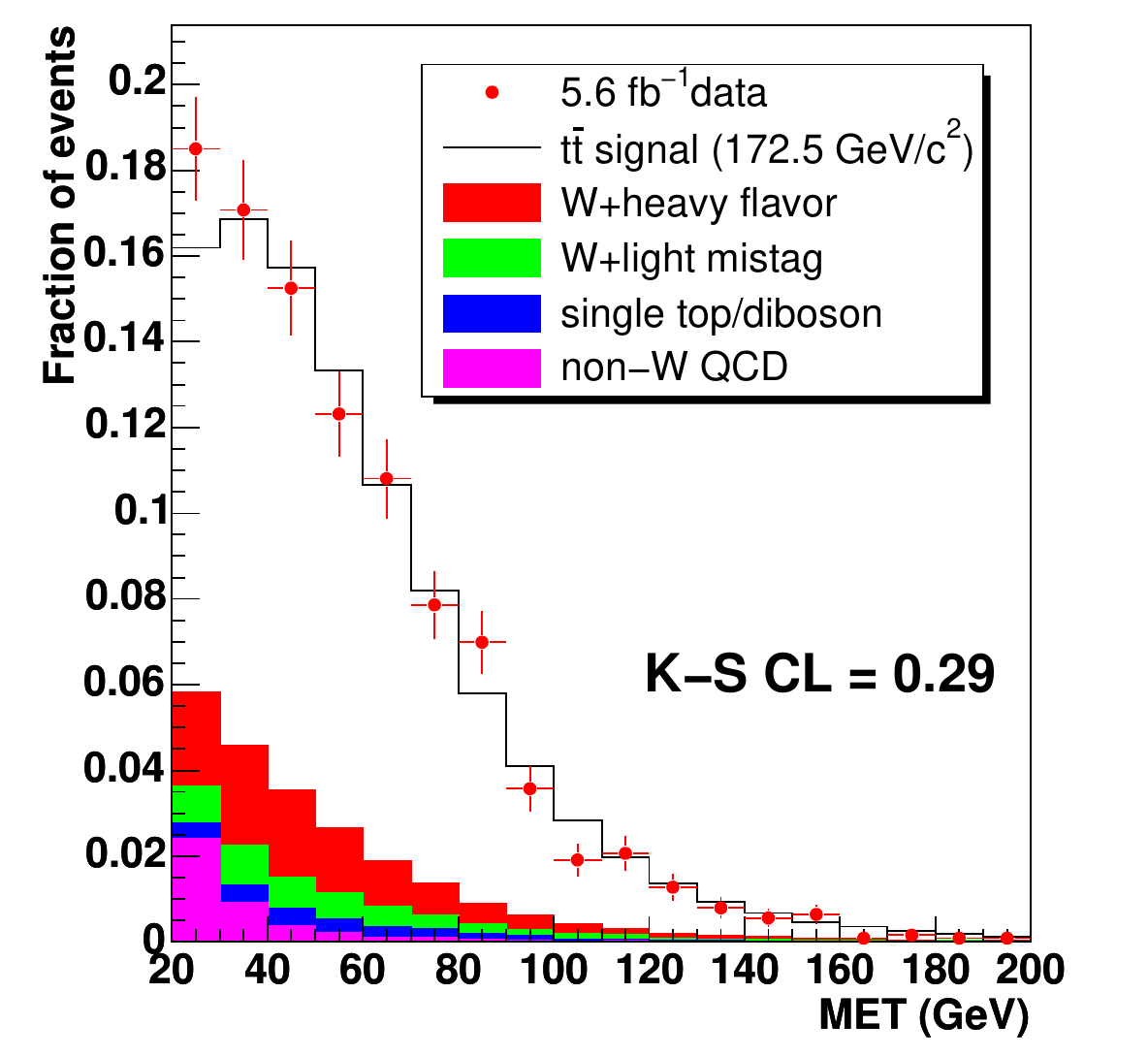} 
\includegraphics[width=.44\textwidth]{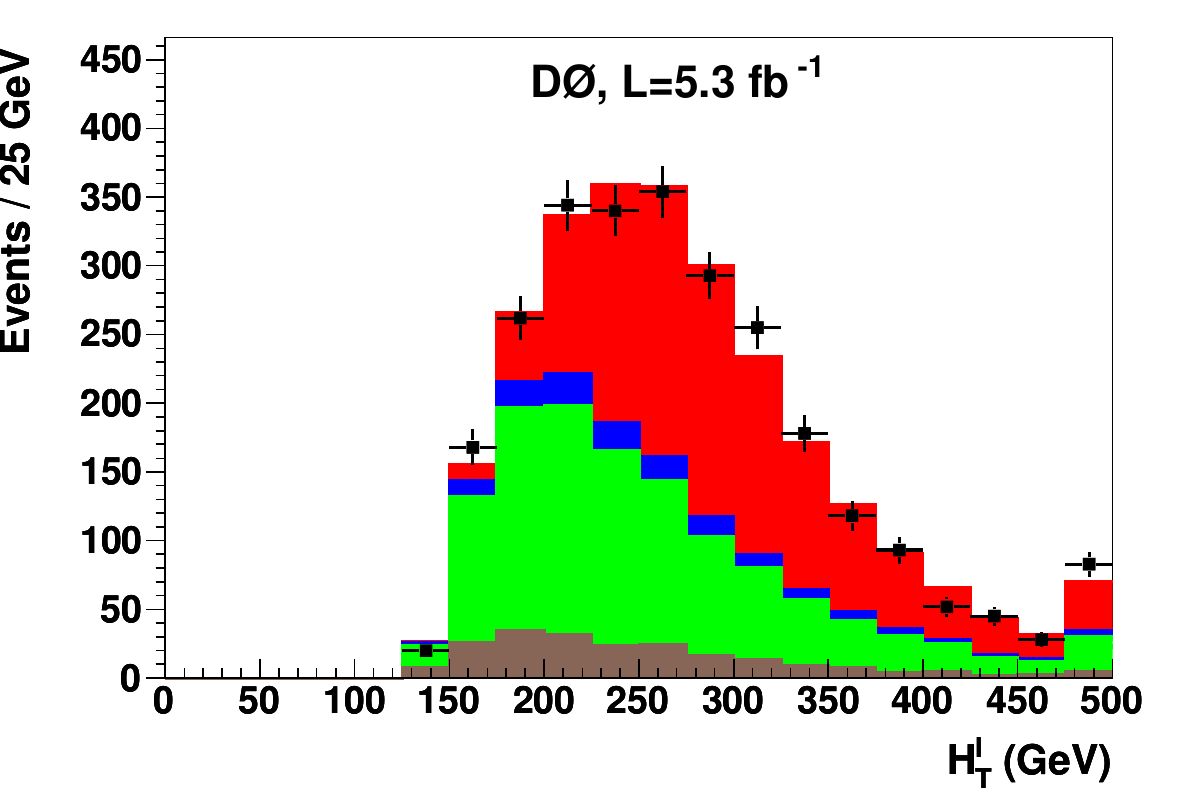} 
\caption{Distribution of the $\met$ (left plot) modeled in the CDF top mass 
measurement~\cite{ref:plujanThesis} and $\mathrm{H_T^l}$ (right plot) in 
top-antitop events in the $\ell$+jets final state in the \D0 \  top cross 
section measurement. The $\mathrm{H_T^l}$ variable
is defined as the scalar sum of 
the transverse energies of the jets
and of the charged lepton~\cite{Abazov:2011mi}.}
\label{fig:kin} 
\end{figure}

The measurements described in this Review use up to 5.8~fb$^{-1}$ of $p \bar p$ collision data.
This corresponds to about $4 \times 10^4$ $t \bar t$ events produced at each detector site. The size of this dataset allows for a thorough
validation 
% of the quality 
of the signal kinematics and acceptance modeling. 
Measurements of the inclusive $t \bar t$ cross section~\cite{Aaltonen:2010ic, Abazov:2011mi, Aaltonen:2010bs, Abazov:2009si, Aaltonen:2010pe, Abazov:2009ss, Aaltonen:2011tm} indicate that applying an NLO-to-LO K-factor to the PYTHIA/ALPGEN 
simulations results in a very good agreement between predicted
and observed rates of the $t \bar t$ signal and background processes.
% See Figure~\ref{sig-mass} in Section~\ref{sec:top-sig}.
Measurements of $t \bar t$
differential cross sections~\cite{CDFdiff,D0diff} attest to the correctness of
recent Monte Carlo predictions which incorporate
NLO and approximate NNLO computations,
as illustrated in Figure\,\ref{fig:diff}. 
%CDF and \D0 \  also measured several other properties of top quark events with varying sensitivity and found good-to-excellent agreement with the Standard Model predictions~\cite{Aaltonen:2010js,Aaltonen:2010nz,Abazov:2011qu,Aaltonen:2010ea, Abazov:2010tm,Aaltonen:2008ei,Abazov:2010jn,Acosta:2005hr,Abazov:2008yn}. 
Consistency between observations and Monte Carlo predictions has been
verified for a large number of experimental distributions, with two examples
 provided in Figure~\ref{fig:kin}.
%
%There are only two observables where the current Monte Carlos fails to describe the data: the distribution of the transverse momentum of the $t \bar t$ system, and the forward backward asymmetry of top-antitop events~\cite{Aaltonen:2011kc}. The former is due to a lack of a tree-level description of the initial state radiation that reflects into an insufficient boost in the transverse plane of the top quark pairs; NLO Monte Carlo provides a better agreement than the nominal PYTHIA/ALPGEN. In the latter case, LO theory predicts zero asymmetry, and at NLO an asymmetry of about 5\% has been computed, but data show an asymmetry much larger than the NLO (and approximate NNLO) calculations. This anomalous kinematic effect is currently under deep investigation. Top quark mass measurements have been found not to be sensitive to a discrepancy between data and Monte Carlo in the two observables above.
%
\subsection{Modeling of the physics backgrounds}
\label{sec:backg}
The CDF and \D0 \  collaborations employ several Monte Carlo generators
to model $t \bar t$ backgrounds processes.
% which are described in this section. The $W$ and $Z$ 
% plus jets production cross sections
% are sufficiently large so that these processes 
% constitute a major background to most top-antitop decay modes. 
% The uncertainty on the theoretical cross section for 
% $W$+jets production is large because calculation beyond NLO are unavailable. 
% Current estimates rely upon a mixture of partial calculation 
% and parton shower Monte Carlo models to extrapolate to 
% large jet multiplicities.
Both collaboration use ALPGEN for simulating $W$+jets production,
%v2.10 prime (with latest updates from MLM)
  with parton showering and hadronization performed by PYTHIA.
The event samples are generated with the 
renormalization and factorization scale, $Q^2$,
set to $M\sub{W}^2  + \sum_{partons} p_T^2$.
% where ${p_T}$ 
% is the transverse momentum of the generated parton. 
%For light partons, $u,d,s,g$ the mass $m$ is approximately zero; $m_b$ is set to 4.7\,GeV/$c^2$ and $m_c$ is set to 1.5\,GeV/$c^2$. 
The overall normalization of the $W$+jets background is obtained from data by subtracting other physics and instrumental backgrounds and the 
$t \bar t$ signal as a function of jet multiplicity. 
The $W$+jets background is subdivided into three exclusive categories 
according to the flavor of the partons produced in association with 
the $W$ boson: 
(i) $W+b \bar b/c \bar c$ is the agglomeration
of all final states which include the $W$ boson, the $b \bar b$ or $c \bar c$ quark pair, and any number of additional jets; (ii) $W + c$ consists of events with a $W$ boson produced with a single charm quark and any number of additional jets; and (iii) events in which the $W$ bosons are produced together with light flavor jets. The relative contributions from these three classes of events are determined using NLO QCD calculations based on the MCFM generator~\cite{Campbell:2002tg}.

% Other physics background are $Z/\gamma^*$+jets production, dibosons, and single top quark production.
The diboson processes ($WW$, $WZ$, $ZZ$) are simulated using PYTHIA.
Single top quark $s$- and $t$-channel production is simulated with
COMPHEP~\cite{Boos:2004kh} for \D0 \  measurements, and with
MadEvent~\cite{MADEVENT} for CDF measurements.
The contributions from these background sources are normalized to the corresponding NLO predictions.

QCD multijet production is the major background to $t \bar t \to q_1 \bar q_2  b q_3 \bar q_4 \bar b$.
The production of six or more partons is poorly understood at the theoretical level. In addition, the large multijet production cross section
combined with the powerful background rejection needed to isolate the all-hadronic $t \bar t$ signal translates into a computationally prohibitive demand
to generate an enormous number of events.
For the above reasons, the multijet background is modeled using data in a region depleted of $t \bar t$ signal.
Application of $b$-tagging techniques is necessary in the all-hadronic channel to achieve a reasonable S/B ratio.
CDF uses collisions with at least six jets in the final state
%before $b$-tagging 
to model the QCD background, 
and corrects for the bias introduced by $b$-tagging using a parametrization of the $b$-tag rate derived with an independent dataset devoid of QCD events. 
%D0 creates a background sample by attaching low-$p_T$ jets selected from events  with six or more jets, to events with four or five jets. 
%In both instances 
The validity of this model is tested in large-statistics control samples deprived of top-antitop events. 

\section{Data analysis techniques}
\label{sec:meas_techs}

In this section we utilize the following notation.
The symbol ${\bf x}$ is used to denote the collection of parton-level
kinematic quantities needed in order to completely define an inelastic
hard scattering reaction at
the leading order perturbation theory ({\it i.e.,} to 
specify a point in the reaction phase space).
In all data analysis methods described, it is always assumed that points
${\bf x}$ in different collisions are independent and identically distributed (i.i.d.).
The symbol ${\bf y}$ refers to one or more
``observed'' quantities inferred from detector data on event-by-event basis.
Both the probabilistic nature of ${\bf x}$ and
the nondeterministic detector response contribute
to the randomness of ${\bf y}$. Even though we call ${\bf y}$
an ``observed'' quantity, it is usually a product
of a sophisticated event reconstruction procedure
which involves pattern recognition, tracking, clustering of jets,
kinematic fitting, {\it etc.} It is commonly assumed that
values of ${\bf y}$ are also i.i.d., and efforts are made to ensure effective i.i.d.
behavior if this is not the case\footnote{For example, both CDF and \D0 \ 
apply jet energy corrections
which depend on the number of primary vertices in the event (and, indirectly, on instantaneous
luminosity).}.
% The symbol $m_t$ denotes an estimate of the top quark mass
% obtained in a single collider event.

\subsection{Major issues for different $t \bar{t}$ topologies}

In the process of analyzing collision data for the purpose of measuring \Mt,
several important decisions have to be made about
statistical and computational techniques applied,
signal and background modeling, measurement calibration,
trade-off between expected statistical 
and systematics uncertainties, {\it etc.}
Different techniques adopted by different authors
resulted in a variety of substantially distinct
data analysis approaches, with both complementary and competing features.
In each approach,
the following major issues have to be considered and
addressed in a consistent manner:
\begin{mylist}
% \item Degree to which the data analysis procedure relies on 
%      the standard model of particle interactions.
\item Choice of the $t \bar{t}$ final state. The traditional
      taxonomy of dilepton, $\ell$+jets and all-hadronic
      final states dictated by distinct event kinematics and different levels
      of background contamination is maintained by most authors, although joint
      measurements which combine events with different final states
      started to appear in the literature~\cite{ref:cdf.template.pub1,
      ref:cdf.joint.3.2fb, ref:cdf.template.latest}.
      The main advantage of such joint measurements is the ability to apply
      the detector jet energy scale calibration obtained in the final
      states with hadronic $W$ decays to the dilepton channel.
\item Event sample selection. Less strict event selection criteria
      can result in improved statistical uncertainty at the cost of
      increased background fraction and more complicated calibration.
\item Degree to which the method depends on the calibration of the
      detector jet energy scale --- the largest source
      of systematic uncertainty in the most precise
      CDF and \D0 \  measurements of \Mt.
\item Level of detail in the statistical model of detector response 
      to jets (for methods which do rely on jet energy reconstruction).
\item Mapping of the objects reconstructed in the detector to
      the leading order parton-level entities. The
      difficulty here is due to the fact that
      the charge and the flavor of jets produced
      in top decays and associated processes ({\it e.g.,} initial
      and final state QCD radiation) can not be measured with certainty.
      This leads not only to the ambiguity of assigning jets to
      the $t \bar{t}$ decay products but also to the problem
      of choosing a correct set of jets when the number of jets
      observed exceeds the number of strongly interacting partons produced in the leading
      order perturbation theory. In the $\ell$+jets
      and dilepton channels, additional
      ambiguities arise when multiple solutions
      of kinematic equations for the $\nu$ momenta exist. In the 
      top mass measurement literature,
      multiple acceptable mappings between partons and detected objects
      are called ``permutations''.
\end{mylist}
% It should be mentioned that all CDF and \D0 \  top mass measurements
% rely heavily on the predictions of the Standard Model
% of particle interactions.
% Due to this built-in model dependence, measured \Mt\ is essentially
% a Standard Model parameter (top quark pole mass) rather than a
% constant of nature which can stand by itself.

\noindent The feasibility of a simple kinematical
analysis of the $p\bar{p} \rightarrow t\bar{t} \rightarrow W^+bW^-\bar{b}$
process can be
determined using the following considerations.
Charged leptons in decays of $W$ bosons
are produced in association with neutrinos.
Lepton presence, while allowing for
powerful discrimination of QCD
background, leads to complications in the
reconstruction of event kinematics due to undetected neutrinos. 
We assume for the moment that the four-momenta of
all charged final state partons are reasonably well measured,
that masses of initial state partons and final state neutrinos are zero,
that the transverse momentum of the incoming partons is zero, and that
all of the event transverse missing energy is taken away by escaping
neutrinos. With these assumptions, the neutrino 3-momenta and the
longitudinal components of the momentum of the incoming partons
are the only kinematical quantities not known, and then
the total number of additional kinematic constraints needed in order
to completely specify the reaction is $2 + 3 n_{\nu}$, where
$n_{\nu}$ is the number of neutrinos in the final state. Four of these
constraints come from the energy-momentum conservation, and three
additional constraints can be introduced by using the narrow width
approximation for top and $W$: the invariant masses of the $W$ decay
products are required to be consistent with the mass of the $W$
and the invariant masses
of top decay products are set to \Mt. \Mt\ itself is considered
unknown, so the latter requirement generates only
one effective constraint for the two top quarks present in the process.
With these seven constraints, the total number of kinematical
degrees of freedom (DoF) is $3 n_{\nu} - 5$.
This gives 1 DoF for the dilepton final state, $-2$ for $\ell$+jets
(the system is overconstrained), and $-5$ for the all-hadronic channel.
This indicates that kinematic fitting techniques based
on $\chi^2$ minimization can be utilized
for the analysis of both $\ell$+jets and all-hadronic final states.
In the dilepton channel kinematical equations are underconstrained,
so this channel has to be treated in a substantially different manner.
Kinematical analysis of $t\bar{t}$ events in
different final states is described in more detail
in Section~\ref{subsec:kinemfit}.

From the statistical point of view, the result of any \Mt\ 
measurement is an interval estimate of this parameter.
Most authors utilize frequentist confidence intervals~\cite{ref:james} by
numerically studying the distribution of the point estimator
with Monte Carlo simulations. Use of Bayesian credible intervals
for top mass estimation has also appeared in the 
literature~\cite{ref:d0.ljets.mem.nature, ref:d0.ljets.matrel.firstrun2,
ref:d0.ljets.matrel.last, ref:d0.ljets.matrel.may2011}.
Techniques used for constructing the point estimator itself are described
in detail in the remainder of this section.

\subsection{Methods based on distribution fitting}
\label{subsec:distrofit}

An unbiased point estimator of \Mt\ can be obtained by fitting
a distribution of some observed quantity
${\bf y}$ to a sum of signal and background contributions.
In principle, any quantity whose distribution depends on the
parameter of interest can be used to build an estimate of
that parameter. In practice, one strives to choose ${\bf y}$
satisfying the following conditions:
\begin{mylist}
\item The sensitivity of the combined (signal plus background) distribution
      of ${\bf y}$ to \Mt\ is high. A quantitative measure
      of such a sensitivity
      can be provided, for example, by Fisher
      information~\cite{ref:james}. All other factors
      being equal, a measurement which uses
      a distribution with higher sensitivity to \Mt\ will have
      lower statistical uncertainty.
\item Modification of distribution nuisance parameters, such as
      detector calibration constants and sample signal fraction,
      does not lead to
      a noticeable change of the 
      \Mt\ estimate ({\it i.e.,} off-diagonal
      elements of the Fisher information matrix which involve \Mt\ 
      should be small)\footnote{It is always assumed
      that point estimators of \Mt\ are properly calibrated and
      unbiased and that the presence of uncertain
      nuisance parameters is the sole reason
      for the systematic uncertainty of the measurement.}.
\end{mylist}
It can be easily appreciated that, in the case of \Mt\ measurements
at CDF and \D0 , these conditions are contradictory. For example,
an \Mt\ estimate with high sensitivity
can be obtained by calculating the invariant mass
of three jets produced in the decay $t \rightarrow Wb \rightarrow jjb$
(assuming for the moment that the
jets can be chosen correctly).
This invariant mass is proportional
to the detector jet energy scale with
coefficient of proportionality close to unity.
As discussed in Section~\ref{sec:jet-rec}, 
a few percent relative uncertainty is typical
for standard jet energy scale calibrations.
This uncertainty immediately translates into a strong limitation
on the precision of the \Mt\ determination in such a measurement.

For any particular \Mt\ measurement technique, the relative
contributions of statistical and systematic uncertainties
vary with the amount of available data.
To first order, the statistical uncertainty
is expected to scale in proportion to $1/\sqrt{\int \mathcal{L} dt}$.
% where $\int \mathcal{L} dt$ is the 
% integrated luminosity accumulated by an experiment.
Due to improvements in detector calibration and Monte Carlo simulations,
the systematic
uncertainty also decreases as $\int \mathcal{L} dt$ increases,
albeit at a slower rate. Combined with the rich set
of possible ${\bf y}$ choices, this variability resulted,
over time, in a large number of studies which use distribution fitting
for the \Mt\ determination. Both signal and background
distributions of ${\bf y}$ are constructed
from Monte Carlo samples, for a set of discrete values of \Mt. Recent
works also include the jet energy scale as the second parameter, so that
the set of distributions is initially defined on a 2-d parameter grid.
These Monte Carlo-derived 
distributions are called ``templates'', and the whole technique
of distribution fitting is referred to as ``the template method''
in the top mass measurement literature.
The following ${\bf y}$ quantities have been utilized:
\begin{mylist}
\item Event-by-event top mass estimates 
      $m_t$.
      Due to its relative simplicity and adequate sensitivity, this is
      by far the most popular 
      approach~\cite{ref:cdf.ljets.run1prl,
      ref:d0.ljets.run1prl,
      ref:cdf.dilep.run1prl,
      ref:cdf.ljets.run1prd,
      ref:d0.ljets.run1prd,
      :2007qf,
      Abazov:2004ng,
      ref:cdf.dilep.3templates,
      ref:cdf.dilep.nuphiweight,
      ref:d0.dilepton.matrixweighting1,
      ref:d0.dilepton.matrixweighting2,
      ref:cdf.dilep.xsecconstrained,
      ref:d0.dilep.run2a,
      Aaltonen:2010pe,
      LucaNote}. Kinematical techniques 
      used to obtain these estimates are presented in
      Section~\ref{subsec:kinemfit}. An example
      set of $m_t$ templates is shown in Figure~\ref{mt_templates}, taken 
      from~\cite{ref:cdf.ljets.mtmw}.
\begin{figure}
\centerline{
\epsfig{file=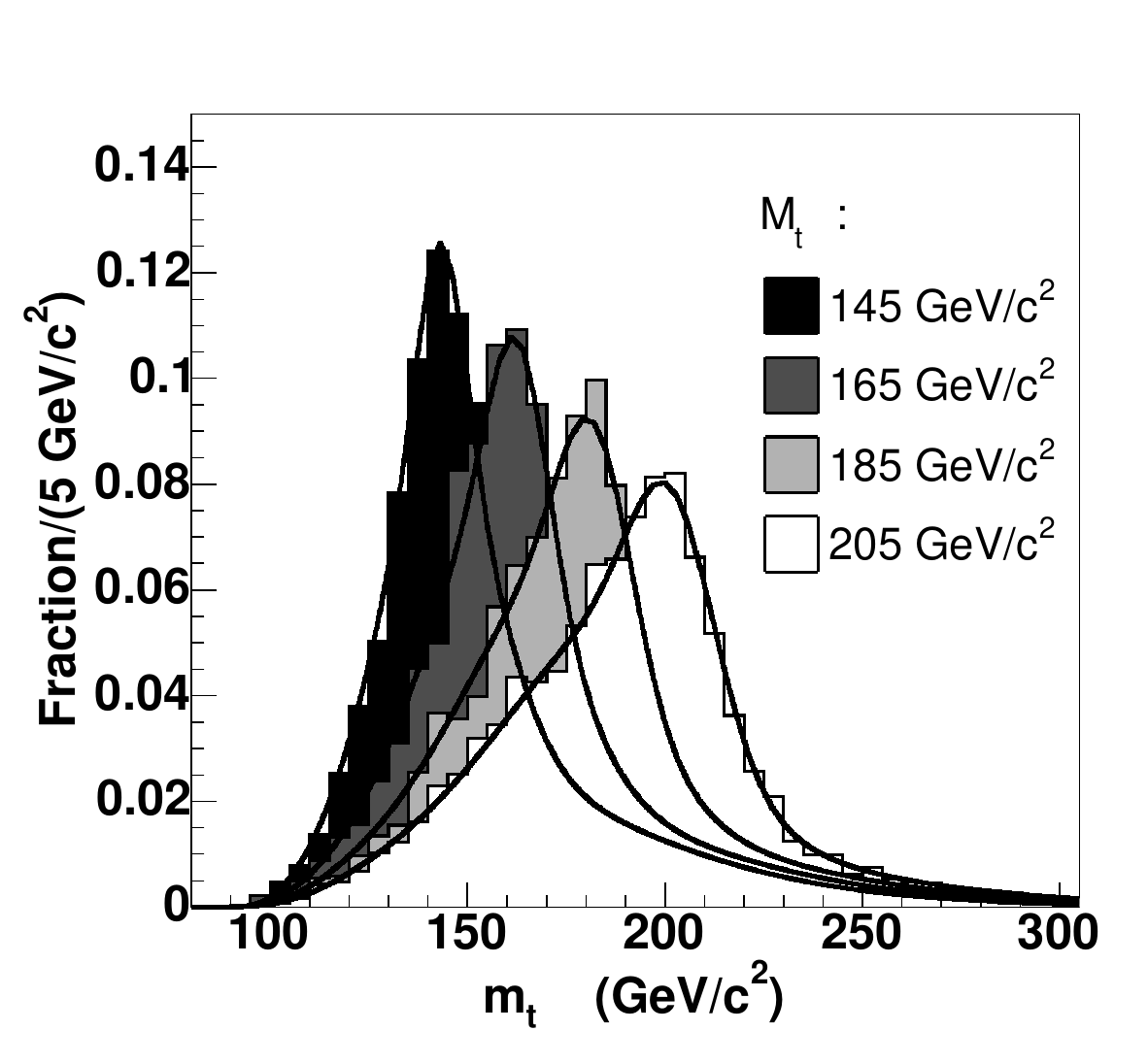,width=.3\textwidth}
}
\caption{Reconstructed $m_t$ distributions from simulated $p \bar{p} \rightarrow t \bar{t}$
events, for several different \Mt\ values. Events in the samples
used to build these templates have exactly one $b$-tagged jet.
The overlayed curves are continuous template
parameterizations~\cite{ref:cdf.ljets.mtmw}.}
\label{mt_templates}
\end{figure}
\item The transverse mass $m$\sub{T2}~\cite{ref:cdf.dilep.mt2}.
      This kinematic variable approximates $m_t$ in a manner appropriate
      for use with two missing neutrinos in the dilepton final state.
\item Combination of $m_t$ and $W$
      mass~\cite{ref:cdf.ljets.mtmwanddlm, ref:cdf.ljets.mtmw, Aaltonen:2008bg, ref:cdf.template.pub1}.
      The $W$ mass is included in order
      to decrease the sensitivity to changes in the jet energy
      scale. Two $m_t$ values (which correspond to the best and second
      best jet-to-parton assignment) are used
      in~\cite{ref:cdf.template.latest, ref:cdf.etmissplusjets}
      which results in a 3-dimensional template.
\item $m_t$ and the scalar sum of the $p_T$ of the four leading
        jets~\cite{ref:cdf.multivariate.note1}.
\item $m_t$ and $H_T$~\cite{ref:cdf.template.pub1}. 
      % $H_T$ is the scalar sum of
      % transverse energies of charged leptons, jets,
      % and missing transverse energy.
\item $m_t$ and $m$\sub{T2}~\cite{ref:cdf.template.latest}.
\item $m_t$ and its estimated resolution,
      $\sigma$~\cite{ref:d0.dilep.run2a, ref:d0.dilep.run2b}.
\item Invariant mass of the charged lepton and the $b$ quark or energy of
      the two highest $E_T$ jets (early dilepton channel measurement \cite{ref:cdf.dilep.run1xsecandm}).
\end{mylist}
Several template techniques have been developed with the explicit goal
of minimizing the dependence of the \Mt\ estimate on the calorimeter
jet energy scale. These techniques rely on the
tracking information alone which eliminates
the jet-related systematic uncertainty but also results in a substantial
degradation of the measurement sensitivity. The following 
quantities have been employed for this purpose:
\begin{mylist}
\item The transverse decay length of $b$-tagged jets,
      L\sub{xy}~\cite{ref:cdf.ljets.lxy, ref:cdf.ljets.lxyandlpt}.
      Average L\sub{xy} is increased with \Mt\ because $b$ jets receive stronger
      transverse boost from heavier top quarks.
\item The transverse momentum of leptons ($e$ and $\mu$) from $W$
      decays~\cite{ref:cdf.dilep.oldlpt, 
      ref:cdf.ljets.lxyandlpt, ref:cdf.dilep.lpt, ref:cdf.ljets.lpt}. Example
      distributions of this and the previous quantity are shown
      in Figure~\ref{tracking_templates}.
% \begin{figure}
% \centerline{
% \epsfig{file=tracking_templates.pdf,width=1.0\textwidth}
% }
% \caption{The transverse decay length of $b$-tagged jets (a) and
% the transverse lepton momentum (b) for two
% different values of \Mt\ separated
% by 50 GeV/c$^2$~\cite{ref:cdf.ljets.lxyandlpt}.}
% \label{tracking_templates}
% \end{figure}
\begin{figure}
\includegraphics[width=7.5cm]{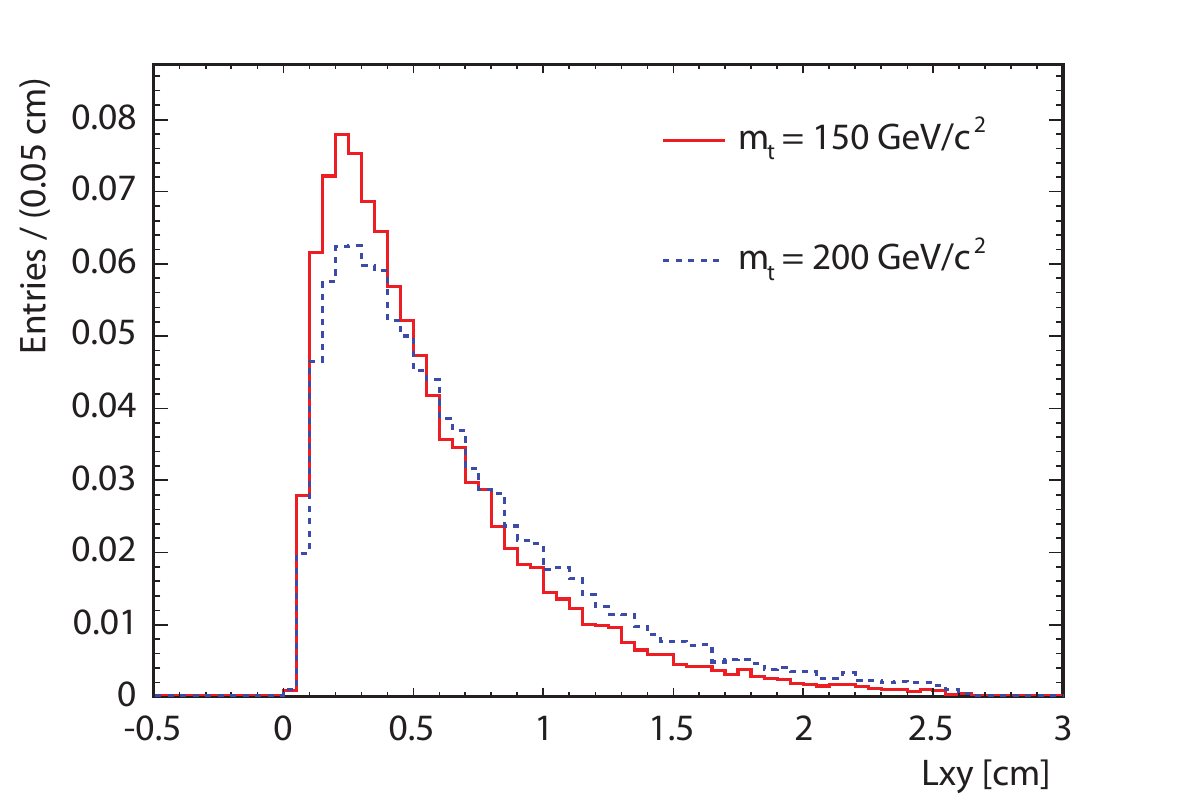}
\includegraphics[width=7.5cm]{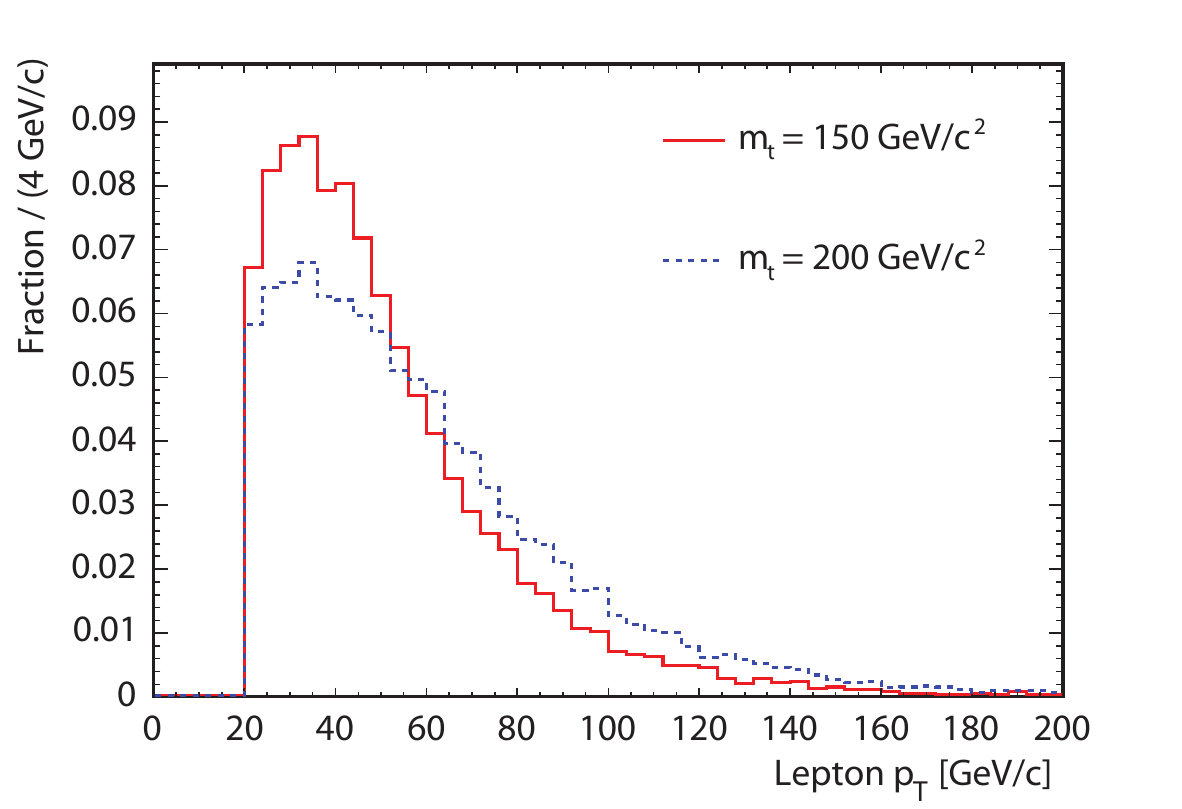}
\caption{The transverse decay length of $b$-tagged jets (left) and
the transverse lepton momentum (right) for two
different values of \Mt\ separated
by 50 GeV/c$^2$~\cite{ref:cdf.ljets.lxyandlpt}.}
\label{tracking_templates}
\end{figure}
\item The invariant mass of the charged lepton from the $W$ decay and the
      muon from the semileptonic decay of the $b$
      quark~\cite{ref:cdf.ljets.linvmass}.
\end{mylist}
A number of measurements of the top quark mass in the dilepton channel
by the \D0 \ collaboration~\cite{ref:d0.dilep.run1prl, ref:d0.dilep.run1prd,
ref:d0.dilepton.matrixweighting2}
utilized a unique template approach which does not quite fit in any
of the categories already described. In this approach, weights
are assigned to all $m_t$ values in each event according to the $\nu$WT
algorithm (see Section~\ref{subsec:kinemfit} for the description of this procedure).
In this algorithm, the $t{\bar t}$ production and decay processes are not modeled
in sufficient detail and background contamination is not accounted for,
so in practice the weights can not be treated as $m_t$ likelihoods.
Instead, a method has been developed in which the probability of the 
complete $m_t$ weight distribution is estimated for each event
as a function of \Mt.
For the purpose of dimensionality reduction, the normalized weight curve
is split into $N_b$ bins, weights are integrated in each bin,
and the ($N_b - 1$)-dimensional vector of integrated weights is used as ${\bf y}$
(due to the imposed normalization, $N_b - 1$ weights are independent).
According to~\cite{ref:d0.dilep.run1prl}, this approach improves
the statistical sensitivity of the measurement by $\approx$25\% in
comparison with $m_t$ templates.

For any choice of ${\bf y}$,
initial template construction is performed by nonparametric density
estimation techniques. Simple histogramming is sufficient
in case ${\bf y}$ is one-dimensional, while multivariate templates
are usually built by kernel density estimation~\cite{ref:kde}.
Fitting of the template distributions to the observed data
is performed by the maximum likelihood
method, with MINUIT~\cite{ref:minuit} being
the common choice of the optimization
and error analysis engine. To ensure proper fit convergence,
it is highly desirable for the likelihood to have at least two continuous
derivatives as a function of each parameter. However,
Monte Carlo-derived templates are only defined for a discrete parameter set.
Two distinct solutions to this problem have been identified: 
\begin{mylist}
\item A continuous parametric statistical model is fitted to the templates
      so that they become continuous functions, together with their
      first few derivatives, of both their parameters
      and ${\bf y}$. In this approach, the likelihood for the data
      can be determined for arbitrary parameter values, and
      it is continuous.
\item The log-likelihood is initially determined for the parameter values
      for which
      templates were defined. Then the log-likelihood itself is interpolated
      to other parameter values using, for example, local polynomial
      regression~\cite{ref:localreg}, or by
      simply fitting the log-likelihood to a second or third degree
      polynomial near the peak.
      The smoothness assumptions are thus introduced directly
      into the likelihood curve.
\end{mylist}
The second solution is also widely employed by the phase space integration
methods described in Section~\ref{subsec:mem}. There, direct calculation
of the likelihood for arbitrary parameter values becomes too expensive 
CPU-wise, so one has to resort to interpolation.

Let's assume that a continuous representation of the signal and
background templates,
$S({\bf y} | M_{\mbox{\scriptsize t}}, {\bf \theta})$ and
$B({\bf y} | M_{\mbox{\scriptsize t}}, {\bf \theta})$,
respectively, is available. Here, ${\bf \theta}$ represents
nuisance parameters of the measurement 
other than the sample signal fraction (typically, jet energy scale).
Then a simple likelihood for the observed
sample can be written as
\begin{equation}
\label{eq:simple_likelihood}
\begin{array}{ll}
\fl L(M_{\mbox{\scriptsize t}}, {\bf \theta}, f_s) = \prod_{j=1}^N 
& {\!\!\!\!\!}
\left[ f_{s\,} S({\bf y}_j | M_{\mbox{\scriptsize t}}, {\bf \theta}) \right. \\
& {\!\!\!\!\!}
\left. + \, (1 - f_s) B({\bf y}_j | M_{\mbox{\scriptsize t}}, {\bf \theta}) \right], \\
\end{array}
\end{equation}
where $j = 1, 2, ..., N$ is the 
event number in the sample and $f_s$ is the sample
signal fraction. The subsequent elimination of the nuisance parameters
${\bf \theta}$ and $f_s$ is performed either by profiling or by marginalization.
The profiling procedure consists in maximizing the
likelihood~\eref{eq:simple_likelihood} with respect to
${\bf \theta}$ and $f_s$ for each value of \Mt. Marginal likelihood
is instead calculated by introducing a prior distribution
for the nuisance parameters (such a prior distribution
usually takes into account uncertainties of existing
estimates of these parameters) and integrating the likelihood
over ${\bf \theta}$ and $f_s$ with the prior. Mixed treatments
have been utilized as well, in which the likelihood is initially
profiled over $f_s$ and then ${\bf \theta}$ is marginalized.
Relative merits of likelihood profiling
and marginalization are discussed, for example,
in~\cite{ref:likelihoodProfAndMarg}.

Various modifications of the basic likelihood~\eref{eq:simple_likelihood}
have been employed in a number of \Mt\ measurements.
These modifications take into account uncertainties in the
$S({\bf y} | M_{\mbox{\scriptsize t}}, {\bf \theta})$ and
$B({\bf y} | M_{\mbox{\scriptsize t}}, {\bf \theta})$ shapes due
to the limited number of Monte Carlo events
available~\cite{ref:cdf.ljets.run1prd, ref:cdf.dilep.3templates}
as well as binning effects in case histogramming is used
to build the templates~\cite{ref:d0.ljets.run1prd,
ref:bayestemplate}. In addition,
the observed event sample is often split into
several non-overlapping subgroups (which differ, {\it e.g.,} by the number
of $b$-tagged jets) with different template shapes and 
expected background fractions. This allows for more precise
template modeling in each subgroup as well as for the introduction
of separate $f_s$ priors. After elimination of the nuisance parameters
by profiling or marginalization,
the \Mt\ log-likelihoods from different
subgroups are added together to obtain the final result.

The ``ideogram'' technique is a modification of the template method
which attempts to take into account the resolution of the $m_t$
estimate on event-by-event basis~\cite{ref:d0.ljets.ideogram, Aaltonen:2006xc,
ref:cdf.allhad.latestideogram, ref:d0.ljets.runiiearly}. All \Mt\ measurements
which utilized this technique so far lacked a consistent likelihood
formulation owing to the absence of the
resolution prior, as discussed in detail in~\cite{ref:punzi}.
Perhaps, this is the reason why such measurements did not
demonstrate convincing uncertainty improvements over results
obtained with templates.

The top quark mass can also be estimated
indirectly from the measurement of the $t{\bar t}$ production cross 
section~\cite{ref:d0.dilep.mtfromxsec, ref:cdf.dilep.xsecconstrained,
ref:d0.ljets.fromxsec}, assuming
that the cross section dependence on \Mt\ can be determined
with sufficient precision from the Standard Model
theoretical calculations.

\subsection{Kinematic reconstruction of the top mass}
\label{subsec:kinemfit}

In the $\ell$+jets and all-hadronic channels the number of 
available kinematic constraints exceeds the number of unknown
quantities (including \Mt\ itself), therefore these channels
are amenable to kinematic fitting using $\chi^2$ minimization.
The $\chi^2$ is constructed using transverse momenta of
detected leptons and jets as well as the missing transverse energy:
\begin{equation}
\label{eq:chisq}
\begin{array}{rl}
\fl \chi^2 = & \sum_{i = \ell, 4jets} \frac{(p_T^{i,fit} - p_T^{i,meas})^2}{\sigma_{i}^2} \\
& + \sum_{j = x, y} \frac{(p_j^{UE,fit} - p_j^{UE,meas})^2}{\sigma_{UE}^2} \\
& + \, \frac{(M_{\ell \nu} - M_W)^2}{\Gamma_W^2}
+ \frac{(M_{j j} - M_W)^2}{\Gamma_W^2} \\
& + \, \frac{(M_{b \ell \nu} - m_t)^2}{\Gamma_t^2}
+ \frac{(M_{b j j} - m_t)^2}{\Gamma_t^2}.
\end{array}
\end{equation}
This particular
expression is appropriate for the  $\ell$+jets channel
(the notation follows~\cite{ref:cdf.ljets.mtmw}).
Symbols with superscript {\it fit} denote fitted parton-level
variables which determine leading order $t\bar{t}$ production and
decay kinematics,
while superscript {\it meas} refers to quantities measured in the detector.
The quantities used to form this expression are:
\begin{mylist}
\item $p_T^{i}$ are transverse momenta of leptons and quarks (fitted)
      or jets (measured). 
      % The measured jet momenta are calibrated using the
      % procedure described in Section~\ref{sec:jet-rec}.
\item $p_j^{UE}$ are the transverse components of the unclustered energy.
      The transverse momenta of the jets, neutrino, and unclustered energy
      are related at the parton level by
      ${\vec p}_T^{\,\nu} = - \left({\vec p}_T^{\,\ell} + \sum {\vec p}_T^{\,jet} + 
       {\vec p}_T^{\,UE}\right)$, assuming that the initial transverse momentum
       of the colliding particles is ${\vec 0}$.
\item $M_{\ell \nu}$, $M_{j j}$ are invariant masses of $W$ decay products
      constructed using parton-level quantities.
\item $M_{b \ell \nu}$, $M_{b j j}$ are invariant masses of top quark
      decay products constructed using parton-level quantities.
\end{mylist}
The $\chi^2$ for the all-hadronic channel is similar,
with light jets used
everywhere instead of $\ell$ and $\nu$, the first sum
running over 6 jets, and the $p_j^{UE}$ terms
omitted~\cite{Aaltonen:2010pe, :2007qf}.

\begin{figure}
\centerline{
\epsfig{file=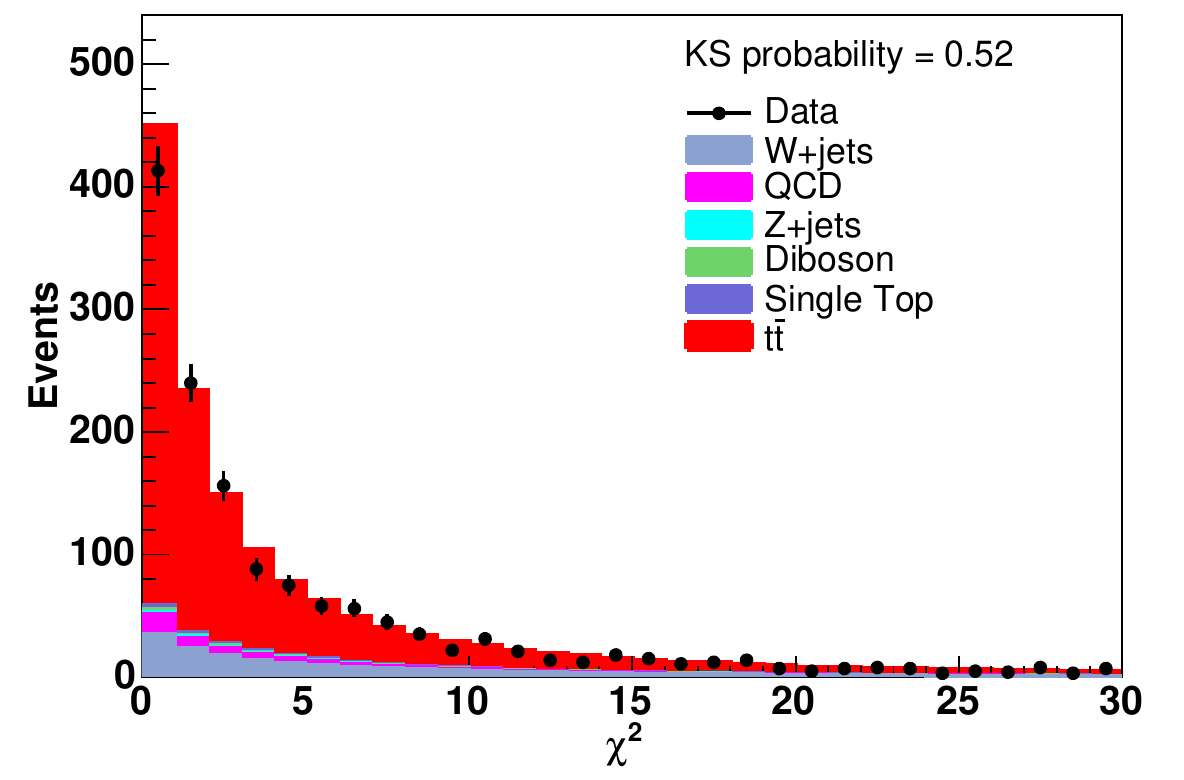,width=.4\textwidth}
}
\caption{Minimum $\chi^2$ distribution for events
         with at least one $b$~tag in the CDF 5.6~fb\supers{-1} event 
         sample~\cite{ref:cdf.template.latest}.}
\label{fig:cdf_chisq_distro}
\end{figure}

In~\eref{eq:chisq},
Gaussian resolution functions are employed for 
magnitudes of all transverse momenta,
missing energy, as well as for the mass constraints.
The method does not take into account angular resolutions of lepton
and jet directions (these directions are assumed to be perfectly measured),
asymmetry of jet $p_T$ resolution functions, or
expected event population in the reaction phase space.
The $\chi^2$ is minimized with respect to all parton-level
kinematic quantities (magnitudes of the transverse momenta
of $b$ and light quarks,
3-momentum of the neutrino) as well as $m_t$. This minimization
is performed separately for all jet-to-parton assignments
(permutations) consistent with the observed set of $b$ tags.
Typically, two solutions are obtained in the $\ell$+jets channel when kinematic
equations are solved for the $t \rightarrow W b \rightarrow \ell \nu_{\ell} b$
decay sequence, thus increasing the number of effective permutations by a factor of two.
The value of $m_t$ at the $\chi^2$ minimum over all permutations and parton-level
kinematic variables is usually taken as the top mass estimate
for the event.
Distributions of the $\chi^2$ minimum values are illustrated
in Figure~\ref{fig:cdf_chisq_distro} for the $\ell$+jets channel.

Several modifications of this basic approach were explored,
typically resulting in a~more complicated method with
only a slight improvement in the measurement sensitivity:
\begin{mylist}
\item Use of $m_t$ estimates from more than one
       permutation~\cite{ref:d0.ljets.run1prd, ref:cdf.ljets.3chisq,
       ref:cdf.template.latest}.
\item Inclusion of the jet angular resolution 
       terms~\cite{ref:d0.ljets.run1prl, ref:d0.ljets.run1prd, ref:cdf.ljets.run1prd,
                   ref:cdf.multivariate.note1}.
\item Use of a separate jet energy scale factor for each permutation~\cite{ref:cdf.multivariate.note1}.
\item Addition of a term which models the $b$ tagging probability~\cite{ref:cdf.ljets.run1prd}.
\item Replacement of Gaussian mass constraints with Breit-Wigner
       constraints~\cite{ref:d0.ljets.run1prd}.
\end{mylist}
Another possible modification that has not been tested yet in the context of
\Mt\ measurements consists in replacing the mass constraints
in~\eref{eq:chisq} with the term $-2 \ln [\,p({\bf x}|m_t)\,J^{-1}\,p_{tag}]$,
where $p({\bf x}|\Mt)$ represents the phase space density
of the process normalized by the total observable cross section,
$J = \left| \frac{\partial\,{\bf p}^{fit}}{\partial\,{\bf x}} \right|$ is the Jacobian
which relates the fitted quantities as well as the quantities which
are assumed to be perfectly measured to the phase space variables, and $p_{tag}$
is the permutation-specific probability of the observed $b$ tag configuration
(see Section~\ref{subsec:mem}).
The goal is to find the most probable values of all fitted quantities
taking into account theoretical assumptions
about the reaction and thereby to profile the $m_t$ likelihood over the complete
parton-level phase space\footnote{A very similar approach, albeit without proper
normalization, is discussed in~\cite{ref:kopylov}.}. 
% To the best of
% our knowledge, this modification of $\chi^2$ has never been realized
% in practice.

% A simple kinematical method without parton $p_T$ adjustment
% was utilized in the \D0 \  Run~I \Mt\ measurement in the all-hadronic
% channel~\cite{Abazov:2004ng}. $\chi^2$ minimization was used
% for permutation selection only. Invariant mass
% constraints were imposed on the reconstructed jets in the form
% \begin{equation*}
% \fl \chi^2 = \frac{(m_{t1} - m_{t2})^2}{4 \,\sigma_{m_t}^2}
%       + \frac{(M_{W_1} - M_{W_0})^2}{\sigma_{M_W}^2}
%       + \frac{(M_{W_2} - M_{W_0})^2}{\sigma_{M_W}^2},
% \end{equation*}
% where $M_{W_1}$, $M_{W_2}$, $m_{t1}$, and $m_{t2}$ are invariant
% masses of jets assigned to $W$s and top quarks.
% The average dijet invariant mass, $M_{W_0}$,
% and resolutions used in these constraints were
% estimated from Monte Carlo simulations, employing
% known correct jet permutation only.
% The jet-to-parton assignments
% considered were required to be consistent with the available
% $b$-tagging information from soft lepton tagging.
% The permutation with smallest $\chi^2$ was chosen,
% and the variable $m_t = (m_{t1} + m_{t2})/2$
% was used to construct the templates.

For the dilepton channel, the number of constraints is insufficient
to determine $m_t$. It would be possible to infer $m_t$ if at
least one additional kinematic quantity or constraint was available. This suggests
the following method of $m_t$ determination: values of
some unobserved kinematic quantities $\xi$ are assumed, kinematic
equations are solved, and allowed
values of top mass, $m_t'(\xi, {\bf y})$, are determined (in addition to $\xi$,
$m_t'(\xi, {\bf y})$ depends on some or all quantities ${\bf y}$ measured in the detector).
$\xi$ values are scanned within their kinematic
limits. Obtained $m_t'(\xi, {\bf y})$ values are assigned weights proportional
to the probability density, $\rho(\zeta)$, of some other kinematic
quantities $\zeta(\xi, {\bf y})$ (variable sets
$\xi$ and $\zeta$ may overlap partially, completely, or not at all).
$\rho(\zeta)$ is determined in advance, either
from theoretical considerations or by estimating this density numerically
using large samples of Monte Carlo events.
While it is not strictly necessary, this method operates in the most
transparent manner if $\rho(\zeta)$ has little or no
dependence on \Mt.
The distribution of possible $m_t$ values, $\rho(m_t)$,
is thus formed, proportional to 
$\int \rho(\zeta(\xi, {\bf y})) \,\delta(m_t'(\xi, {\bf y}) - m_t) \,d\xi$.
The location parameter of this distribution (usually the mode)
is used as the top mass estimate for this particular event. 
This approach is very general, and a judicious
choice of $\xi$ and $\zeta$ variables will produce an estimate
with good sensitivity to \Mt\ while maintaining a relatively
simple statistical model for $\rho(\zeta)$ (inclusion of {\em all}
parton-level phase space variables in both $\xi$ and $\zeta$
leads to techniques described in Section~\ref{subsec:mem}).

The following choices of $\xi$ and  $\zeta$ have been explored:
\begin{itemize}
\item $\xi = (\eta_{\nu}, \eta_{\bar{\nu}})$: a two-dimensional
      variable, where $\eta_{\nu}$ and $\eta_{\,\bar{\nu}}$ are pseudorapidities
      of the neutrinos. $\zeta = (\eta_{\nu}, \eta_{\bar{\nu}}, {\not}{\vec{E}_T})$,
      where ${\not}{\vec{E}_T}$ is the two-dimensional missing transverse energy.
      This method is called {\it neutrino weighting algorithm}
      ($\nu$WT)~\cite{ref:d0.dilep.run1prd, ref:cdf.dilep.run1prl, ref:cdf.dilep.3templates,
      ref:d0.dilepton.matrixweighting2, ref:d0.dilep.run2a, ref:d0.dilep.run2b,
      ref:cdf.template.latest, ref:d0.latest.nuwt}. Weight distributions for the six dilepton
      events from~\cite{ref:d0.dilep.run1prd} are shown in Figure~\ref{fig:d0DilepNuW}.
\begin{figure}
\centerline{
\epsfig{file=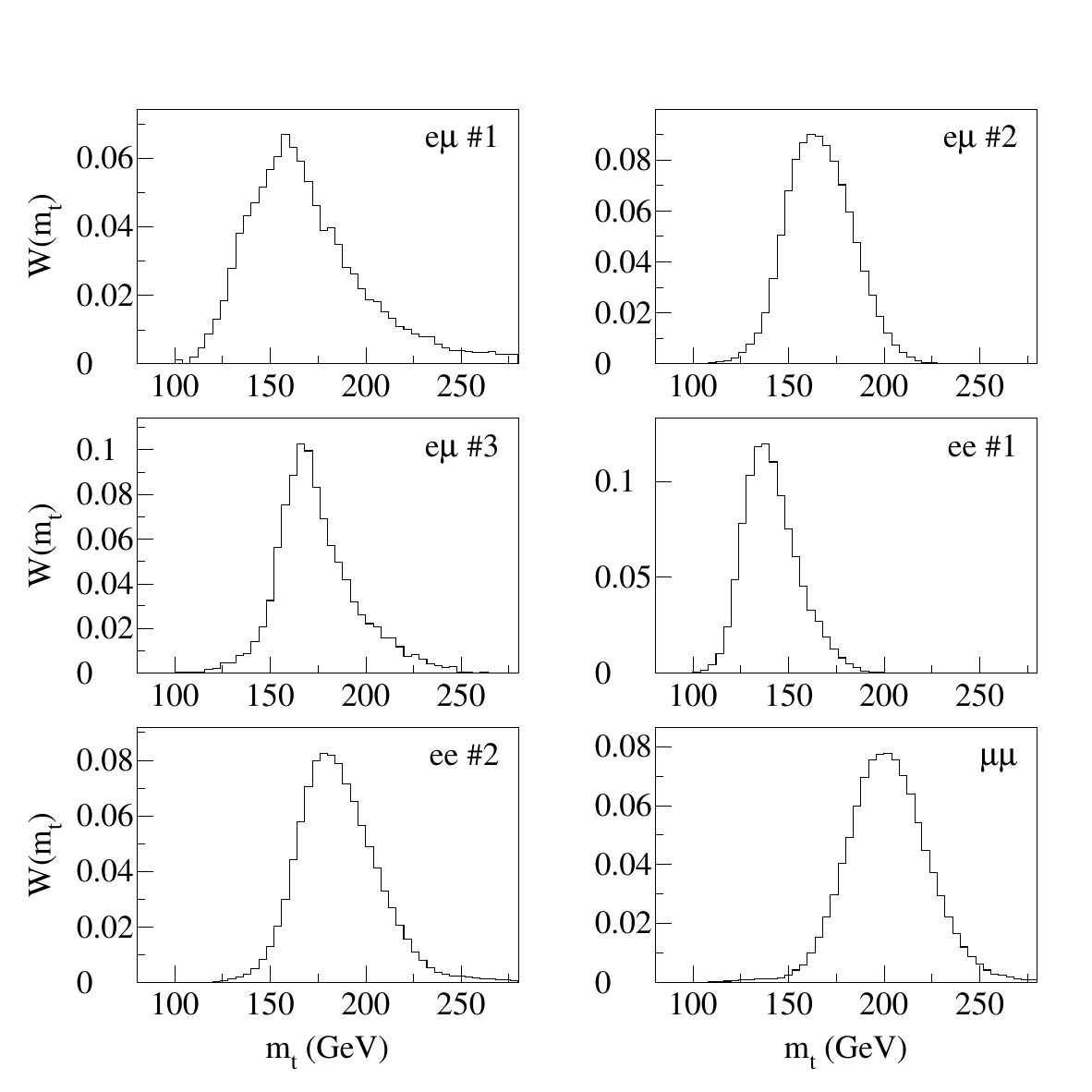,width=.45\textwidth}
}
\caption{Weights for the individual events in the \D0 \ Run~I
dilepton $\nu$WT analysis as a function
of $m_{t}$~\cite{ref:d0.dilep.run1prd}.}
\label{fig:d0DilepNuW}
\end{figure}
\item $\xi = (\varphi_{\nu}, \varphi_{\bar{\nu}})$, where $\varphi_{\nu}$ and
      $\varphi_{\bar{\nu}}$ are neutrino azimuthal angles in the
      plane transverse to the beam. A $\chi^2$ variable formed in a manner similar
      to \eref{eq:chisq} is utilized as $\zeta$ (since values of two variables are
      assumed, the number of constraints exceeds the number of unknowns by 1, so
      it becomes possible to construct a meaningful $\chi^2$). Weights proportional
      to $e^{-\chi^2/2}$ are assigned to $m_t(\xi, {\bf y})$ solutions. This
      method is referred to as the {\it neutrino $\varphi$ weighting algorithm}
      (PHI)~\cite{ref:cdf.dilep.3templates, ref:cdf.dilep.nuphiweight}.
\item $\xi = (m_t', p_{\ell^+}, p_{\ell^-}, p_{b}, p_{\,\bar{b}})$,
      $\zeta = {\bf x}$, where ${\bf x}$ represents all leading order
      parton-level variables which determine the reaction kinematics. 
      Directions of the charged leptons and $b$ quarks are assumed to be perfectly
      measured, while magnitudes of their momenta (inverse magnitude in the case of $\mu$)
      are varied within their
      expected Gaussian resolutions. Neutrino momenta are calculated
      from kinematic constraints in the narrow width
      approximation for $t$ and $W$ and zero transverse momentum assumption for the
      $t\bar{t}$ system. Values of
      $m_t = m_t'$ are weighted by the phase space density of the process
      at ${\bf x}$,
      % (not properly normalized)
      summed over all possible kinematical solutions for the neutrinos
      and jet-to-quark assignments. This approach is known as
      the {\it matrix element weighting algorithm}
      (${\mathcal M}$WT)~\cite{ref:dalitzgoldstein, ref:d0.dilep.run1prd,
      ref:d0.dilepton.matrixweighting2, ref:d0.dilep.run2a}.
      It is a direct precursor
      of the phase space integration methods described in Section~\ref{subsec:mem}.
      % Note that the weights in~\cite{ref:d0.dilep.run1prd} were normalized
      % so that their average over $t\bar{t}$ Monte Carlo events which passed the selection
      % cuts was made independent of the top quark mass.
      % This is hardly meaningful: the $t\bar{t}$
      % Monte Carlo events (generated in this case by HERWIG~\cite{ref:herwig})
      % are themselves distributed in the phase space according to these weights,
      % so the phase space density factor enters into such a
      % calculation twice. Proper normalization (by total observable cross section)
      % is described later in this article.
% \item $\xi = (m_t', p_{\ell^+}, p_{\ell^-}, p_{b}, p_{\,\bar{b}})$,
%      $\zeta = (x, \bar{x}, E^{*}_{\ell^+}, E^{*}_{\ell^-})$.
%      $x$ and $\bar{x}$ are the fractions of $p$ and $\bar{p}$ momenta,
%      respectively, carried
%      by colliding partons. $E^{*}_{\ell^+}$ and $E^{*}_{\ell^-}$
%      are the lepton energies in the rest frames of their corresponding
%      ancestor top quarks.
%      We will call this scheme {\it modified Dalitz-Goldstein algorithm}
%      (MDG)~\cite{ref:dalitzgoldstein, ref:d0.dilep.mdg1, ref:d0.dilep.mdg2}.
%      It uses a simplified phase space weight originally employed in~\cite{ref:dalitzgoldstein},
%      with added smearing of momentum magnitudes of the observed particles.
\item $\xi = \zeta = (p_z^{t\bar{t}}, E_{b}, E_{\bar{b}}, {\not \!E}_T)$, where 
      $p_z^{t\bar{t}}$ is the z component (along the beam) of the $t\bar{t}$
      system, while $E_{b}$ and $E_{\bar{b}}$ are the energies of the $b$ and $\bar{b}$
      quarks, respectively.
      The probability density of $\zeta$ is factorized into the product of probabilities
      for individual components, with all marginals
      represented by Gaussian distributions and correlations neglected.
      This method is known as the {\it full kinematic analysis}
      (KIN)~\cite{ref:cdf.dilep.3templates}.
\end{itemize}
It appears that the \Mt\ sensitivity of the ${\mathcal M}$WT and $\nu$WT algorithms is
similar~\cite{ref:d0.dilep.run2a} and slightly exceeds that of both PHI and KIN
methods~\cite{ref:cdf.dilep.3templates}. In order to increase the overall 
sensitivity, results obtained
in the dilepton channel on the same data with multiple template methods are often combined using
the ``best linear unbiased estimator'' (BLUE)
approach~\cite{Lyons:1988, Valassi:2003}. In this case correlation
coefficients between different methods  are determined using common pseudo-experiments.

\subsection{Phase space integration methods}
\label{subsec:mem}

The technique of calculating event observation probabilities
by integrating over all parton-level quantities in the
reaction phase space was introduced into the high energy physics
data analysis practice in the pioneering Run~I measurement of the top
quark mass by the \D0 \ collaboration~\cite{ref:d0.ljets.mem.nature}.
This method can be understood as an application of the Bayesian
principle of integrating over all unobserved degrees of 
freedom
% \footnote{Multiple imputation~\cite{ref:multimput} is a similar statistical technique
% which can be interpreted in frequentist terms.}
with a well-motivated {\em informative} prior provided by the
Standard Model theory ({\it i.e.,} event-by-event 
marginalization of ${\bf x}$).
The prior is proportional to
the matrix element squared of the process, so the technique
is often referred to as the ``matrix element method'' (MEM).
The probability of observing quantities  ${\bf y}$ in a particle
detector, $P\sub{ev}({\bf y} | {\bf a})$, is determined according to
\begin{equation}
\label{eq:prob_integral1}
P\sub{ev}({\bf y} | {\bf a}) = \sum_{i} f_{i} P_{i}({\bf y} | {\bf a}),
\end{equation}
where ${\bf a}$~is the set of model parameters. This set
includes \Mt\ and can also include
detector jet energy scale as well as
other theoretical and instrumental quantities.
$f_{i}$ are the fractions of different
non-interfering production channels consistent with ${\bf y}$
and constrained by $\sum_{i} f_{i} = 1$.
In the context of \Mt\ measurements, the probability to measure
${\bf y}$ in each channel $i$ is most often estimated from
\begin{equation}
\label{eq:prob_integral2}
\begin{array}{ll}
\fl P_{i}({\bf y} | {\bf a}) = 
&  {\!\!\!\!\!}
\frac{\Omega({\bf y})}{\sigma_{i}({\bf a}) A_{i}({\bf a})} \\
\fl &  {\!\!\!\!\!}
\times \int_{\Phi_i} W_{i}({\bf y} | {\bf x}, {\bf a})\,
|M_{i} ({\bf x}, {\bf a})|^{2} \,T_{i}({\bf x}, {\bf a}) \,d {\bf x},
\end{array}
\end{equation}
where the following notation is utilized:

$\Omega({\bf y})$ --- Indicator function
for the analysis acceptance (1 for events which pass the
event selection criteria, 0 otherwise). This term can be
replaced by 1 in case only accepted events are considered.

${\bf x}$ --- Variables which uniquely specify a point in
              the channel phase space $\Phi_i$, ${\bf x} \in \Phi_i$.

$d {\bf x}$ --- Differential element of the phase space $\Phi_i$.

$\sigma_{i}({\bf a})$ --- Channel cross section:
\[
\sigma_{i}({\bf a}) = \int_{\Phi_i} 
|M_{i} ({\bf x}, {\bf a})|^{2} \,T_{i}({\bf x}, {\bf a}) \,d {\bf x}.
\]

$A_{i}({\bf a})$ --- Overall experimental acceptance for channel~$i$.

$W_{i}({\bf y} | {\bf x}, {\bf a})$ --- Detector transfer function (TF).
This is the probability density for observing detector response ${\bf y}$ 
from the space of possible measurements $Y$ when
the ``true'' phase space coordinate of the event is ${\bf x}$.
This function should be normalized for every value of $i$, ${\bf x}$, and ${\bf a}$
either by 
$\int_{Y} W_{i}({\bf y} | {\bf x}, {\bf a})\, d {\bf y} = 1$
or by $\int_{Y} \Omega({\bf y}) W_{i}({\bf y} | {\bf x}, {\bf a})\, d {\bf y} 
= \epsilon_{i} ({\bf x}, {\bf a})$, where $\epsilon_{i} ({\bf x}, {\bf a})$
is the efficiency to detect an event
with phase space coordinate~${\bf x}$. The difference between
these TF normalization conditions and their effect on TF
modeling is discussed in~\cite{ref:memeff}.

$|M_{i} ({\bf x}, {\bf a})|^{2}$ --- Squared matrix element of the process.

$T_{i}({\bf x}, {\bf a})$ --- Other factors which do not depend 
on ${\bf y}$
({\it e.g.,} flux of colliding beams, parton distribution functions).

In a compact symbolic notation, \eref{eq:prob_integral2} can be represented as
\begin{equation}
P_{i}({\bf y} | {\bf a}) = \frac{1}{\sigma_{i, obs}({\bf a})} \frac{d \sigma_{i, obs}({\bf y} | {\bf a})}{d {\bf y}},
\label{eq:prob_integral3}
\end{equation}
where $\frac{d \sigma_{i, obs}({\bf y} | {\bf a})}{d {\bf y}}$ is the differential
observable cross section for channel $i$. Parameter set ${\bf a}$ is estimated
by maximizing the likelihood which is constructed
by multiplying probabilities for all events in the sample:
$L({\bf a}) = \prod_{j=1}^N P\sub{ev}({\bf y}_j | {\bf a})$.
The denominator term $\sigma_{i, obs}({\bf a}) = 
\sigma_{i}({\bf a}) A_{i}({\bf a}) = \int_{Y} \frac{d \sigma_{i, obs}({\bf y} | {\bf a})}{d {\bf y}} d {{\bf y}}$ (total observable cross section)
ensures proper likelihood normalization.

The MEM statistical model of a particle physics
process is significantly more precise than the models employed by
the distribution fitting methods, as it takes into account the process
phase space density. Detailed shapes of detector transfer functions 
can be utilized, while techniques based on $\chi^2$ minimization
essentially assume Gaussian resolutions. These and other
advantages of MEM~\cite{ref:memeff}
result in an improved statistical precision of the parameter estimates.
On the other hand, efficient calculation of the phase space
integral~\eref{eq:prob_integral2} is nontrivial, as it can not
be performed with standard phase space sampling schemes developed
for Monte Carlo event generation because of the presence of transfer
functions which alter the peak structure of the integrand.
This situation resulted in a development of a number of {\it ad~hoc}
dimensionality reduction and
phase space sampling schemes for calculating~\eref{eq:prob_integral2}
in the context of the $t\bar{t}$ production and decay. A recent
study~\cite{ref:madweight} attempts to address this problem in a
systematic manner for a number of processes.

All \Mt\ measurements performed so far integrate~\eref{eq:prob_integral2}
 over the leading
order parton phase space, so that
the final state ``soft QCD'' processes (parton showering and
hadronization) are combined together with the detector response,
and this combination is subsequently modeled
empirically by the detector transfer function. For each channel, the
transfer function of the process is factorized into a product of transfer
functions for $K$ individual partons traced in the detector (neutrinos are excluded): 
\begin{equation}
\fl W({\bf y} | {\bf x}, {\bf a}) = 
\sum_{\pi[m_1, ..., m_K]} \prod_{k=1}^K p_{m_{k}k} W_{k}({\bf y}_{m_k} | {\bf x}_{k}, {\bf a}),
\label{eq:tfmodel}
\end{equation}
where ${\bf x}_{k}$ is the phase space coordinate of the $k$\supers{th} parton,
${\bf y}_{m_k}$ are the measured quantities for the ``physics object''
with index $m_k$ (jet, lepton candidate) obtained in the event
reconstruction and pattern
recognition process, $p_{m_{k}k}$ are the prior probabilities
for associating parton $k$ with physics object $m_k$, and the channel
index $i$ is omitted for brevity. The sum is performed
over $K!$ possible assignments of indices $[m_1, ..., m_K]$
to permutations of $[1, ..., K]$.
For $t\bar{t}$ signal modeling, it can be assumed that
leptons are identified unambiguously
({\it i.e.,} $p_{m_{k}k}$ is reduced to Kronecker
$\delta_{qk}$ in case the lepton has the index $q$ in the list of physics objects),
but jet charge and flavor can not be determined with
complete certainty. A number of
measurements~\cite{ref:freeman, ref:plujan, ref:d0.ljets.matrel.last,
ref:d0.ljets.matrel.may2011}
utilized
the following $p_{m_{k}k}$ assignment scheme:

$p_{m_{k}k} = p\sub{tag}({\bf y})$ if $k$ refers to a $b$ or $\bar{b}$ quark in the final state and $m_{k}$ refers to a $b$-tagged jet reconstructed in the detector.  $p\sub{tag}({\bf y})$
is the probability to flavor tag a $b$ jet as a function of its reconstructed
momentum and other measured quantities (such as the number of tracks).

$p_{m_{k}k} = 1 - p\sub{tag}({\bf y})$ if $k$ refers to a $b$ or $\bar{b}$ quark in the final state and $m_{k}$ refers to an untagged jet.

$p_{m_{k}k} = p\sub{mistag}({\bf y}, f)$ if $k$ refers to a light ($f = u$, $d$, or $s$)
or charm ($f = c$) quark in the final state and $m_{k}$ refers to a $b$-tagged jet.
$p\sub{mistag}({\bf y}, f)$ is the probability to 
mistag a jet originating from a quark with flavor $f \ne b$ as a $b$ jet.

$p_{m_{k}k} = 1 - p\sub{mistag}({\bf y}, f)$ if $k$ refers to a light or charm
quark in the final state and $m_{k}$ refers to an untagged jet.

\begin{figure}
\centerline{
\epsfig{file=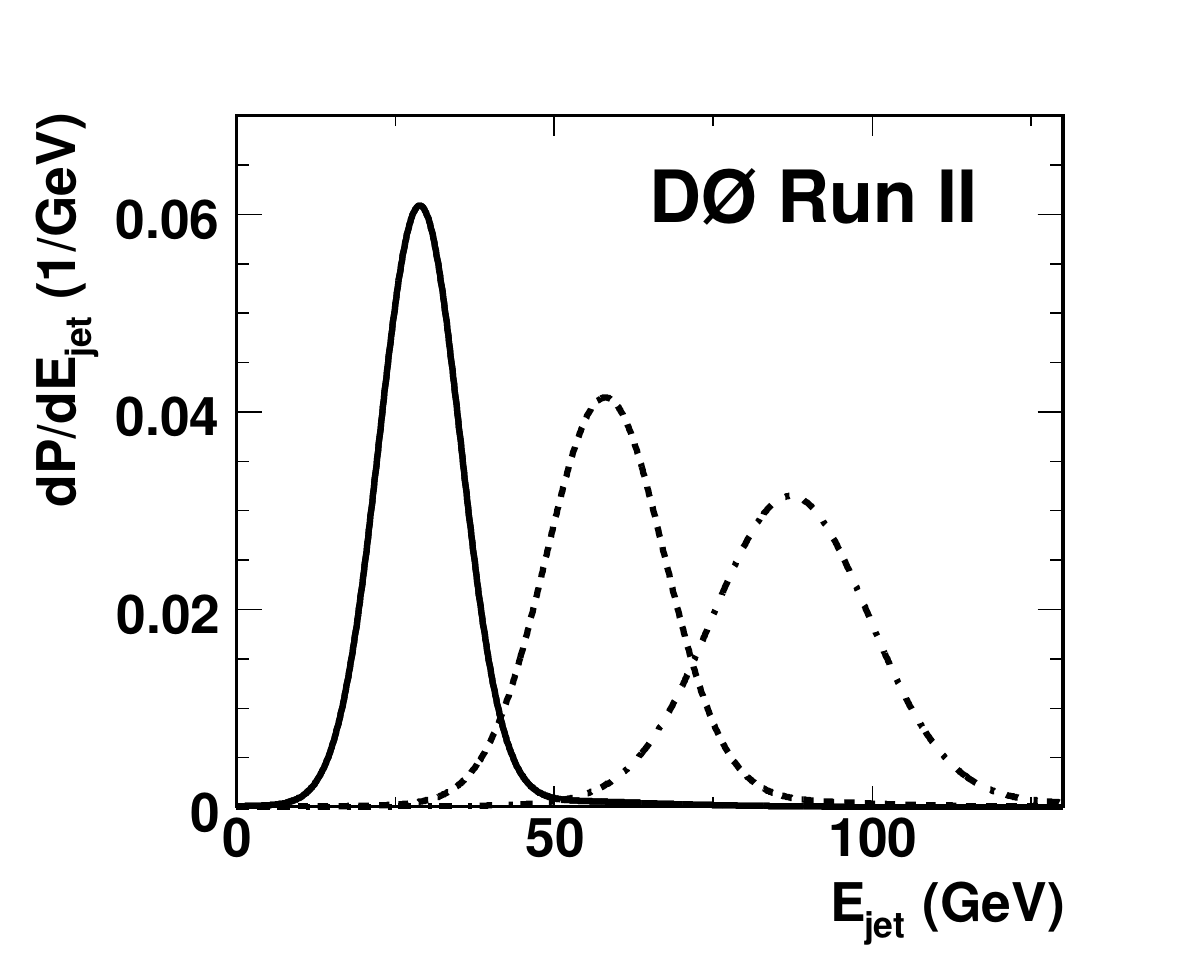,width=.3\textwidth}
}
\caption{Jet transfer functions for light quark jets, $|\eta| < 0.5$, for
parton energies $E_p = 30$~GeV (solid curve), 60 GeV (dashed
curve), and 90 GeV (dash-dotted curve). Perfect angular
resolution is assumed~\cite{ref:d0.ljets.matrel.firstrun2}.}
\label{tf_funct_d0}
\end{figure}

Although in practice the priors
$p_{m_{k}k}$ are derived as functions of
${\bf y}_{m_k}$, their dependence on ${\bf y}$ is usually mild and can be neglected
in comparison with the 
fast variation of $W_{k}({\bf y}| {\bf x}_{k}, {\bf a})$.
It is also commonly assumed that the overlap between 
phase space regions which contribute to different permutations
is negligible. With these approximations, the
transfer functions for  individual partons can be normalized
either by $\int_{Y_k} W_{k}({\bf y}_k | {\bf x}_k, {\bf a})\, d {\bf y}_k = 1$
or by $\int_{Y_k} \Omega_k({\bf y}_k) W_{k}({\bf y}_k | {\bf x}_k, {\bf a})\, d {\bf y} _k = \epsilon_{k} ({\bf x}_k, {\bf a})$.
Both of these normalization conditions assume (either implicitly or explicitly)
that the factorization model
can also be applied to acceptance and efficiency.
An example jet energy transfer function is shown in Figure~\ref{tf_funct_d0}.

Appropriate normalization for $p_{m_{k}k}$
depends, in general, on the precise meaning of these coefficients.
For the flavor tagging scheme outlined above, it becomes
$\sum p_{tag} = 1$,
where $p_{tag} = \prod_{k=1}^K p_{m_{k}k}$ is the permutation-specific
probability
of a certain heavy flavor tag assignment and 
the sum is performed over all $2^K$ possible 
assignments. Note the absence of the sum over permutations,
contrary to the convention
used by some authors~\cite{ref:d0.ljets.matrel.firstrun2}.
Different permutations correspond
to different phase space points, and the overall transfer
function has to be normalized for each phase space point separately.

On top of the product model~\eref{eq:tfmodel} for the 
detector transfer function,
additional correction terms were utilized in a number of \Mt\ measurements:
\begin{mylist}
\item Simplifying assumptions about jet transfer functions
      ({\it e.g.,} perfect angular resolutions and
      independence of jet response from proximity to other jets)
      were partially compensated for either by
      adjusting the process phase space density
      (the ``effective propagators'' approach of~\cite{ref:freeman})
      or by introducing transfer functions which
      model smearing of the angular distances between jets~\cite{ref:meat2009}.
\item Presence of the initial state QCD radiation was modeled 
      either with a prior on the transverse momentum of the $t\bar{t}$
      system~\cite{ref:freeman, ref:plujan, ref:cdf.dilep.matrel.prd, Aaltonen:2008bg}
      or with an explicit transfer function which takes
      into account additional energy visible in the detector and not associated
      with the $K$ most energetic physics objects expected 
      at the leading order~\cite{ref:cdf.dilep.matrel.isrtf}.
\end{mylist}

Out of the 32 variables needed to specify the phase space
point for the $p\bar{p} \rightarrow t\bar{t} \rightarrow 6$~partons
reaction, only the masses of initial partons and outgoing charged
leptons and neutrinos can be considered exactly known.
Taking into account the energy-momentum conservation,
this leads to 22, 24, and 26-dimensional phase space integrals in the dilepton,
$\ell$+jets, and all-hadronic channels, respectively.
In case the integral~\eref{eq:prob_integral2}
is evaluated by Monte Carlo integration, 
the relative precision of the result
is $\delta_{rel} = (\sigma_{I}/\left < I\right >) \,N^{-1/2} $,
where $\left < I \right >$ and $\sigma_{I}$ are, respectively,
the average value and the standard deviation
of the integrand, and $N$ is the number of integrand evaluations.
$I$ is sharply peaked over small regions of phase space which
leads to a large prefactor $\sigma_{I}/\left< I \right >$.
For example, for the $W$ mass Breit-Wigner distribution
integrated between 0 and \Mt, $\sigma_{I}/\left < I\right >\approx 3.5$,
so that reaching relative precision of $10^{-3}$ requires $\sim10^7$ integrand
evaluations.
Efficient Monte Carlo calculation of such integrals can be performed by
applying convergence acceleration techniques, such as importance sampling
and stratified sampling. These techniques, however, only work well
if the peak structure of the integrand is factorizable in terms
of integration variables. In~\eref{eq:prob_integral2}, 
the peaks are not
aligned with the momenta of final state particles in terms
of which the phase space integration is formally defined. Instead,
the transfer functions are concentrated around the observed
directions of leptons and jets, while propagators in the matrix
element are strongly peaked around $t$ and $W$ masses. Thus
a change to a new, efficient variable set is necessary.

Another technique commonly used to make the prefactor smaller
is dimensionality reduction: a constraint is imposed on the
integration variables, and the integrand is evaluated only for
the subspace defined by that constraint\footnote{Prefactor
dependence on the integral dimensionality, $D$, is determined
by the exact nature of the integrand and the chosen subspace.
In the mildest case, when the
peaks are fully factorized, prefactor behaves as ${\cal{O}}(D)$.}. A number of assumptions
have been introduced by different authors for this purpose, in various combinations:
\begin{enumerate}
\item Assume that some or all of the quarks are on shell ({\it i.e.},  masses
      of the light and $b$ jets are fixed). This assumption removes
      $n_q$ integration variables, where $n_q$ is the number of quarks
      in the final state.
\item Ignore individual transverse momenta of the incoming partons
      and consider only the transverse momentum of the $t\bar{t}$ system
      ($-2$ integration variables).
\item Assume that $t\bar{t}$ transverse momentum is 0 (this assumption
      is stronger than the previous one, $-4$ variables).
\item Assume that jet directions are perfectly measured in the
      detector. In this case, jet transfer functions are proportional
      to $\delta^2(\Omega - \Omega_{meas})$ ($-2\,n_q$ variables).
\item Assume that charged lepton momenta are perfectly measured ($-3\,n_\ell$ variables).
\item Use narrow width approximations for $t$ ($-2$ variables) and/or $W$ (also $-2$).
\end{enumerate}
In principle, any such assumption reduces the fidelity of the
statistical model, but practical considerations (in particular, the
need to evaluate a large number of integrals for a set of parameter
values using limited CPU resources) often take priority. For example,
in the original MEM \Mt\ measurement by
\D0 \  in the $\ell$+jets channel~\cite{ref:d0.ljets.mem.nature},
(i), (iii), (iv), and (v) were assumed which leads to a 5-d integral.
Masses of $W$ and top resonances, as well as the momentum magnitude
for one of the partons in the hadronic $W$ decay were chosen as
integration variables. This resulted in a reasonably well factorized
phase space peak structure, so the integrals could be subsequently 
evaluated automatically by the VEGAS~\cite{ref:vegas}
adaptive numerical integration program.
In a recent $\ell$+jets \Mt\
measurement by CDF~\cite{ref:plujan},
only (ii) and (v) were assumed which allowed
the authors to employ detailed transfer function models and
resulted in a 19-d integral. The convergence was accelerated by appropriate
variable transformations (importance sampling), 
by early pruning of jet permutations with small 
contributions into the overall integral, and by use of low-discrepancy
sequences~\cite{ref:Niederreiter} for points in which the
integrand was evaluated (Quasi-Monte Carlo).

The {\em Dynamical Likelihood Method} (DLM)~\cite{ref:firstdlm, ref:kondodlm} is a phase
space integration technique which evaluates~\eref{eq:prob_integral2}
in a particular manner. Note that the integral can be formally
written as
\begin{equation}
\label{eq:dlmintegral}
\begin{array}{l}
\fl P_{i}({\bf y} | {\bf a}) = \frac{\Omega({\bf y})}{\sigma_{i}({\bf a}) A_{i}({\bf a})}
\left< |M_{i} ({\bf x}, {\bf a})|^{2} \,T_{i}({\bf x}, {\bf a}) \right>_{G_i} \\
\times \int_{\Phi_i} W_{i}({\bf y} | {\bf x}, {\bf a}) d {\bf x},
\end{array}
\end{equation}
where $\left< ... \right>_{G_i}$ stands for averaging over points ${\bf x} \in \Phi_i$
distributed in the phase space with density 
$G_{i}({\bf x} | {\bf y}, {\bf a}) \equiv \frac{W_{i}({\bf y} | {\bf x}, {\bf a})}{\int_{\Phi_i} W_{i}({\bf y} | {\bf x}, {\bf a}) d {\bf x}}$.
Following the naming convention of~\cite{ref:kondodlm}, 
we refer to $G_{i}({\bf x} | {\bf y}, {\bf a})$ as
{\em posterior transfer function} (PTF). 
Phase space sampling according to PTF efficiently takes into account all structure
present in the integrand due to detector resolution (peaks in the
matrix element still require special treatment). In principle, PTF can be derived
from Monte Carlo simulations of the physics process and detector, 
using joint distributions
of ${\bf x}$ and ${\bf y}$ in the region defined by $\Omega({\bf y}) = 1$,
where every event enters with the weight
$\left( |M_{i} ({\bf x}, {\bf a})|^{2} \,T_{i}({\bf x}, {\bf a}) \right)^{-1}$.
In practice, however, due to large dimensionalities of ${\bf x}$ and ${\bf y}$,
some kind of a PTF product model must be employed.
When the terms in the PTF product expansion,
such as jet posterior transfer functions, are derived,
correlations between different ${\bf x}$ dimensions
(caused, {\it e.g.,} by energy-momentum conservation and
complicated phase space boundaries) are ignored.
During Monte Carlo integration, lepton and jet variables
are sampled first, while proper kinematics is subsequently
enforced by imposing constraints on invisible particles (neutrinos).
In addition, the factor
$\int_{\Phi_i} W_{i}({\bf y} | {\bf x}, {\bf a}) d {\bf x}$ which depends on ${\bf a}$
and normalizes $P_{i}({\bf y} | {\bf a})$ in~\eref{eq:dlmintegral} is 
usually neglected
({\it i.e.,} assumed to be constant). Compounded, these approximations
result in some degradation of the statistical model and 
a biased maximum likelihood estimator of \Mt ~\cite{ref:cdf.ljets.dlm2, ref:cdf.ljets.dlm3,
ref:cdf.dilep.dlm}. The estimator bias
is subsequently removed by a calibration procedure
(see Section~\ref{sec:calibration}).

\subsection{Background handling}

In addition to optimizing event sample selection requirements
and heavy flavor tagging, a number
of methods have been developed for the purpose of strengthening
the background suppression
and for taking into account background contamination in the \Mt\ estimates.
Increase in the signal fraction $f_s$ improves sample sensitivity to \Mt.
At the same time, background discrimination should preferably be implemented
in such a manner that this fraction (a nuisance parameter in the measurement)
is not correlated strongly with the \Mt\ estimate.
The relative importance
of these arguments in the context of any particular \Mt\ measurement
method is not obvious {\it a priori}.
As the results (with the notable exceptions
of~\cite{:2007qf, ref:cdf.dilep.neuroevolution}) are usually presented 
without describing the extent to which different background
suppression options were explored,
relative merits of different background handling
techniques are difficult to compare.

\begin{figure}
\centerline{
\epsfig{file=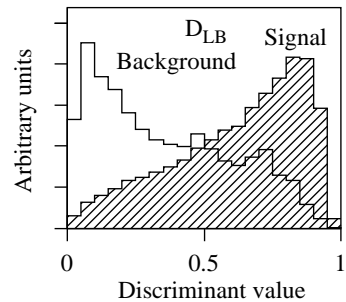,width=.25\textwidth}
}
\caption{
The D\sub{LB} discriminant plotted
for the $\Mt = 175$~GeV/$c^2$ $t\bar{t}$ sample (hatched) and the simulated
background (not hatched). The histograms are normalized to
the same area~\cite{ref:d0.ljets.run1prd}.}
\label{fig:d0dlb}
\end{figure}

Correlation reduction between $f_s$ and \Mt\ was emphasized
in~\cite{ref:d0.ljets.run1prl, ref:d0.ljets.run1prd}.
With this purpose in mind, one can not rely upon the most
distinguishing properties of $t\bar{t}$ events,
the invariant masses of various jet combinations
and the large amount of energy visible in the detector,
for background rejection. Topological event characteristics have
to be exploited instead, such as differences between angular distributions of
$t\bar{t}$ decay products and particles produced in background processes. 
Even though there appears to
be no simple kinematic variable of this kind with high discriminating
power, optimized combinations of multiple variables have been used
in practice with considerable success. For example,
the ``low bias discriminant'' technique of~\cite{ref:d0.ljets.run1prl,
ref:d0.ljets.run1prd} developed for the $\ell$+jets channel
employed a likelihood ratio discriminant in the form
of a cut on D$\sub{LB}({\bf y}) = \frac{p_s({\bf y})}{p_s({\bf y}) + p_b({\bf y})}$,
where $p_s({\bf y})$  and $p_b({\bf y})$ are the quasi-probability densities
of the signal and background, respectively.
The set of observed variables {\bf y} (missing transverse energy,
aplanarity, centrality, and one other variable which characterizes
average angular separation between jets) was chosen
and the shapes $p_s({\bf y})$ and $p_b({\bf y})$
were tuned in such a manner that the distributions of D$\sub{LB}({\bf y})$ were
independent from \Mt. The discriminant distributions for simulated samples
of signal and background events are shown in Figure~\ref{fig:d0dlb}.
The D$\sub{LB}({\bf y})$ cut was then optimized
for the best overall statistical uncertainty while 
encumbering little correlation between $f_s$ and \Mt. The use of a neural network with the same input variables was explored as well, with similar results.

Neural networks have also found their use in the \Mt\ measurements which
do rely, at least partially, upon energy-dependent variables for background
discrimination. According to this approach, better signal/background separation
is valued over reduced
correlation between $f_s$ and \Mt. These studies
include~\cite{ref:freeman, ref:plujan} in the $\ell$+jets channel
and~\cite{Abazov:2004ng, Aaltonen:2010pe, :2007qf, LucaNote}
in the all-hadronic channel.

An advanced investigation of the effect of background
suppression on the precision of \Mt\ measurement in the dilepton channel
has been carried out 
in~\cite{ref:cdf.dilep.neuroevolution}. In this study,
a neural network with six input variables was directly optimized 
for the best statistical uncertainty of the \Mt\ estimator.
The estimator itself was obtained by MEM in which the background probability
was calculated for the $Z/\gamma^{*} + 2$~jets and $WW + 2$~jets production
as well as for jets faking leptons in $W + 3$~jets.
The NN optimization algorithm (neuroevolution) allowed for
modification of both NN topology and weights. As a result,
the space of potential selection requirements was explored
in considerable detail. Counter to prior expectations, the best discriminator
permitted a significant fraction of background events (about 65\%) to remain in the sample.
The loose selection resulted in $\approx$20\% improvement in the {\it a priori}
statistical uncertainty in comparison with the previous dilepton sample definition
employed in~\cite{ref:cdf.dilep.matrel.prd}.

A number of \Mt\ measurements performed with MEM did not evaluate
background probability explicitly. The event observation probability
was formed using the signal hypothesis only,
so background did not contribute to~\eref{eq:prob_integral1}
and had to be taken into account by other means.
It is common for such measurements to discard events whose
peak \Mt\ likelihood magnitude is below a certain cutoff.
This requirement discriminates the signal not only against the background
but also against $t\bar{t}$ events in which an incorrect set of
jets was selected to form the $t\bar{t}$ system (this can happen,
for example, due to a presence of an energetic initial state radiation jet).
Several distinct approaches can be employed in order to form
a consistent \Mt\ estimate. One can fall back onto the
template method~\cite{Aaltonen:2008bg}, adjust the likelihood so that the
background contribution is removed on
average~\cite{ref:freeman, ref:plujan}\footnote{A similar approach
called ``pseudolikelihood method'' was applied as a crosscheck
in the template-based measurement~\cite{ref:d0.ljets.run1prd}.},
or just correct for background presence
in the calibration procedure~\cite{ref:cdf.ljets.dlm2}. The template method
can handle an arbitrary background mix and therefore it was used
in the background-dominated all-hadronic channel; however,
event-by-event resolution information was discarded by
using only the peak mass instead of the complete likelihood
curve. The other approaches can function in signal-dominated
samples due to an important feature of the Breit-Wigner mass
shape of the top quark (which limits the \Mt\ resolution that can be achieved
in a single event even by a perfect detector): the slope of
its log-likelihood is bounded. Because of this, any noise event,
no matter how far away from the signal, can
have only a limited impact on the estimate. In other words,
if the MEM analysis does not employ the narrow width approximation
for the top quark, the maximum likelihood estimate of \Mt\ is 
guaranteed to be robust and can tolerate certain deficiencies
in the background modeling.

\subsection{Calibration and statistical uncertainty}
\label{sec:calibration}

Due to the complexity of the physical processes both in the $t\bar{t}$
production/decay
and in the detector response, their statistical descriptions
invariably involve a number of approximations and simplifications.
The assumptions built into such descriptions, combined with the limited
amount of available data, result in biased point estimators of \Mt\
and/or incorrect interval estimator coverage prior to calibration.
Fortunately, the consequences 
of such model misspecifications can be studied by testing the
estimator performance with artificial event samples
produced by combinations of physics Monte Carlo generators and
detector simulation packages. 
% While this approach {\it per se} 
% simply utilizes the most detailed model available at the time and does not
% take account all the differences between data and Monte Carlo, it
% nevertheless can lead to important insights about the consequences
% of various approximations. 
The problem of the \Mt\ estimator calibration
is thereby split into two parts which could be analyzed independently:
derivation of corrections with simulated samples and
characterization of the
differences between simulated samples and observed data. While the first
part of the problem is specific to a particular measurement, the
second part 
is studied intensively over the whole lifetime of the
experiment (about 25 years for CDF and \D0 ) which eventually
results in
a well tuned, highly detailed detector response simulation model.
Therefore, authors of all \Mt\ measurements described in this article 
assume that their detector simulations can accurately predict
both the average detector response and its inherent randomness
for the purpose of evaluating the bias and the statistical
uncertainty of their \Mt\ estimators from the simulated samples.

When the expected statistical
uncertainty of the measurement is small, it is sufficient
to utilize a linear approximation to the behavior of a biased
point estimator, ${\hat {\bf a}}$, 
as a function of the ``true'' estimated parameter ${\bf a}$ (which can
include, for example, \Mt\ and $JES$): 
${\hat {\bf a}}({\bf a})  = {\hat {\bf a}}({\bf a}_{\,0}) + 
J_0 ({\bf a} - {\bf a}_{\,0})$. The Jacobian matrix
$J_0 \equiv \left( \frac{\partial {\hat {\bf a}}}{\partial {\bf a}} \right)$
is evaluated numerically, with simulated samples,
for the parameter value ${\bf a}_{\,0}$
close to the expected measurement result.
Assuming that
$J_0$ is not singular, a consistent point estimator
is obtained by ${\hat {\bf a}}\sub{c}  = {\bf a}_{\,0} + 
J_0^{\,-1} ({\hat {\bf a}} - {\hat {\bf a}}({\bf a}_{\,0}))$. This
transformation is often referred to as the ``mapping function''.
As a rule, the estimator ${\hat {\bf a}}$ is accompanied by an uncalibrated
estimate of the parameter covariance matrix, ${\hat V}\sub{u}$ (which can be
obtained, for example,
from the Hessian matrix of the sample log-likelihood at the maximum).
After application of the mapping function, this matrix
is adjusted according to the standard
error propagation
formula: ${\hat V} = J_0^{\,-1} {\hat V}\sub{u} J_0^{\,-1T}$.
Even if
the resulting ${\hat V}$ estimate is asymptotically correct, its finite sample
behavior is typically not well understood. In the
\Mt\ measurements, the parameter correlation coefficients
which can be extracted
from ${\hat V}$ are usually not of interest, and the complete calibration
of ${\hat V}$ is not performed. Instead, the standard deviations of
individual parameters are scaled so that the 
pull\footnote{``Pull'' is the difference between the estimated value
of a parameter and its true value, divided by the estimated standard deviation.}
distributions of
these parameters (with all other parameters eliminated by profiling
or marginalization) have unit width. These pull distributions 
are obtained with a large number of ``pseudo-experiments'' in which
${\hat {\bf a}}\sub{c}$ is evaluated for simulated event samples whose
number of entries and sample composition
is consistent with that expected in the data.
Resampling techniques are commonly utilized to increase the effective
number of pseudo-experiments without simulating additional
events~\cite{ref:resampling}.

\section{Sources of systematic uncertainties}
\label{sec:systematics}

In addition to the actual event samples used to determine the mass of
the top quark, the \Mt\ measurement results depend on a number of
inputs, both theoretical and experimental. A number of limitations on
the expected precision is thus imposed, either due to the
finite size of data samples which could be used for detector calibration
or due to the incomplete description of perturbative and non-perturbative
parts of the relevant QCD processes. To characterize such a lack of knowledge,
possible sources of systematic uncertainties are divided into a number of
categories which, for all practical purposes, can be considered
independent. Systematic uncertainty contributions from each
category are subsequently added in quadrature.

Several different methods
are used to evaluate the effect of the independent uncertainty
sources on \Mt\ estimates.
The first step usually consists in
identifying a nuisance parameter associated with a certain particular
source in the relevant theoretical or detector
model. It is commonly assumed that this parameter
has a Gaussian prior with a range of variations, $\sigma$, determined from
a number of consistency checks between the observations and the model.
One of the following approaches is employed afterwards:
\begin{itemize}
\item The event sample likelihood
      is evaluated as a function of this parameter. The parameter is
      then marginalized or the likelihood is profiled. This procedure
      results in a statistically efficient treatment of the uncertainty
      source under study:
      the events are combined in an optimal manner which takes into
      account the system response to the nuisance parameter variations
      for the particular kinematic configuration encountered. 
      This method is CPU-intensive,
      and so far
      it has been applied only to the most important nuisance parameter
      in the \Mt\ measurement: the detector jet energy scale.
\item Large simulated event samples are produced with
      $- 1$, $0$, and $+ 1 \sigma$ changes in the parameter, and
      with the input top mass close to the obtained result.
      Corresponding \Mt\ estimates
      are obtained: $M\sub{t}^{-}$, $M\sub{t}^{0}$, and $M\sub{t}^{+}$.
      These estimates are then sorted in increasing order.
      If, after sorting, $M\sub{t}^{0}$ lies between 
      $M\sub{t}^{-}$ and $M\sub{t}^{+}$ then
      the estimator systematic uncertainty associated with this source
      is taken to
      be $|M\sub{t}^{+} - M\sub{t}^{-}|/2$. If, on the other hand,
      $M\sub{t}^{0}$ ends up outside the interval defined by the
      $M\sub{t}^{-}$  and $M\sub{t}^{+}$ endpoints then the uncertainty is
      evaluated as 
      $\mathrm{max}(|M\sub{t}^{+} - M\sub{t}^{0}|/2,
                    |M\sub{t}^{-} - M\sub{t}^{0}|/2)$.
      The main advantage of this method is its simplicity.
      It works well on average 
      if the \Mt\ estimator can be reasonably expected to
      change linearly as a function of the nuisance parameter
      and if the $|M\sub{t}^{+} - M\sub{t}^{-}|$ shift is much
      larger than the statistical uncertainty of the $M\sub{t}^{+}$
      or $M\sub{t}^{-}$ determination.
\item Instead of generating three separate simulated samples,
      an event reweighting
      scheme is applied to the $0\sigma$ sample in such a manner that
      the effective value of the parameter under study is shifted in
      either direction while other parameters are kept intact.
      The changes in the \Mt\ estimator response
      are processed as in the previous method. 
      While designing such a reweighting scheme can be
      a complicated problem in itself, this technique can be used to study
      very small systematic shifts
      because the statistical uncertainty of the $M\sub{t}^{+}$
      and $M\sub{t}^{-}$ determination mostly cancels out in the
      $M\sub{t}^{+} - M\sub{t}^{-}$ difference.
\end{itemize}
It is not always possible to identify
a limited set of nuisance parameters or to define reasonable
ranges of their variations. For example, when there are only two
reasonable hypotheses available for comparison, the full extent
of the change
in the \Mt\ estimator between these hypotheses is used
as the systematic uncertainty.

% Systematic sources have been split whenever possible in detector, simulation, physics effects in order to minimize the overlap between the various effects. For this reason all systematic sources are considered as independent and thus their effect on each top mass measurements is summed in quadrature. The ideal situation is to perform a systematic variation in the definition of each object in the event. Whenever this is not possible, systematic shifts are performed event-by-event. In some instances, the Monte Carlo samples used to model signal and/or backgrounds are reweighted to account for possible systematic variations. Whenever none of the above is possible, a new Monte Carlo sample containing a systematic variation of the observable under study is generated. The last instance is the less ideal because small systematics shifts can be covered by the limited size of the MC sample generated.
%The reader should be aware that this understanding might change with more collisions analyzed and more refined theoretical computation for perturbative effects and increasingly complex modeling of non-perturbative effects.
%	WE NEED A COMMENT ON THE FACT THAT SOME UNCERTAINTIES ARE OVERLAPPING AND WE ARE WORKING ON MINIMIZING THE OVERLAP
%	LET'S ADD A STATEMENT THAT SAYS THAT WHENEVER WE CAN WE 1) VARY THE SYSTEMATIC EVENT-BY-EVENT 2) VARY THE SYSTEMATIC IN THE WHOLE SAMPLE BY REWEIGHTING IT 3) PRODUCE A NEW SAMPLE WITH THE SYSTEMATIC VARIATION

\subsection{Jet energy scale}
\label{sec:jesUncert}

% Jets are identified and their energies measured by the CDF and D0 collaboration through the calorimeter. 
%The single particle response of the calorimeter has been studied using test beam data and {\em in situ} using minimum bias, $J/\Psi \to e^+ e^-$ and $Z \to e^+ e^-$ data. 
% Jet response in data and Monte Carlo have been studied in $\gamma$+jet, jet-jet, $Z$+jet data. 
% The $Z\to b\bar b$ events are much harder to isolate because of the irreducible QCD $b\ bar b$ background. Still, $Z\to b\bar b$ data have been studied in order to investigate specifically the $b-$jet energy scale~\cite{Donini:2008nt}. Results suggest that the $b-$jet energy scale is under control by using the corrections and systematic derived in a light quark/gluon dominated samples. 
%The latter study is not used in this paper as it makes use of the larger $R=0.7$ cone size for jets.
%The jet energy scale systematic is split into its several subcomponents in order to describe the correlations among the several CDF and D\O\ results that will be an important ingredient in the combination described in Section~\ref{sec:combo}. 

% Depending on the decay modes of the $W$ bosons,
As the $t \bar t$ signature contains at least two jets in the final state and
most measurement techniques make explicit use of the jet transverse momenta,
imperfect calibration of the detector 
jet energy scale has traditionally resulted in the single largest
systematic uncertainty of the top quark mass estimates.
%all the measurements that make use of the information 
%coming from the jet $E_T$s and do not explicitely 
%use the {\in situ} $W \to q \bar q^{\prime}$, 
In the analysis of Tevatron Run~I data it was
found that 1\% change in the jet energy scale
corresponds, as a rule of thumb, 
to a 1-1.5~GeV/$c^2$ shift in the \Mt\ estimate. Given the 2-3\% 
jet energy calibration uncertainties the \D0 \ and CDF collaborations quote,
this translates into about 3~GeV/$c^2$ limitation on the \Mt\ measurement
precision~\cite{ref:cdftdr}.

Recent \Mt\ measurements have by far surpassed this precision by
calibrating the jet energy scale {\it in situ} with the mass of the
$W$ boson. Both top quark and $W$ boson masses can be simultaneously
estimated in $t\bar{t}$ events in which at least
one of the $W$s decays hadronically.
If the correlations between these
estimates are known, a constraint imposed on the $W$ mass estimate
% to its well-known value
leads to an improvement in the top mass determination.
Technically, this idea is more conveniently realized by introducing
an overall factor, $JES$, which is used to scale
four-momenta of all jets in the
event sample with respect to their reference values obtained with
a standard calibration~\cite{ref:cdf.multivariate.note1, ref:cdf.ljets.mtmw}.
The $t\bar{t}$ samples accumulated by CDF
and \D0 \ are sufficient to estimate $JES$ defined in this manner
with a significantly higher precision than that provided by
the standard jet calibration.

% This precision has been by far surpassed thanks to the fact that $t \bar t$ events produce at least one hadronically decaying $W$ boson in the final state in $\sim 90\%$ of the decays. The $W \to q \bar q^{\prime}$ decay(s) can then be used to measure the JES systematic for light quark jets, by constraining the reconstructed dijet mass to be equal to the precisely measured $W$ mass. The light quark JES uncertainty is thus statistical in nature, and is uncorrelated among the measurements performed. Often, the - correlated - prior knowledge of the JES is used in the measurements. Still, the present statistics is sufficient to set the scale to higher precision. A number of caveats prevent the CDF and D0 collaborations from using this calibration in analyses other than the very same dataset where the calibration is derived.
% First, $t \bar t$ statistics at the Tevatron is not sufficiently large to probe the expected jet $p_T$ and $\eta$ dependence of this systematic. The large jet multiplicity and momentum scale of top-antitop events make any extrapolations to datasets with different jet multiplicity and $p_T$ merely speculative. Finally, there are 

Although the automatic adjustment of the overall $JES$ factor improves
the most important jet-related systematic uncertainty, notable
differences remain in the parton fragmentation and hadronization
dynamics between light quarks produced in the $W$ boson decays and
$b$ quarks. 
%
% , leads to
% a number of additional considerations. 
%
%There are notable differences in the parton fragmentation and hadronization dynamics whenever the parton is a light quark as in the $W$ decays under study, or a $b$ quark or a gluon. For the above reason, the JES as measured {\em in situ} with light flavor jets cannot does not fully represent the $b-$jets energy scale. 
%Equally, it is hard to be used to other samples where generally the quark/gluon composition of the jets is unknown.
%Since the jet energy corrections are derived on data samples deprived of heavy flavors, an additional uncertainty comes from considering the different properties of $b$ quarks. 
%
% Three different sources of systematic uncertainties have been considered: 
%
Four additional sources of systematic uncertainties have
been considered to account for these differences:
1) the uncertainties on the branching fraction of semi-leptonic decays of $b$ and $c$ quarks as measured by the LEP experiments; 2) the uncertainties on $b$ quark fragmentation parameters~\cite{Peters:2006zz} computed as the difference between the parameters measured by the SLD collaboration~\cite{Abe:1999ki} and the ALEPH, DELPHI, and OPAL collaborations~\cite{Heister:2001jg, BenHaim:2004kn, Abbiendi:2002vt}; 3) the uncertainty in the calorimeter response difference between light and heavy flavor quarks; and 4) the calorimeter response non-linearity
which contributes an uncertainty due to different transverse
energy spectra of $b$ quarks and $W$ decay products.

%There is an intrinsic uncertainty due to the non-perturbative nature of quark and gluon fragmentation and in the amount of jet energy that flows outside of the clustering cone. 
%HERE I WANTED TO TALK ABOUT L7 SYSTEMATIC. THIS IS PROBABLY NOT NECESSARY AS LINA TALKS ABOUT IT. STILL, IF WE CHOOSE TO TALK ABOUT IT WE NEED TO MENTION THE FACT THAT PYTHIA-HERWIG IS INCLUDED IN L7

%A component of the JES uncertainty originates from limitations in the data samples used for calibrations. This  corresponds to uncertainties associated
%with the $\eta$-dependent JES corrections which are estimated using di-jet data events. For D\O\ this includes uncertainties in the calorimeter response for light jets and from the sample dependence of using $\gamma$+jets data samples to derive the JES. 
%MAYBE THIS SHOULD GO IN THE COMBINATION

In a number of recent \Mt\ measurements~\cite{ref:cdf.ljets.mtmw,
ref:cdf.template.pub1, ref:plujan, ref:cdf.template.latest},
$JES$ is related to the fractional systematic
uncertainty from prior calibration, $\sigma_{jet}$, by
\begin{equation}
JES = 1 + \Delta_{JES} \cdot \sigma_{jet},
\label{eq:jesfromsigma}
\end{equation}
where $\Delta_{JES}$ is the underlying nuisance
parameter which represents the relative jet energy shift 
in units of $p_T$-dependent
$\sigma_{jet}$ (see Figures~\ref{jet-sys-cdf} and \ref{jet-sys-d0}).
This formula explicitly manifests the
assumption that energies of jets with different transverse momentum 
or flavor are completely correlated.
Subsequent treatment of the $\Delta_{JES}$ parameter ({\it e.g.,} elimination
by profile likelihood) minimizes the dependence of the \Mt\ estimate on
this fully correlated component of the jet energy scale uncertainty.
It is worth mentioning that
this treatment can be extended so that it takes
into account all important uncertainties related to jet reconstruction
in a uniform and consistent manner.
The prior knowledge of the jet energy systematic uncertainty
can be quantified not only with a $p_T$-dependent $\sigma_{jet}$ but
with a complete covariance function which depends on jet $p_T$ and flavor.
      According to the Karhunen-Loeve theorem~\cite{ref:loeve},
such a covariance function permits a decomposition
of the jet energy systematic uncertainty into orthogonal
independent components. Components corresponding
to several largest eigenvalues of the covariance function
could then be treated as
measurement nuisance parameters, by {\it in situ} calibration or
by other techniques. Another important advantage of this decomposition
stems from the simplicity with which various jet-related measurements
performed with the same detector
can be cross-calibrated ({\it i.e.,} component constraints obtained
in one measurement can be immediately applied in another).
We therefore encourage derivation and use of
such covariance functions (or covariance matrices in case
the jet $p_T$ range is binned) for representing jet energy scale
uncertainties in future measurements of \Mt\ and other jet-related
quantities.

% add the residual JES sys here
After calibration of an overall JES factor as described above,  the next step 
is to evaluate the systematic uncertainty due to the uncertainties on
the individual jet corrections. These are 
shown in Figures~\ref{jet-sys-cdf} and \ref{jet-sys-d0} for CDF 
and \D0 , respectively.
The $p_T$ dependence of the systematic uncertainties on the individual
 corrections are very different. 
A ``Residual JES'' systematic uncertainty (see
 Tables~\ref{table:sysbestcdf} and \ref{table:sysbestd0} 
in Section~\ref{sec:lepjets}) 
is evaluated by adding in quadrature the uncertainties 
for each component. These are obtained in the usual manner, {\it i.e.},
by generating 
samples with the correction shifted by $\pm 1 \sigma$ to obtain 
the corresponding mass shifts. The procedure described earlier is then 
followed to obtain a final uncertainty.

\subsection{Lepton-related uncertainties}
\label{sec:lepuncert}

The response of the detectors to leptons is calibrated using $J/\Psi$
decays and $Z \rightarrow \ell^+ \ell^-$ events.
% Similarly to the jet energy response, there are multiple
% effects to be considered. 
The uncertainty on the overall lepton $p_T$ scale in $t\bar{t}$ 
events is conservatively estimated to be $1\%$.
As in the case of jet energy response determination, there
are additional effects to consider.
% The difference between lepton $p_T$ resolutions in data and
% in simulated samples is evaluated and its effect on
% the \Mt\ estimates is determined.
The muon momentum resolution is found to be better in MC than in data.
Appropriate extra smearing is applied to the muon $p_T$ in the
simulated samples in order to estimate the corresponding \Mt\ systematic
uncertainty.
Further, for $\ell$+jets and dilepton channels,
the complete data acquisition chain is triggered
by the presence of a charged lepton in the detector.
The trigger turn-on curve is computed with its uncertainty, and the
uncertainty is propagated to
the \Mt\ estimate. 
These additional \Mt\ variations are small, and they are
neglected if their magnitude is below 100~MeV/$c^2$.

\subsection{Uncertainties from Monte Carlo generators}
\label{sec:mcgenuncert}

%\begin{itemize}
%\item[parton shower/hadronization]
% Move this comment to section 4.  
% The PYTHIA parton showering and hadronization model tuned 
%to LEP and Tevatron
%data provides a good description of jet formation and development
%at the Tevatron~\cite{Acosta:2005ix}. 
%Agreement between HERWIG modeling of collider jet shapes
%and data improves once the energy flow from minimum bias
%events is properly taken into account using the JIMMY subroutine.
%PYTHIA  uses mass ordering, while HERWIG uses angular ordering) 

\hspace*{\parindent} {\bf Signal Modeling}.
CDF evaluates the signal modeling systematics by comparing the \Mt\
measurement calibrations obtained with PYTHIA and HERWIG. 
\D0 \ uses the difference in 
reconstructed $M_t$ values obtained with their default generator, ALPGEN 
interfaced to PYTHIA, and with PYTHIA proper.
%\item[MC@NLO]
Recently, \D0 \ has evaluated the uncertainty due to
ignoring higher order Feynman diagrams for $t \bar t$ production
% I see they have hadronization sys, but where is the parton shower term? 
by  comparing  $t \bar t$ samples generated by 
MC@NLO and by ALPGEN + HERWIG~\cite{ref:d0.ljets.matrel.may2011} 
(see Table~\ref{table:sysbestd0} in Section~\ref{sec:lepjets}). 
%CDF has evaluated this component of the signal systematics comparing
%MC@NLO with HERWIG itself~\cite{ref:cdf.template.latest}. It is not
%in the publication! 

%(I CAN'T UNDERSTAND WHAT THIS STATEMENT IS SAYING. DOES IT SAY THAT
%CDF COMPARES MC@NLO WITH HERWIG? DO WE REALLY DO THIS?yES Hyun Su did
%this)
%(THERE IS A DUPLICATED DESCRIPTION OF THIS LATER IN THIS SECTION,
%WHEN ISR/FSR IS DISCUSSED.)
%\item[Color Reconnection (CR):] 

{\bf Color Reconnection (CR)}. As the top quark and $W$ bosons decay
 quickly on the timescale associated with the parton shower and
 fragmentation processes ({\it i.e.}, $1/\Gamma_t, 1/\Gamma_W \ll
 1/\Lambda\sub{QCD}$), it is possible that
 the decay  products from the different top quarks interact with
 each other via color reconnections. 
Color reconnection effects were first investigated at LEP~\cite{Azzurri:2006un} as a possible source of systematic uncertainty in precision measurements of the $W$ mass in $WW$ events; no evidence of color reconnections was found in these studies.
The partons emerging from the Tevatron $p \bar p$ initial state carry color charge, thus phenomenological color reconnection modeling is significantly more complicated than for the $e^+ e^-$ initial state at LEP. Recently, 
%Monte Carlo models providing an adequate description of Tevatron $p \bar p$ 
a new version of PYTHIA (v6.4) with improved description of parton 
shower  ($p_T$-ordered), multiple parton interactions, underlying 
event, and color 
reconnection (CR) effects has been released~\cite{Skands:2007zg} 
and subsequently tuned to collider data~\cite{Skands:2009zm}. 
The CR systematic uncertainty is calculated
as the difference between the \Mt\ estimators obtained with
the event samples generated by PYTHIA~6.4 tune ``A-pro" and 
PYTHIA~6.4 tune ``ACR-pro". These tunes differ in the color
reconnection model but use the same mass-ordered parton shower
model as the one used in the default PYTHIA~6.2.
This approach was cross-checked with the $p_T$-ordered parton shower,
and compatible top mass difference was found.
%
%Alternative tunes in PYTHIA~6.4 that use a $p_T$-ordered parton shower have been tested with and without the color reconnection model included and the \Mt\  variation has been found in agreement with the former prescription. 
%
The difference between the mass-ordered and the $p_T$-ordered parton shower 
prescriptions is not taken into account in the signal systematics,
as preliminary versions of the latter
did not match CDF jet shape evolution~\cite{BenLinaJay}.
More studies of the $p_T$-ordered parton shower model are in progress.

%
%  \item[Multiple Hadron Interactions (MHI):] 
{\bf Multi Hadron Interactions (Pile-up)}.
The luminosity profile, {\it i.e.}, the number of interactions per
bunch crossing, is different in the data and in the simulated
samples used to calibrate the \Mt\ measurements. This disagreement
leads to imperfect pile-up modeling. To first order, appropriate dependence
of the jet energy corrections on the number of primary vertices (NPV)
found in the event cancels this effect out, so the corresponding systematic
uncertainty is expected to be small. Nevertheless, a conservative
estimate of the uncertainty due to multiple interactions
is made by studying the dependence
of the \Mt\ measurements on NVP in the simulated samples as well
as by comparing the jet response as a function of NPV
in the simulated $t\bar{t}$ samples and in minimum bias events.

% The difference is due to the practical difficulty of
% continually updating the Monte Carlo simulation in
% order to reproduce the steady increase of the Tevatron
% collider daily luminosity during data-taking.

%\item[ISR/FSR]
{\bf {Initial and Final State Radiation (ISR/FSR)}}.
Default Monte Carlo generators used by the CDF and \D0 \ collaborations 
to simulate $t\bar{t}$ production and decays utilize LO matrix element
of the process. NLO effects appear in the form of additional initial (ISR) and
final state radiation (FSR) as well as loop corrections. To evaluate the
systematic uncertainty associated with the imprecise radiation modeling,
additional $\ttbar$ samples are generated by
PYTHIA with increased/decreased amount of ISR and FSR.
% + and - one sigma.
% A data-driven method is used to estimate an uncertainty on the initial
% and final state radiation contributing to the collision.
The uncertainty on the parameters which control the strength of the ISR
was estimated
by CDF using  $p \bar p \to Z^*/\gamma^{*}+$jets~$\to \mu^+ \mu^-$+jets
collisions~\cite{ref:cdf.ljets.mtmw}. Just as $t\bar{t}$ events, these are
produced predominantly via $q \bar q$ annihilation. 
% Similarly to $t \bar t$ production at the Tevatron, 
% $p \bar p \to Z^*/\gamma^*$ plus jets $\to \mu^+ \mu^-$ plus jets 
% collisions are produced mainly through $q \bar q$ annihilation. 
The transverse momentum of the muon-antimuon system, $p_T^{\mu^+
\mu^-}$, depends on the ISR presence in the event. The $p_T^{\mu^+
\mu^-}$ distribution was used to constrain the generator
parameters affecting the ISR modeling. As both the initial and the
final state radiation (FSR) processes are subject to DGLAP
evolution, the same parameters determine the amount of FSR in the
generated samples.  Thus, the $\pm 1 \sigma$ changes to these
parameters, 
which are used to derive corresponding $M\sub{t}^{+}$ and $M\sub{t}^{-}$,
affect both ISR and FSR simultaneously.
% 
% DO WE NEED TO SAY THAT THERE IS ANOTHER METHOD THAT WE DO NOT USE? I
% I WOULD TAKE THIS OUT
% An alternative method of estimating the 
% uncertainty due to ISR/FSR consists in
% comparing PYTHIA/HERWIG inclusive $t \bar t$ production simulation in which the
% extra radiation is added by the parton showering with a simulation
% in which the extra radiation is produced at the tree level.  \D0 \ and
% CDF use ALPGEN and MADGRAPH~\cite{MADEVENT}, respectively, both linked 
% to PYTHIA for
% showering and hadronization, to produce samples with $t\bar t + 0/1/2$
% or more additional hard partons in the final state.
% For these samples, the decision which partons to add at the tree level
% depends on a $p_T$ cutoff. This cutoff is varied to test the 
% robustness of the method. The \Mt\ uncertainty determined by this method
% is smaller than that obtained using $\mu^+ \mu^-$+jets data. So we
% use the first method. 

%\item[PDF]
{\bf  Parton Distribution Functions (PDFs)}.
The PDFs used to model $t\bar{t}$
production are determined by fitting multiple observations performed
at a number of $e p$ and $p \bar p$ collider experiments.
Statistical uncertainties of these fits, together with the approximations used to
derive the PDF functional representations, contribute to the \Mt\
measurement systematic uncertainty. CTEQ5L~\cite{CTEQ5L} is used as the default
PDF set for the CDF $t\bar{t}$ Monte Carlo generators. This set is compared to the
MRST72~\cite{Martin:1998sq} set in order to estimate the
uncertainty arising from the possibility of multiple functional representations.
The PDF sets MRST72 and MRST75~\cite{Martin:1998sq} are compared
to each other in order
to estimate the \Mt\ uncertainty due to the imprecisely
known $\alpha_s$ value.
%
% CDF also calculates the shift in \Mt\ that would result from
% variation in the $\alpha_s$ coupling, expressed as the difference
% between the MRST72 and the MRST75 sets.
%
The CTEQ6M PDF sets are produced together with the 
eigenvectors of the parameter covariance matrix
({\it i.e.,} the 20 principal components)~\cite{Pumplin:2002vw}.
Both CDF and \D0 \ evaluate the \Mt\ uncertainty due to
the imprecisely known PDF parameters by using the CTEQ6M sets
in which parameters are shifted along these eigenvectors
and by combining in quadrature the resulting shifts of
the \Mt\ estimate.

% provides eigenvectors representing $90\%$ coverage variations of 
% twenty uncorrelated parameters affecting the resulting PDFs. 
% CDF and \D0 \ estimate the difference in the measured \Mt\ 
% associated with the variation of each eigenvector in the 
% nominal Monte Carlo sample. The resulting PDF-induced 
% systematic uncertainty is the quadratic sum of the twenty individual shifts.

{\bf Background Modeling}. 
For simulation of the $W$+jets samples, there is an ambiguity associated with the choice of factorization and renormalization scales, represented in the literature as $\mu_F$, $\mu_R$, and/or $Q^2$, at which to evaluate the relevant matrix elements. 
%CDF and D0 chose to describe $W$ + jets processes using $Q^2 = M_W^2 + \sum_i p^2_{T,i}$, where the sum runs over all the outgoing partons at a given vertex. 
A set of additional $W$+jets samples is generated by doubling and halving
the $Q^2$. The background fraction and kinematic distributions are
re-evaluated using these samples. Corresponding \Mt\ estimator shifts
are included in the systematic uncertainty of the measurement.
% affecting the $W$+jets background processes.

%\item[Background normalization] 
{\bf{Background Normalization}}.
A number of SM processes involving high mass particles appear as
backgrounds to $t \bar t$ production, as described in
Section~\ref{sec:backgrounds}. As a rule, backgrounds are generated
at LO. It was found that background kinematic distributions
relevant for \Mt\ 
measurements do not change appreciably between LO and NLO simulations.
The LO shapes of such distributions ({\it e.g.}, $m_t$ in
the template-based measurements) are therefore normalized to NLO (or NNLO
if available) theoretical cross sections. The \Mt\ systematic
uncertainty due to background normalization is obtained by
studying the response of the \Mt\ estimator as these
normalizations are varied within their theoretical uncertainties.
For $W$+jets production, an additional uncertainty is contributed by
the imprecisely known fraction of $b$ jets in the sample.
This fraction is measured in the $W$ + 1 jet data and then extrapolated
to the $t \bar t$ event selection together with the uncertainties
obtained~\cite{Aaltonen:2010ic,d0-btag-algo}. 
The QCD multijet background kinematics and normalization is evaluated
using an independent dataset, and the uncertainty on its normalization
is propagated to the \Mt\ estimate~\cite{Aaltonen:2010pe}. 

% The total background normalization systematic uncertainty is  
% computed by adding in quadrature contributions from two different sources: 
% the overall background normalization is changed by the 
% estimated $\pm1 \sigma$ to obtain the corresponding \Mt\ shifts;
% also, the background composition is changed by varying each component 
% normalization within its uncertainties to
% obtain the corresponding \Mt\ shifts.
% 
% THIS IS ALREADY DESCRIBED WHEN WE SAY THAT WE STUDY THE RESPONSE
% OF THE MT ESTIMATOR TO CHANGES IN THE NORMALIZATION

%The uncertainties from modeling of the QCD multijet background determined 
%rom data and dominated by limited statistics. 
%The overall background normalization systematic is obtained as
%follows. This systematic is computed by adding in quadrature two
%different sources: first the CDF and \D0 \ collaboration change the
%overall background normalization up and down by the estimated $1
%\sigma$ and compute the corresponding \Mt\  shifts. Then the
%background composition is changed by varying each component
%normalizations within uncertainties, and computing the corresponding
%\Mt\  shifts. 

% \end{itemize}
\subsection{Other uncertainties in detector modeling}

%\begin{itemize}
%\item[jet ID and resolution]   
The \D0 \ collaboration finds a small difference in
the jet reconstruction efficiency for data and Monte Carlo samples.
The corresponding uncertainty on \Mt\ 
was found to be very small~\cite{ref:d0.ljets.matrel.may2011}. 
% ID efficiency work is mostly by D0, CDF does not find the difference 

%difference in resolution
The consistency in the jet energy resolution
between data and simulated events has been studied by \D0 \
for the $\gamma$+jet sample. It was found that 
the resolution is a few percent better in the simulation.
% The jet energy resolution difference between events generated
% with PYTHIA and data has been studied for $\gamma$+jet samples
% and was found to be a few percent better in the MC. 
\D0 \ smears the jet energies to match the resolution observed in the $\gamma$+jet events, and then estimates a systematic shift on \Mt\  by varying the 
smearing function within its uncertainty~\cite{ref:d0.ljets.matrel.may2011}. 
%CDF finds that that HERWIG reproduces the jet energy resolution quite well, so assumes this uncertainty is contained in the PYTHIA - HERWIG difference. As a cross-check, the analyses attempted smearing PYTHIA jets to match the discrepancy observed in the control region data, and repeated the measurement in this configuration, finding a negligible systematic shift.
For CDF, this effect is of the order of a few tens of MeV/$c^2$ and is 
thus neglected.

%\item[trigger efficiency uncertainty] 
Top-antitop events are collected by triggering data acquisition and recording using a certain set of requirements. The trigger efficiency uncertainties are small and their effects on \Mt\  measurements are usually neglected, 
as the physics objects requirements in the data 
analysis procedures are chosen to be sufficiently
above the trigger thresholds. 
% D0 has studied sys effects from trigger and found that they are small

%\item[b-tagging efficiency  again it is a D0 study]  
The $b$-tagging algorithm efficiency has been 
studied in QCD  $b \bar b$ events.
The systematic uncertainty has been assigned for
the difference between the efficiency found in the data
and in simulated events. This uncertainty has been subsequently 
propagated to \Mt\  measurements.
% 
% and a systematic uncertainty between the results obtained
% with data and with simulated samples
% has been assigned and propagated to \Mt\  measurements. 
% Furthermore, data shows a possible trend in this offset as a function of the jet $E_T$. This trend has been parametrized and the uncertainty on the parametrization assigned. The \Mt\  measurements are then redone varying this function and the corresponding systematic estimated. 
%
This effect is usually negligible in comparison with 
other systematic uncertainties~\cite{ref:d0.ljets.matrel.may2011}. 
%\end{itemize}

\subsection{Uncertainties from the measurement method}

This category includes all uncertainties stemming from the finite size
of simulated event samples used to calibrate the measurement, such as 
the uncertainty in the parameters of the mapping function.
For template-based measurements
in which templates are represented by continuous probability densities,
this also includes the template parameterization uncertainties estimated from 
a $\chi^2$ or maximum likelihood fit of the template distributions.

%\end{itemize}

\section{Tevatron Run II measurements}
\label{sec:tev_runii}

From the first Tevatron collisions in October 1985
and until the start-up of LHC 2.36~TeV center-of-mass
collision operations in December 2009, the Tevatron collider
at Fermilab was the only place on Earth where top quarks were
copiously produced. It is thus not surprising that until
2011 all direct measurements of the top quark mass utilized
the data collected
by the two main Tevatron experiments, CDF and \D0 . About 160~pb\supers{-1}
of integrated luminosity was delivered to these experiments
during the 1.8~TeV 
% center-of-mass energy
collision operation period known as ``Run~I'' (1992-1996).
By the time of this
writing (June 2011), about 11~fb\supers{-1} of data were accumulated
at $\sqrt{s} = 1.96$~TeV
during the ``Run~II'' period which commenced in March 2001 and is still
in progress. In this
section we discuss in greater detail the most advanced Tevatron measurements
of \Mt\ performed so far with the Run~II data.

\subsection{Lepton+jets topology}
\label{sec:lepjets}

Measurements performed in the $t\bar{t} \rightarrow \ell$+jets channel,
$\ell = e$ or $\mu$, have traditionally
resulted in the most precise estimates of the top quark mass
at the Tevatron. This particular channel is favored by its
relatively large branching fraction ($\approx 34$\%) and presence
of a lepton and missing energy in the final state which allows
for efficient background suppression. In addition, $W$ boson decays
into two jets provide an important reference point for detector
jet energy calibration.

At the time of this writing, the most precise single measurement
of \Mt\ is
performed with 5.6~fb\supers{-1} of data collected by the CDF detector
during Run II of the Fermilab Tevatron~\cite{ref:plujan}.
Most of the events used in this measurement
are collected with high transverse momentum lepton triggers that require
a well-reconstructed electron or muon candidate with $p_T > 18$~GeV/$c$
in the central detector region~\cite{ref:cdf_hiptleptrig}.
The lepton $p_T$ cutoff in the sample selection
criteria is increased to 20~GeV/$c$ in order to avoid
the difficult to model trigger turn-on curve. 
The sample also contains a fraction of events 
with 
% non-central 
loosely identified muons collected with a 
missing transverse energy trigger.

Exactly
four jets with transverse energy $E_T > 20$~GeV are required within the
pseudorapidity region $|\eta| < 2.0$, with at least one jet tagged
as a $b$ jet using a secondary vertex tagging algorithm. The presence
of the neutrino in the final state 
is exploited by imposing the requirement $\Etmiss > 20$~GeV.
The overall $t\bar{t}$ selection
efficiency with these criteria 
(including the $\ell$+jets branching fraction and trigger acceptance)
is about 2\%, while the resulting S/B ratio
in the selected sample is 3.6. 
% Further details on lepton identification,
% jet energy reconstruction, and heavy flavor tagging used in
% this measurement can be found in~\cite{ref:plujanThesis}.
The expected sample composition is shown in
Table~\ref{table:ljetssample}.
\begin{table}
\caption{Expected sample composition for the CDF $\ell$+jets
\Mt\ measurement with 
$\int \mathcal{L} dt = 5.6$~fb\supers{-1}. The $t\bar{t}$ contribution
is estimated using a cross section of 7.4 pb and
$\Mt = 172.5$~GeV/$c^2$~\cite{ref:plujan}.}
\label{table:ljetssample}
% \begin{indented}
% \item[]\begin{tabular}{@{}lcc}
\begin{tabular}{@{}lcc}
\br
Event Type & 1 $b$ Tag & $\ge$ 2 $b$ Tags \\
\mr
$W$ + Heavy Flavor & $129.5 \pm 42.1$ & $15.7  \pm  5.5$ \\
Non-$W$ QCD & $50.1 \pm 25.5$ & $5.5 \pm 3.8$ \\
$W$ + Light Flavor Mistag & $48.5 \pm 17.1$ & $1.0 \pm 0.4$ \\
Diboson ($WW$, $WZ$, $ZZ$) & $10.5 \pm 1.1$ & $1.0 \pm 0.1$ \\
Single Top & $13.3 \pm 0.9$ & $4.0 \pm 0.4$ \\
$Z \rightarrow \ell \ell$ + jets & $9.9 \pm 1.2$ & $0.8 \pm 0.1$ \\
Total Background & $261.8 \pm 60.6$ & $28.0 \pm 9.6$ \\
$t\bar{t}$ Signal & $767.3 \pm 97.2$ & $276.5 \pm 43.0$ \\
Total Expected & $1029 \pm 115$ & $304.5 \pm 44.1$ \\
Events Observed & 1016 & 247 \\
\br
\end{tabular}
% \end{indented}
\end{table}

The subsequent analysis of the
event sample is performed by the matrix element method.
A highly detailed statistical
model of the
$p\bar{p} \rightarrow t\bar{t} \rightarrow \ell$+jets signal process
and of the detector response is developed and utilized.
The process density in the parton phase space, $\Phi_0$, is described
by a leading order matrix element which includes
both $q\bar{q} \rightarrow t\bar{t}$ and
$gg \rightarrow t\bar{t}$ production processes, as well as
$t\bar{t}$ spin correlations~\cite{ref:kleissstirling}.
Lepton momenta are assumed to be well-measured.
The detector response to jets is modeled by nonparametric
statistical techniques as a function of parton transverse momentum
and jet mass. A dedicated kinematical mapping is used to relate
the leading order parton-level phase space to a~more realistic
phase space, $\Phi_1$, in which jets can be massive. Both angular
and transverse momentum jet resolutions are represented, and corresponding
degrees of freedom are integrated over $\Phi_1$ together with the
matrix element evaluated in $\Phi_0$ with the aid of
the $\Phi_1 \rightarrow \Phi_0$ mapping. 
To improve the convergence rate, the 19-dimensional
numerical integration is performed by Quasi-Monte Carlo
(the dimensionality reduction assumptions leading to
this integral were
described in Section~\ref{subsec:mem}).

\begin{figure}
\centerline{
\epsfig{file=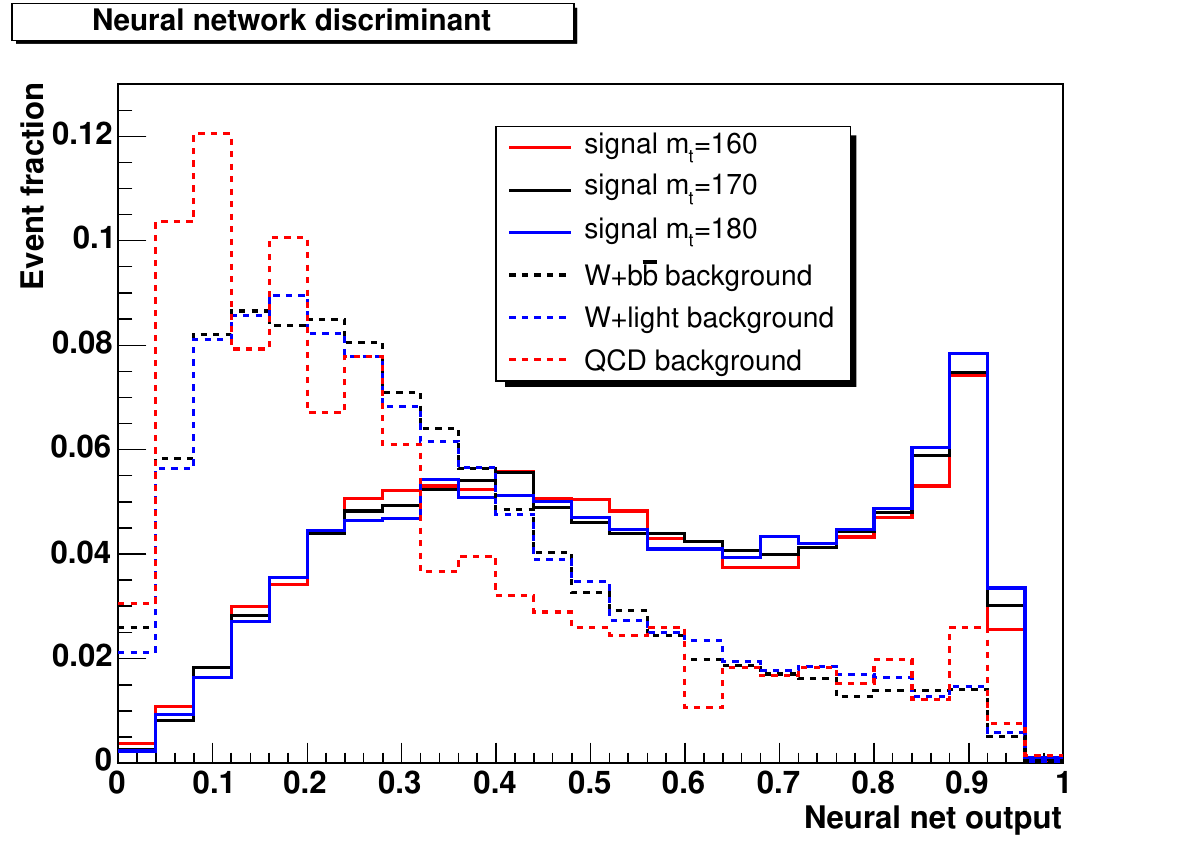,width=.5\textwidth}
}
\caption{Neural network output, $q$,
for simulated signal and background samples. The network
is trained with 10 kinematic variables~\cite{ref:plujanThesis}.}
\label{cdf_ljets_nn}
\end{figure}

The detector transverse momentum jet response is calibrated
{\it in situ} with $W$ decays by employing a special parameterization
of the transfer functions:
\begin{equation}
\fl W_f(u | {\bf x}, \Delta_{JES}) = \varrho_f(u \cdot JES | {\bf x}) \left( JES + u \frac{d\, JES}{d u} \right),
\label{eq:tfjes}
\end{equation}
where $u = p_{T,jet} / p_{T,parton}$, $JES$ depends on
$\Delta_{JES}$ according to \eref{eq:jesfromsigma}, and
${\bf y}_{m_k} \equiv u$, ${\bf a} \equiv \Delta_{JES}$
in relation to \eref{eq:tfmodel}.
The reference transfer functions, $\varrho_f(u | {\bf x})$, are
derived separately for light and heavy flavor jets in several different pseudorapidity
regions
from well-tuned Monte Carlo detector simulations assuming
$\Delta_{JES} = 0$. The subscript $f$ enumerates jet flavors and $\eta$ regions used. 
The transfer functions are normalized by the corresponding phase
space jet reconstruction efficiencies:
$\int_{u\sub{min}({\bf x}, \Delta_{JES})}^{\infty} W_f(u | {\bf x}, \Delta_{JES}) d u = \epsilon_f({\bf x}, \Delta_{JES})$, where the lower integration limit
$u\sub{min}({\bf x}, \Delta_{JES})$ depends on $p_{T,parton}$ and on the
effective $p_{T,jet}$ cutoff for the given $\Delta_{JES}$.
The efficiencies $\epsilon_f({\bf x}, \Delta_{JES})$ are derived
from Monte Carlo simulations by nonparametric logistic regression.
The background contamination remaining
after event selection
is estimated on event-by-event basis with a neural network discriminant.
The network classification performance is illustrated in
Figure~\ref{cdf_ljets_nn}.
Compared to calculating the background contribution
into the observed event probability using matrix element techniques,
this method results in a~significantly simpler background treatment
at the cost of some degradation of the \Mt\ estimator
statistical uncertainty.
The overall sample likelihood is adjusted to remove
the expected background contribution according to
\begin{equation*}
\begin{array}{ll}
\fl \ln L\sub{adj}(M_t, \Delta_{JES}) =
\sum_{j} 
& {\!\!\!\!\!}
\left[\ln L({\bf y}_j | M_t, \Delta_{JES}) \right. \\
& {\!\!\!\!\!}
- \left. f\sub{b}(q_j)\,
\overline{\ln L\sub{b}(M_t, \Delta_{JES})} \, \right].
\end{array}
\label{eq:adjlogli}
\end{equation*}
Here, $L\sub{adj}$ is the adjusted total likelihood for a given set of
events, $L({\bf y}_j | M_t, \Delta_{JES})$ is the signal likelihood
calculated for each event $j$, and
$f\sub{b}(q_j)$ is the background fraction
for a given event estimated from the neural network output:
$f\sub{b}(q_j) = B(q_j)/(S(q_j) + B(q_j))$.
In this ratio, the $B(q)$ and $S(q)$ distributions 
are normalized to
the overall expected background and signal fractions, respectively.
A simulated signal sample with $\Mt = 170$~GeV/$c^2$
is used to define $S(q)$, as shown in
Figure~\ref{cdf_ljets_nn}.
$\overline{\ln L\sub{b}(M_t, \Delta_{JES})}$
is the average background log-likelihood estimated from a large sample
of simulated Monte Carlo events.
\begin{figure}
\centerline{
\epsfig{file=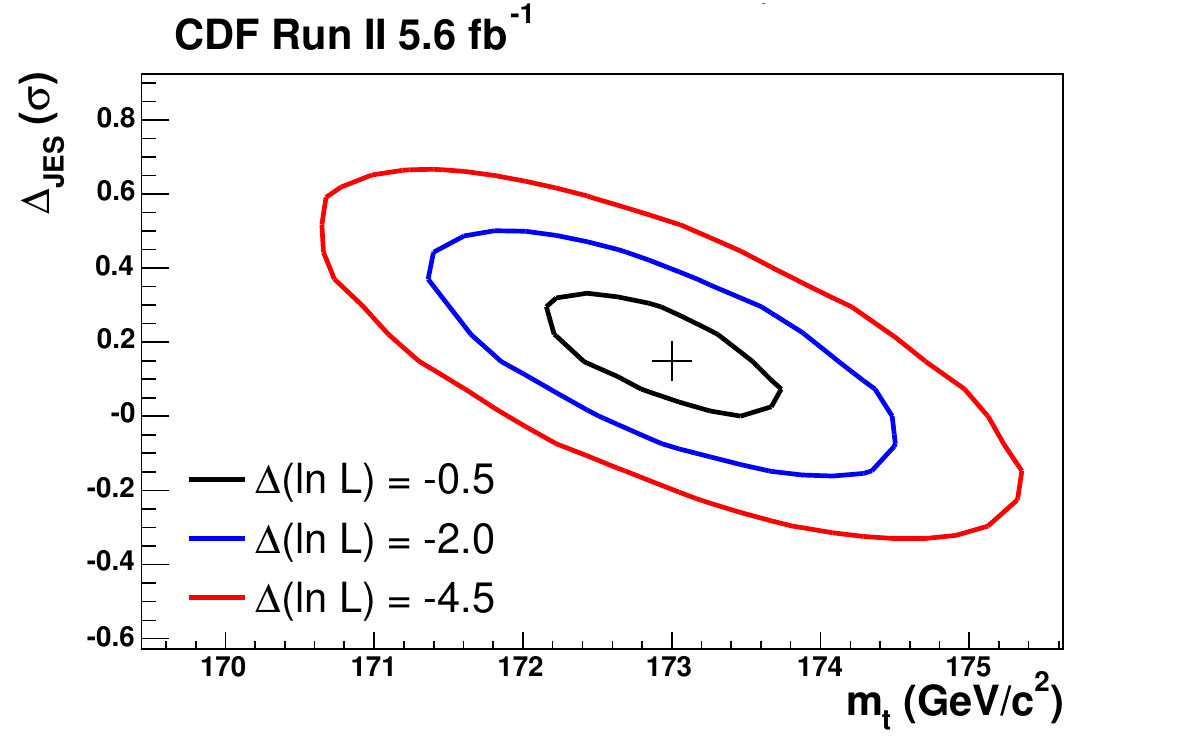,width=.5\textwidth}
}
\caption{Measured 2D likelihood on the CDF data events, with the
contours corresponding to a 1-, 2-, and 3-$\sigma$ uncertainty in the
final \Mt\ measurement from the profile method. The marker
shows the point of maximum likelihood. For consistency
with other results, the sign of $\Delta_{JES}$ plotted is reversed
w.r.t.~\eref{eq:jesfromsigma},~\eref{eq:tfjes}~\cite{ref:plujan}.}
\label{cdf_ljets_likelihood}
\end{figure}
A calibration (mapping function) is applied to $L\sub{adj}$
in order to remove the remaining bias and to ensure proper
frequentist coverage of the interval estimate.
The calibrated adjusted likelihood
is shown in Figure~\ref{cdf_ljets_likelihood}.
The $\Delta_{JES}$ nuisance parameter is eliminated by profiling.

% The final \Mt\ estimate is constructed
% by profiling $L\sub{adj}$ over the
% $\Delta_{JES}$ nuisance parameter, and by
% applying a calibration which removes the remaining
% bias and adjusts the pulls to ensure proper frequentist coverage.
% The adjusted likelihood is shown in Figure~\ref{cdf_ljets_likelihood}.

The systematic uncertainty of the CDF measurement
takes into account a number of sources,
according to the discussion presented in Section~\ref{sec:systematics}.
The breakdown of the assigned systematic uncertainty
in a set of independent components combined in quadrature
is reproduced in Table~\ref{table:sysbestcdf}.
\begin{table}
\caption{Systematic uncertainties of the CDF $\ell$+jets
         \Mt\ measurement with 5.6~fb\supers{-1} of data~\cite{ref:plujan}.}
\label{table:sysbestcdf}
% \begin{indented}
% \item[]\begin{tabular}{@{}lc}
\begin{tabular}{@{}lc}
\br
Systematic Source & Uncertainty (GeV/$c^2$) \\
\mr
Calibration                         & 0.10 \\
Monte Carlo Generator               & 0.37 \\
ISR and FSR                         & 0.15 \\
Residual JES                        & 0.49 \\
$b$-JES                             & 0.26 \\
Lepton $p_T$                        & 0.14 \\
Multiple Hadron Interactions        & 0.10 \\
PDFs                                & 0.14 \\
Background Modeling                 & 0.33 \\
Color Reconnection                  & 0.37 \\
Total                               & 0.88 \\
\br
\end{tabular}
% \end{indented}
\end{table}
The overall result is
$\Mt = 173.0 \pm 0.7 \,(\mbox{stat.}) \pm 0.6\, (\mbox{JES}) \pm 0.9\, (\mbox{syst.})$~GeV/$c^2$, with a total uncertainty of 1.2~GeV/$c^2$. 

The current most precise \D0 \ measurement of the top quark mass
employs the matrix element method to analyze 3.6~fb\supers{-1} of
Run~II data in the $\ell$+jets channel~\cite{ref:d0.ljets.matrel.may2011}.
The presence of one well-reconstructed charged lepton is required by the
event sample selection criteria, together with
exactly four high energy jets and substantial missing transverse momentum.
A neural network is utilized for heavy flavor tagging~\cite{nim-d0-btag}.
At least one $b$-tagged jet is required resulting in
about 70\% signal content in the final sample.

A parametric model is employed for the jet energy
transfer functions, while jet angles are assumed
to be well-measured. The parameterizations are derived
for a number of pseudorapidity regions
separately for light jets, $b$ jets, and $b$ jets with
a soft muon tag (in the latter case, a fraction of the energy of the jets
is carried away by a neutrino which escapes detection).
The jet energy scale is included in the transfer functions
as an overall multiplicative factor, $JES$, which is not related
to the prior systematic uncertainty:
\begin{equation*}
\fl W_f(E\sub{jet} | E\sub{parton}, JES) = 
\frac{1}{JES} \,\varrho_f \! \left(\left.\frac{E\sub{jet}}{JES} \right| E\sub{parton}\right).
\end{equation*}
In this analysis, the reference transfer functions
$\varrho_f(E\sub{jet} | E\sub{parton})$
are represented by the double Gaussian formula:
\begin{equation}
\label{eq:doublegauss}
\begin{array}{l}
\fl \varrho_f(E\sub{jet} | E\sub{parton})  = \frac{1}{\sqrt{2 \pi} (p_2 + p_3 p_5)} \\
\times \left[ \exp\left(-\frac{(E\sub{jet} - E\sub{parton} - p_1)^2}{2\,p_2^2}\right) \right. \\
\left. + \,p_3 \exp\left(-\frac{(E\sub{jet} - E\sub{parton} - p_4)^2}{2\,p_5^2}\right) \right].
\end{array}
\end{equation}
These transfer functions are normalized by 
\[
\int_{-\infty}^{\infty} \varrho_f(E\sub{jet} | E\sub{parton}) \, d E\sub{jet} = 1
\]
for all possible values of $E\sub{parton}$ and of parameters $p_k$, $k = 1, ..., 5$.
The parameters $p_k$ are linear functions of the parton energy: $p_k = a_k + E\sub{parton} \cdot b_k$.
The coefficients $a_k$ and $b_k$ are determined
by fitting the jet response in fully simulated Monte Carlo events. 
It remains unclear how faithfully this model represents the energy response
of low $p_T$ jets: as emphasized in~\cite{ref:memeff},
transfer functions normalized in this manner
are unavoidably affected by
the jet reconstruction inefficiencies which alter
the reference jet energy distributions.
% (jet transfer functions used in the CDF measurement described above
% do not suffer from this problem due to a different
% lower limit in the employed normalization).
Transfer functions are also constructed for the electron energy and muon 
$p_T^{-1}$. The latter quantity is proportional to the muon track curvature 
in the detector magnetic field (curvature resolution determines the $p_T$
uncertainty of the reconstructed muon).

A number of assumptions are introduced to reduce the
dimensionality of the signal phase space integral to ten.
% and to permit the 
% use of VEGAS~\cite{ref:vegas} adaptive numerical
% integration program.
The following variables are chosen for integration:
the transverse momenta of the colliding partons (for which priors
are derived from PYTHIA),
the energy associated with one of
the quarks from the hadronic $W$ boson decay, the masses of
the two $W$ bosons and the two top quarks, and either the energy
of the electron or $1/p_T$ of the muon.
The signal matrix element models the dominant
$q\bar{q} \rightarrow t\bar{t}$ production process~\cite{ref:d0matrel},
while $t\bar{t}$ spin correlations are ignored.

The background contribution into the event probability is calculated
using the $W$ + 4~jets matrix element provided by VECBOS~\cite{ref:vecbos}.
It is assumed that this contribution alone is sufficient to model
the shape of the background likelihood, while other backgrounds
(mainly QCD multijet events) are not treated explicitly. The
background normalization is adjusted so that the correct $t\bar{t}$
signal fraction can be reproduced in the analysis of simulated event samples.
% Details of the background normalization procedure are provided
% in~\cite{ref:d0.ljets.matrel.firstrun2}.
The overall event sample
likelihood is profiled over the signal fraction nuisance parameter.
$JES$ is marginalized with a Gaussian prior consistent with
the $JES$ uncertainty from jet energy calibrations.

\begin{figure}
\centerline{
\epsfig{file=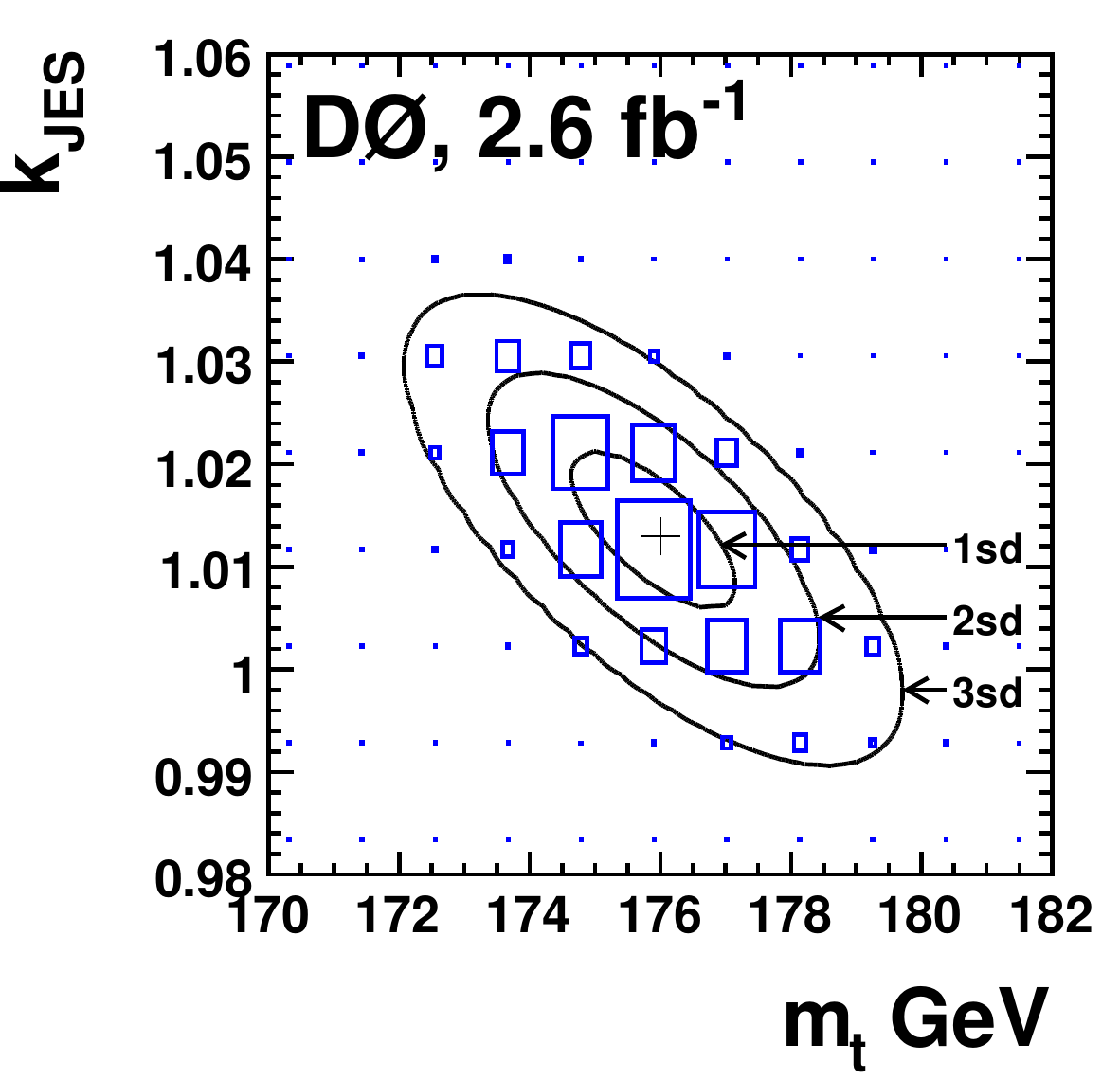,width=.3\textwidth}
}
\caption{Calibrated \D0 \ result obtained with the Run~IIb
data in the $\ell$+jets
channel. The JES prior
factor is included into the likelihood.
The boxes, representing
the bins in the two-dimensional histograms of
the likelihoods, have areas proportional to the
bin contents~\cite{ref:d0.ljets.matrel.may2011}.}
\label{d0_ljets_likelihood}
\end{figure}

Systematics uncertainties of the \D0 \ \Mt\ measurement with Run~IIb data
are shown
in Table~\ref{table:sysbestd0}. The source breakdown is substantially
different from that utilized by CDF. The generator systematic uncertainty, 
here called Higher Order Effects, was calculated comparing ALPGENv2+HERWIG with 
MC@NLO.
\begin{table}
\caption{Systematic uncertainties of the \D0 \ $\ell$+jets \Mt\
measurement with 2.6~fb\supers{-1} of 
Run IIb data~\cite{ref:d0.ljets.matrel.may2011}.}
\label{table:sysbestd0}
% \begin{indented}
% \item[]\begin{tabular}{@{}lc}
\begin{tabular}{@{}lc}
\br
Systematic Source & Uncertainty (GeV/$c^2$) \\
\mr
Higher Order Effects                & 0.25 \\
ISR and FSR                         & 0.26 \\
Hadronization and UE                & 0.58 \\
% Color Reconnection                  & 0.40 \\
Color Reconnection                  & 0.28 \\
Multiple $p\bar{p}$ Interactions    & 0.07 \\
% Background Modeling                 & 0.03 \\
Background Modeling                 & 0.16 \\
$W$+jets Heavy Flavor               & 0.07 \\
\ \ Scale Factor                        & \  \\
$b$-jet Modeling                    & 0.09 \\
PDF                                 & 0.24 \\
Residual JES                        & 0.21 \\
% 
% Relative $b$/Light Response         & 0.81 \\
% Sample-Dependent JES                & 0.56 \\
% 
Data-MC Jet Response                & 0.28 \\
 \ \ Difference                     &  \ \\
$b$-tagging Efficiency              & 0.08 \\
Trigger Efficiency                  & 0.01 \\
Lepton Momentum Scale               & 0.17 \\
Jet Energy Resolution               & 0.32 \\
Jet Identification Efficiency            & 0.26 \\
QCD Background                      & 0.14 \\
Signal Fraction                     & 0.10 \\
MC Calibration                      & 0.20 \\
Total                               & 1.02 \\
\br
\end{tabular}
% \end{indented}
\end{table}
The overall result (Run~IIa and Run~IIb combined) is
$\Mt = 174.94 \pm 0.83 \,(\mbox{stat.}) \pm 0.78 \, (\mbox{JES}) \pm 0.96\, (\mbox{syst.})$~GeV/$c^2$,
with a total uncertainty of 1.49~GeV/$c^2$.
The sample likelihood as a function of \Mt\ and $JES$
(labelled by m\sub{t} and k\sub{JES}, respectively, on the axes) 
is shown in Figure~\ref{d0_ljets_likelihood}.

A set of representative Tevatron \Mt\
measurements in the $\ell$+jets channel 
is collected in Table~\ref{table:mtljets}
which illustrates the evolution
of analysis techniques and attained precision
with integrated luminosity.
\begin{table*}[hbtp]
\caption{Representative Tevatron \Mt\ measurements in the $\ell$+jets channel.
         If a~reference is marked by~(*) then the
         $JES$ calibration uncertainty is included as a part of statistical
         rather than systematic uncertainty.}
\label{table:mtljets}
% \begin{indented}
% \item[]\begin{tabular}{@{}lcccccc}
\begin{tabular}{@{}lcccccc}
\br
Experiment & Method & $\int \mathcal{L} dt$ & \Mt\ & $\sigma\sub{stat}$ & $\sigma\sub{syst}$ & Ref. \\
 \ & \ & (fb\supers{-1}) & (GeV/$c^2$) & (GeV/$c^2$) & (GeV/$c^2$) & \ \\
\mr
CDF (Run I)   & template & 0.02 & 174   & 10  & $\mbox{}^{+13}_{-12}$ & \cite{CDF94} \\
\D0 \ (Run I)    & template & 0.05 & 199   & $\mbox{}^{+19}_{-21}$ & 22 & \cite{top_dis-d0} \\
CDF (Run I)   & template & 0.07 & 176   & 8   & 10  & \cite{top_dis-cdf} \\
CDF (Run I)   & template & 0.11 & 176.1 & 5.1 & 5.3 & \cite{ref:cdf.ljets.run1prl, ref:cdf.ljets.run1prd} \\
\D0 \ (Run I) & template & 0.13 & 173.3 & 5.6 & 5.5 & \cite{ref:d0.ljets.run1prl, ref:d0.ljets.run1prd} \\
\D0 \ (Run I) & MEM      & 0.13 & 180.1 & 3.6 & 3.9 & \cite{ref:d0.ljets.mem.nature} \\

CDF           & template & 0.16 & 179.6 & $\mbox{}^{+6.4}_{-6.3}$ & 6.8 & \cite{ref:cdf.multivariate.note1} \\

CDF           & DLM      & 0.16 & 177.8 & $\mbox{}^{+4.5}_{-5.0}$ & 6.2 & \cite{ref:cdf.ljets.dlm160} \\

\D0           & template & 0.16 & 170.0 & 6.5 & $\mbox{}^{+10.5}_{-6.1}$ & \cite{ref:d0.ljets.runiiearly} \\
\D0           & ideogram & 0.16 & 177.5 & 5.8 & 7.1 & \cite{ref:d0.ljets.runiiearly} \\
\D0           & template & 0.23 & 170.6 & 4.2 & 6.0 & \cite{ref:d0.ljets.230pb} \\

CDF           & DLM & 0.32 & 173.2 & $\mbox{}^{+2.6}_{-2.4}$ & 3.2 & \cite{ref:cdf.ljets.mtmwanddlm, ref:cdf.ljets.dlm2} \\

CDF           & template & 0.32 & 173.5 & $\mbox{}^{+3.7}_{-3.6}$ & 1.3 & \ \cite{ref:cdf.ljets.mtmwanddlm, ref:cdf.ljets.mtmw}* \\

\D0           & template & 0.32 & 169.5 & 4.4 & $\mbox{}^{+1.7}_{-1.6}$ & \ \cite{ref:d0.ljets.320pb}* \\

\D0           & MEM      & 0.37 & 170.3 & $\mbox{}^{+4.1}_{-4.5}$ & $\mbox{}^{+1.2}_{-1.8}$ & \ \cite{ref:d0.ljets.matrel.firstrun2}* \\
\D0           & ideogram & 0.43 & 173.7 & 4.4 &  $\mbox{}^{+2.1}_{-2.0}$ & \ \cite{ref:d0.ljets.ideogram}* \\

CDF           & MEM      & 1.0  & 170.8 & 2.2 & 1.4 & \ \cite{ref:cdf.ljets.firstmeat}* \\

CDF           & template & 1.0  & 168.9 & 2.2 & 4.2 & \cite{ref:cdf.ljets.3chisq} \\

\D0           & MEM      & 1.0  & 171.5 & 1.8 & 1.1 & \ \cite{ref:d0.ljets.matrel.last}* \\

CDF           & DLM      & 1.7  & 171.6 & 2.0 & 1.3 & \ \cite{ref:cdf.ljets.dlm3}* \\

CDF           & MEM      & 1.9  & 172.7 & 1.2 & 1.8 & \cite{ref:freeman} \\

CDF           & template & 1.9  & 171.8 & 1.9 & 1.1 & \ \cite{ref:cdf.template.pub1}* \\

% \D0           & MEM    & 2.1  & 172.2 & 1.1 & 1.6 & \cite{ref:d0.ljets.matrel.2.1fb} \\
\D0           & MEM      & 2.2  & 172.2 & 1.0 & 1.4 & \cite{ref:d0.ljets.matrel.2.2fb} \\

CDF           & template & 3.2  & 172.2 & 1.1 & 1.5 & \cite{ref:cdf.joint.3.2fb} \\
CDF           & MEM      & 3.2  & 172.4 & 1.4 & 1.3 & \ \cite{ref:meat2009}* \\

\D0           & MEM      & 3.6  & 174.9 & 0.8 & 1.2 & \cite{ref:d0.ljets.matrel.may2011} \\
CDF           & MEM      & 5.6  & 173.0 & 0.7 & 1.1 & \cite{ref:plujan} \\
CDF           & template & 5.6  & 172.2 & 1.2 & 0.9 & \ \cite{ref:cdf.template.latest}* \\
\br
\end{tabular}
% \end{indented}
\end{table*}[hbtp]
Recent CDF and \D0 \ results are in good
agreement with each other.

\subsection{All-hadronic topology}
\label{sec:all-had}

The all-hadronic branching fraction of about 46\% is the
largest among all $t\bar{t}$ decay channels.
% 
% Top pair events appear dominantly as events where both $W$s 
% coming from the top quarks decay hadronically, corresponding
% to about 46\% of the decays. 
%
Due to the absence of neutrinos, this final state is kinematically
constrained, and in principle, the $\ttbar$ event can be
reconstructed. The challenges are reduction of the very large QCD 
background due to multijet production and reduction of the 90 possible 
permutations that can be
made in assigning 2 jets to b quarks and pairing the remaining 4 jets. 

% A difficulty arises due to the large multiplicity of jets
% in the final state and thus due to the combinatorial background. 
%
%The main background in this channel is due to the QCD production of light 
%quarks and gluons which gives rise to multijet final states.
%The production of six or more final state partons is poorly known at the theoretical level. Also, the QCD background cross section is several orders of magnitude larger than the signal cross section, so the background rejection must be very large; these conditions would require prohibitive computing time in producing the relevant Monte Carlo simulation. Therefore, a data-driven modeling of the multijet background is mandatory.
Dedicated trigger requirements are needed in order to enhance the top quark signal contribution over the overwhelming background. 
Even after surviving such trigger requirements, the background is about three orders of magnitude larger than the signal.
In order to reconstruct the $t\bar{t}$ events, at least six high-$p_T$ jets must be present in the detector, while charged and neutral leptons are vetoed.
Identifying jets originated from $b$ quarks helps both in suppressing the QCD background and in reducing the number of possible permutations in reconstructing the final state kinematics according to the $t \bar t \to W^{+} b W^{-} \bar{b} \to q_1 \bar q_2  b q_3 \bar q_4 \bar b$ hypothesis. 
% In order to fully reconstruct the events, at least six high-$p_T$ jets are required in the detector, while presence of charged and neutral leptons is vetoed.
% 
% The small signal-to-background (S/B) ratio requires to
% exploit the unique kinematical and topological properties
% of top quark pair events
% to further suppress backgrounds and be able to observe the signal. 
%
The distinct kinematic properties of $t \bar t$
events are further exploited to bring the S/B ratio
to an acceptable level.

%The first evidence for all-hadronic decays of the top-antitop pairs have been obtained in the Tevatron Run~I more than a decade ago by both CDF~\cite{Abe:1997rh} and D0~\cite{Abbott:1999mr,Abbott:1998nn} collaborations. 
Due to the unique challenges, the observation of this decay mode was achieved only after $t\bar{t}$ events were seen in the $\ell$+jets and dilepton channels, by analyzing the full Tevatron Run~I dataset of 110~pb$^{-1}$. CDF utilized a combination of $b$-tagging and kinematic/topological cuts to select a relatively clean sample of top quark pair events 
subsequently used to estimate \Mt ~\cite{Abe:1997rh}.
\D0 \ used neural networks to exploit the S/B discriminating
power of several kinematic and topological 
observables~\cite{Abazov:2004ng}.
%The excess of events over the background predictions in both measurements is in the order of 3$\sigma$, and the excess is attributed to top quark pair production decaying hadronically. 
%
%table with RunI results
\begin{table*}[hbtp]
\caption{Tevatron \Mt\ measurements in the all-hadronic channel.}
\label{tab:1}
% \begin{indented}
% \item[]\begin{tabular}{@{}llccccc}
\begin{tabular}{@{}llccccc}
\br
Experiment & Method & $\int \mathcal{L} dt$ & \Mt\ & $\sigma\sub{stat}$ & $\sigma\sub{syst}$ & Ref. \\
 \ & \ & (fb\supers{-1}) & (GeV/$c^2$) & (GeV/$c^2$) & (GeV/$c^2$) & \ \\
\mr
CDF (Run~I) & template   &  0.11 & 186      & 10    & 12            & \cite{Abe:1997rh} \\
\D0 \  (Run~I) & template   &  0.11 & 178.5   & 13.7 & 7.7           & \cite{Abazov:2004ng} \\
CDF & ideogram &  0.31 &  177.1   &   4.9      & 4.7     & \cite{Aaltonen:2006xc}	     \\
CDF & template   &  1.0 &  174.0      &   2.2     &  1.8     &  \cite{:2007qf}      \\
CDF & MEM-assisted & 1.0 & 171.1  &  2.8  &  3.2  &  \cite{Aaltonen:2008bg}      \\
%CDF & Run~II &      165.2      &   4.4      & 1.9       & 1~900 (IMM)    &   \cite{IMM2D}     \\
CDF & template   &  2.9 &  174.8   &  1.7      & 2.0      &   \cite{Aaltonen:2010pe}      \\
CDF & template   &  5.8 &  172.5   & 1.4     & 1.4     &   \cite{LucaNote} \\
%?   & ?        & ?.? & ???.? & ?.? & ?.? & \cite{ref:unknown} \\
\br
\end{tabular}
% \end{indented}
\end{table*}[hbtp]
%
%
%\begin{center}
%\begin{table}[hbp]
%\begin{tabular}{ll|cccrc}
% Source      &   &         \Mt\  (GeV/c$^2$) & $\sigma_{stat}$ & $\sigma_{syst}$ & Lumi (pb$^{-1}$) & Ref. \\
%\hline
%CDF & Run~I &                     186             &   10              & 12              & 110             & \cite{Abe:1997rh}  \\
%D0    & Run~I &                     178.5          &   13.7           &  7.7           & 110             & \cite{Abazov:2004ng}    \\
%CDF & Run~II &     177.1          &   4.9      & 4.7         & 310 	   & \cite{Aaltonen:2006xc}	     \\
%CDF & Run~II &      174          &   2.2     &  1.8      & 1\,000                &  \cite{:2007qf}      \\
%CDF & Run~II &      171.1      &    3.7     &  2.1      & 1\,000     &  \cite{Aaltonen:2008bg}      \\
%CDF & Run~II &      165.2      &   4.4      & 1.9       & 1\,900 (IMM)    &   \cite{IMM2D}     \\
%CDF & Run~II &      174.8      &  2.4      & 1.1        & 2\,900    &   \cite{Aaltonen:2010pe}      \\
%CDF & Run~II &      172.5      & 1.7        & 1.2       & 5\,800   &   \cite{LucaNote} \\
%\end{tabular}
%\caption{The table lists the measurements performed in Run~I and Run~II  by the CDF and D0 collaborations.}
%\label{tab:1}
%\end{table}
%\end{center}
%
%After upgrades to the collider and the experiments, the Tevatron collider started taking data again in 2001 in the so-called Run~II.
The $t\bar{t}$ events in the all-hadronic final state were observed again during the Run~II by both CDF~\cite{Abulencia:2006se,:2007qf,Aaltonen:2010pe} and \D0 ~\cite{Abazov:2009ss} collaborations. The new measurements of \Mt\ and 
$t\bar{t}$ production cross section benefited from the larger center-of-mass energy resulting in $\sim 30\% $ higher cross section, from improved detectors, and from better analysis techniques.
For both data taking periods, the QCD multijet background is estimated from data; the much larger event sample collected in Run~II allows for a very accurate control of the background modeling.
% The first cross section measurement by CDF uses a combination of kinematical and topological cuts to enhance the signal to background ratio~\cite{Abulencia:2006se}. (DUPLICATION)
%These events are then reanalyzed to measure the top quark mass~\cite{Aaltonen:2006xc}.
%The D0 collaboration measured the $t \bar t$ production cross section in this channel using a likelihood ratio to discriminate the signal over the background~\cite{Abazov:2009ss}. \\

The Tevatron publications of \Mt\  measurements in the all-hadronic channel are listed in Table~\ref{tab:1}. So far, only the CDF collaboration measured the top quark mass with the Run~II data.

The CDF multijet trigger requires four energetic jets in the event and large energy deposit in the calorimeter in order to select all-hadronic $t \bar t$ decays while suppressing the much more common QCD multijet production. The use of a neural network to obtain a cleaner sample of top quark hadronic decays resulted in a large improvement in the statistical sensitivity of the measurement~\cite{:2007qf}. 
% The measurement in the all-hadronic channel suffers from a very large uncertainty induced by the knowledge of the jet energy scale, that amounts to about 1.5~GeV/$c^2$ per each 1\% of jet energy scale uncertainty. As already done in the lepton+jets decays mode, the presence of 
Similar to the $\ell$+jets channel, hadronic decays of the $W$ bosons are used to measure the jet energy scale {\it in situ} with the available dataset. This technique provides an important reduction in the systematic uncertainty that would otherwise limit the precision
of the \Mt\ estimate in this channel to $\sim 3\%$.  
%A newer analysis made use of this and of matrix element techniques to further improve the overall precision on \Mt\  using 1~fb$^{-1}$ of data\cite{Aaltonen:2008bg}, bringing it down to 2.5\%. This measurement has been used in the 2007 Tevatron top mass average~\cite{:2007bxa}.

\begin{figure}
\includegraphics[width=7.5cm]{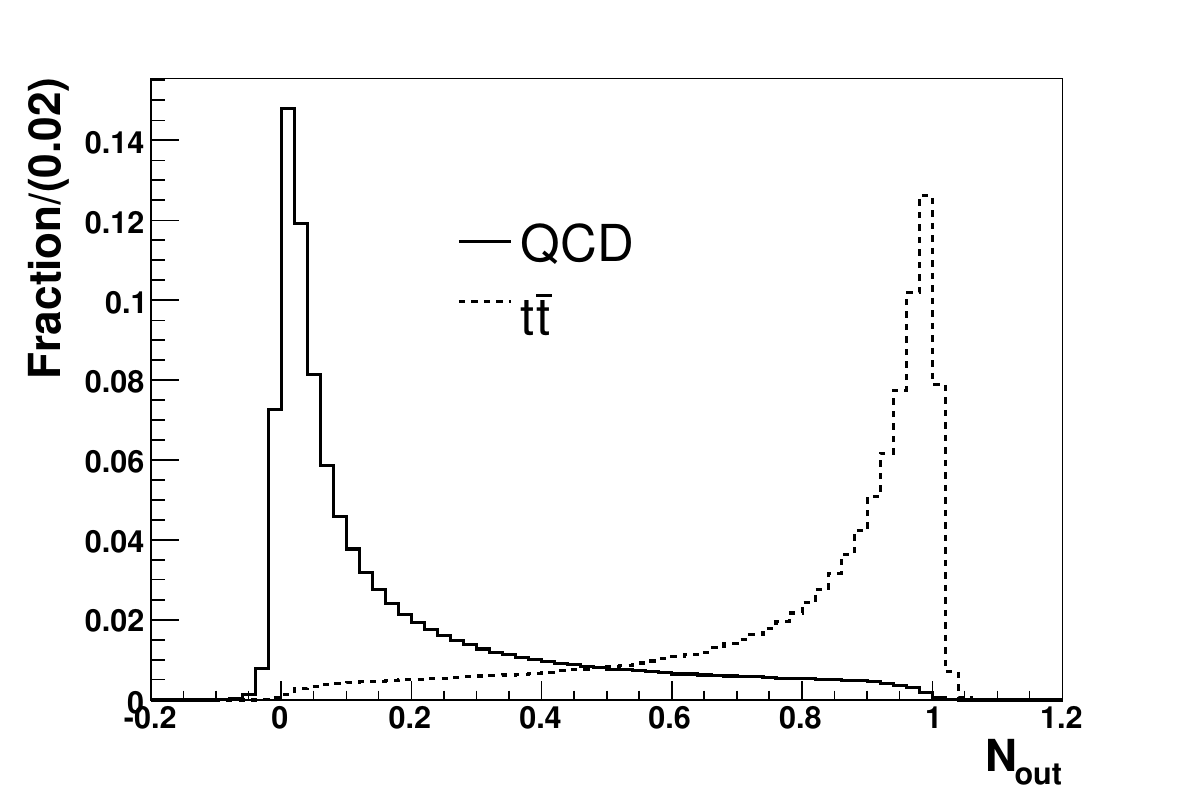}
\caption{Output of the neural network used to discriminate $t \bar t \to q_1 \bar q_2  b q_3 \bar q_4 \bar b$ events from QCD multijet production~\cite{Aaltonen:2010pe}.}
\label{fig:hadNN}
\end{figure}

The latest and most precise \Mt\  measurement in this channel~\cite{Aaltonen:2010pe,LucaNote} uses several kinematic and topological properties of the events as inputs to a neural network to separate the signal from the background. It also exploits the fact that all-hadronic top quark events produce at first approximation only quark jets, while the QCD background produces on average several gluon jets in the final state. 
Signal-like quark jets can be statistically discriminated from background-like gluon jets using calorimeters with sufficient granularity. In fact, observed differences in quark and gluon jet widths have been previously reported by experiments at the KEK $e^+ e^-$ collider (TRISTAN)~\cite{Kim:1989qk} and the CERN $e^+ e^-$ collider (LEP)~\cite{Alexander:1991ce,:1995ima}. 
A cut is placed on the neural network output shown in Figure~\ref{fig:hadNN}, and on the quality of the fit that assigns the jets to the final state partons when the top quark pair decay chain is reconstructed.
Both cuts were chosen to minimize the statistical uncertainty on the top quark mass measurement. The event sample is split into two subsamples: with
one $b$-tagged jet and with two or more $b$-tagged jets. For the latter
subsample, S/B$\,\approx 1$.
%
%This measurement provides a $\sim 1\%$ precision.

\begin{figure}
\includegraphics[width=7.cm]{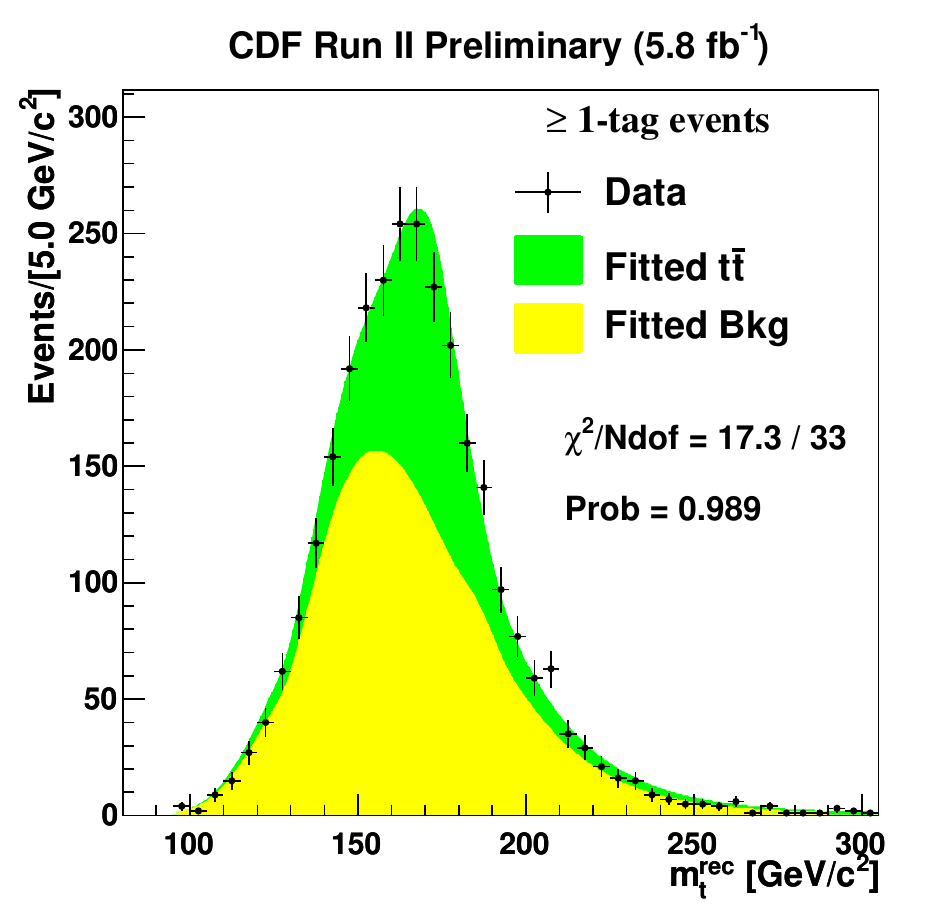}
\includegraphics[width=7.cm]{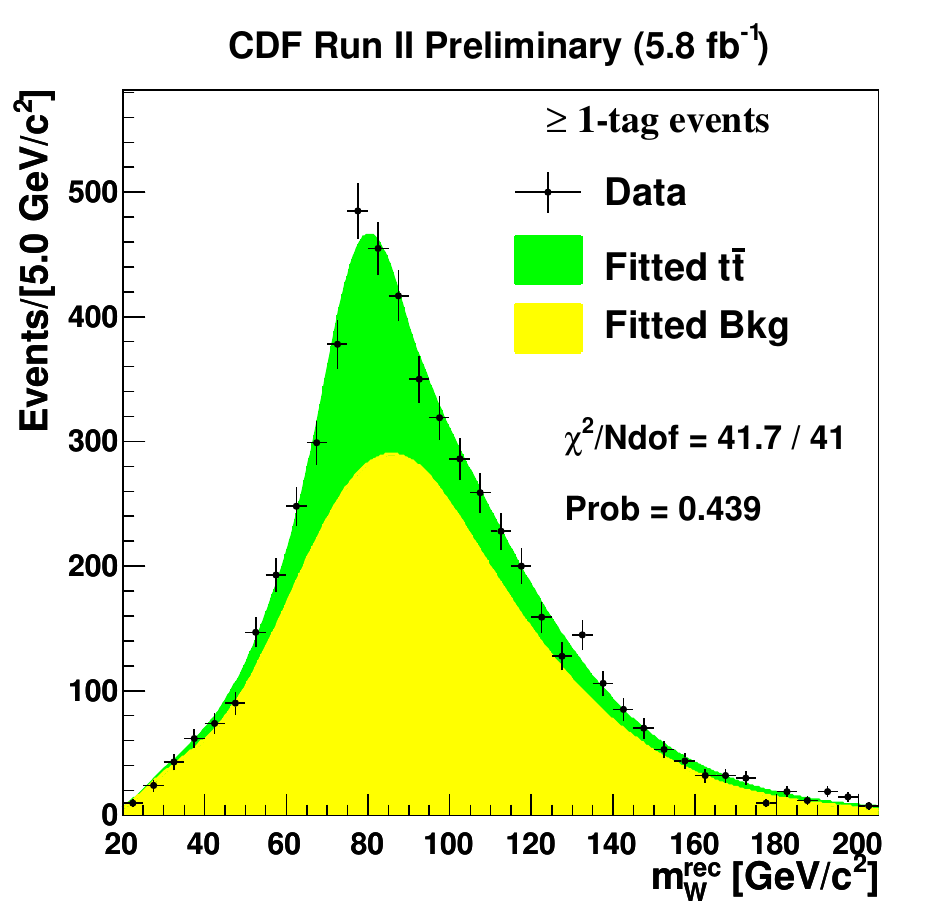}
\caption{Distribution of the reconstructed top quark mass (left) and $W$ mass (right) for events surviving the event selection. The plots show the events 
with at least one $b$-tagged jet passing the neural network event selection and the goodness-of-fit requirement. These two criteria are optimized to reduce the statistical uncertainty on the \Mt\  and $JES$ measurements, 
respectively~\cite{LucaNote}.}
\label{fig:comb0}
\end{figure}

%The trigger requires four energetic jets in the event and large energy deposit in the calorimeter in order to select all-hadronic $t \bar t$ decays while suppressing the much more common QCD multijet production. 
Multiple collisions occurring in the same bunch crossing often produce extra jets, and thus extra energy in the calorimeter. For this reason, as the collider instantaneous luminosity increased over time, so did the trigger rate. The data collected until the end of year 2007 has been obtained with the Tevatron colliding particles with an average instantaneous luminosity at the beginning of the store of about 
% $1 \times 10^{-32}$~cm$^{-2}$/s$^{-1}$. 
$10^{32}$~cm$^{-2}$\,s$^{-1}$. 
Under these conditions, the trigger rate was relatively low and the event selection efficiency is about 4\%, higher than in the $\ell$+jets case but with much higher backgrounds. Starting from 2008, the instantaneous luminosity at the beginning of the store was on average three times larger, thus the trigger rates and as a consequence the multijet background increased by a factor of 3. This was dealt with at the event selection level, by tightening the neural network output cut and the cut on the kinematic fitter goodness-of-fit. This tighter set of cuts provides better reconstructed events at the cost of a lower event selection efficiency. The reconstructed top quark mass and the reconstructed $W$ boson mass distributions can be seen in Figure~\ref{fig:comb0}.
\begin{figure}
\includegraphics[width=8.5cm]{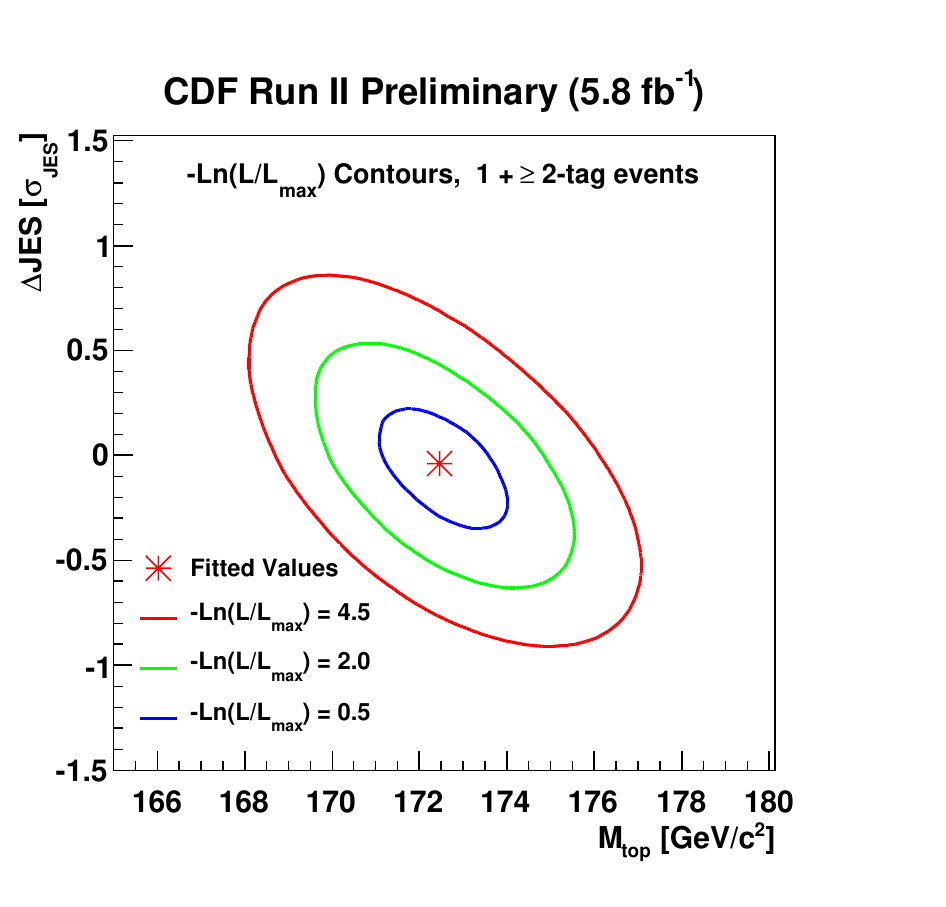}
\caption{Simultaneous measurement of the top quark mass and the jet energy scale in the all-hadronic channel. The $\Delta_{JES}$ parameter is related to the jet energy scale according to~\eref{eq:jesfromsigma}~\cite{LucaNote}.
%The uncertainty on the top quark mass is expressed in GeV while the uncertainty in the JES measurement is expressed in number of times the prior systematic of $\sim 3\%$.
}
\label{fig:had2D}
\end{figure}
Owing to the improved analysis techniques, the statistical precision of the \Mt\ measurement scaled better than with $1/\sqrt{\int \mathcal{L} dt}$, while the dominant systematic uncertainty due to the jet energy scale has been linked to $1/\sqrt{\int \mathcal{L} dt}$. 
The most precise determination of the top quark mass in the all-hadronic channel as of this writing gives $\Mt = 172.5 \pm 1.4 \,(\mbox{stat.}) \pm 1.0 \,(\mbox{JES}) \pm 1.2 \,(\mbox{syst.})$~GeV/$c^2$, or $\Mt = 172.5 \pm 2.0 \,(\mbox{total})$~GeV/$c^2$. The ultimate precision that can be reached in this channel is now limited by 1.1~GeV/$c^2$ due to the systematic uncertainties other than the overall jet energy scale. 
% The $\Delta_{JES}$ template parameter is related to the jet energy scale according to~\eref{eq:jesfromsigma}. 
It can be appreciated from Figure~\ref{fig:had2D} that the measured value of $\Delta_{JES}$ is consistent with the prior expectation (mean value close to zero) with the precision a factor of three better than the prior constraint. 

% The systematics affecting the $\Delta_{JES}$ measurement 
% amount to an additional 1\% uncertainty.
% Given that the signal jets that match to the $W$
% boson are always quark-originated jets, and that
% the background modeling is completely data-driven
% and is thus not affected by the JES systematic,
% the result can be interpreted as a control to slightly
% above the 1\% level on the quark jet energy scale.

\subsection{Events with taus}
\label{sec:metjets}

About 35\% of decays of the tau lepton are pure leptonic, with
an electron or a muon and two neutrinos in the final state.
Charged leptons produced in such decays often pass standard
electron or muon selection requirements.
Therefore, the $t\bar{t}$ event selection criteria in
the $\ell$+jets and dilepton channels actually accept a fraction of events
which contain leptonically decaying taus.
% 
% Therefore, the event selection requirement of one charged electron
%  or muon, large missing transverse momentum, and jets typical of
% the $\ell$+jets event selection, actually accepts a fraction
% of events which contain leptonically decaying taus.
%
On the other hand,
explicit identification of tau leptons which decay hadronically
is difficult: these decays resemble common hadronic jets
originating from quarks and gluons. For this reason, hadronic
tau identification algorithms typically have to operate at 
low efficiency. For signals with many jets in the final state, as
the $t \bar t$ decays under consideration here, the explicit
tau identification is often not sufficient to suppress the QCD
background in the sample.
% The kinematic properties of the $t \bar t$ events and
% the multijet background are sufficiently distinct, so that it is
% possible to distinguish them even without the explicit identification
% of charged leptons in the signal events. 
One can instead rely upon
the experimental signature characterized by multiple jets and large
$\Etmiss$ in the final state as an efficient way to collect the $t \bar t$
events with tau leptons.

The first measurement of the $t \bar t$ cross section which utilized
events with large $\Etmiss$ and multiple jets, with at least one $b$-tagged
jet, was performed by the CDF collaboration using 310~pb$^{-1}$ of
Tevatron Run~II data~\cite{Abulencia:2006yk}.  A more recent CDF study of
this experimental signature utilized 2.2~fb$^{-1}$ of Run~II
data~\cite{Aaltonen:2011tm}. 
In both measurements, a veto on the presence
of reconstructed electrons or muons was applied in order to avoid
overlap with other channels. The more recent study utilized a neural
network for background rejection achieving a S/B ratio of about 4. 
About 90\% of the signal
sample was composed of $t \bar t
\to W^{+} b W^{-} \bar{b} \to \ell \nu + $4~jets events,
with $\ell=\tau$ in 40\% of the cases.

% The dominant backgrounds for this signature
% are QCD multijet production where the jet
% energy mismeasurement gives rise to large missing transverse energy,
% and $W$+jets production where the lepton is not identified.
%
% Similarly to the all-hadronic channel, modeling the QCD background
% with Monte Carlo simulation tools would require a large CPU effort
% and still suffer from large theoretical uncertainties. 
%
%All backgrounds are modeled in this analysis by using events without
%$b$-tagging (the ``pretag'' sample) and correcting for the kinematic
%bias induced by the $b$-tagging algorithm in a probabilistic
%manner. The probability to $b$-tag a jet is derived in a control
%sample with exactly three jets. The model relies on the assumption
%that events without the $b$-tagging requirement have the same
%background composition as the events with $b$-tagging. Rectangular
%cuts and/or neural networks can be used to isolate a region with high
%signal-to-background ratio.

In this sample, one of the two top quarks is expected to decay into
$\ell \nu b$, with charged lepton not identified.  This decay provides
little information about the top quark mass. Still, the other top quark in
the event produces three jets which can be used to reconstruct its decay
chain and to measure the jet energy scale {\em in situ}. This feature
is exploited in the recent CDF \Mt\ measurement~\cite{ref:cdf.etmissplusjets}.
The kinematic analysis starts by identifying decay products of
the hadronically decaying $W$. All pairwise combinations of jets
without $b$ tags are
considered, and the pair which gives the invariant mass closest to
the world average $W$ mass is chosen. Then the jet with the highest $E_T$
is chosen among the remaining jets and added to the pair in order
to construct a 3-jet system whose invariant mass is correlated
with the top quark mass. Another
such system is built by adding the jet with the second highest $E_T$ to the pair.
3-dimensional templates are constructed using the di-jet and the two
tri-jet invariant masses.
\begin{table}
\caption{Tevatron \Mt\ measurements in the $\Etmiss$+jets channel (all by CDF).}
\label{table:mtmetjets}
% \begin{indented}
% \item[]\begin{tabular}{@{}ccccc}
\begin{tabular}{@{}ccccc}
\br
$\int \mathcal{L} dt$ & \Mt\ & $\sigma\sub{stat}$ & $\sigma\sub{syst}$ & Ref. \\
(fb\supers{-1}) & (GeV/$c^2$) & (GeV/$c^2$) & (GeV/$c^2$) & \ \\
\mr
0.3 & 172.3 & 10.2 & 10.8 & \cite{Aaltonen:2007xx} \\
5.7 & 172.3 & 1.8 & 1.8 & \cite{ref:cdf.etmissplusjets} \\
%?   & ?        & ?.? & ???.? & ?.? & ?.? & \cite{ref:unknown} \\
\br
\end{tabular}
% \end{indented}
\end{table}
%
%
%\begin{center}
%\begin{table}[hbp]
%\begin{tabular}{ll|cccrc}
% Source      &   &         \Mt\  (GeV/c$^2$) & $\sigma_{stat}$ & $\sigma_{syst}$ & Lumi (pb$^{-1}$) & Ref. \\
%\hline
%CDF & Run~II &        172.3            &   10.2              & 10.8              & 311            & \cite{Aaltonen:2007xx}  \\
%CDF & Run~II &        172.3      & 2.4        & 1.0       & 5\,800    &   \cite{JianNote} \\
%\end{tabular}
%\caption{The table lists the measurements performed in Run~I and Run~II  by the CDF collaborations in the missing energy plus jets signature.}
%\end{table}
%\end{center}
%
%
%RUN II
%
\begin{figure}
\includegraphics[width=7.cm]{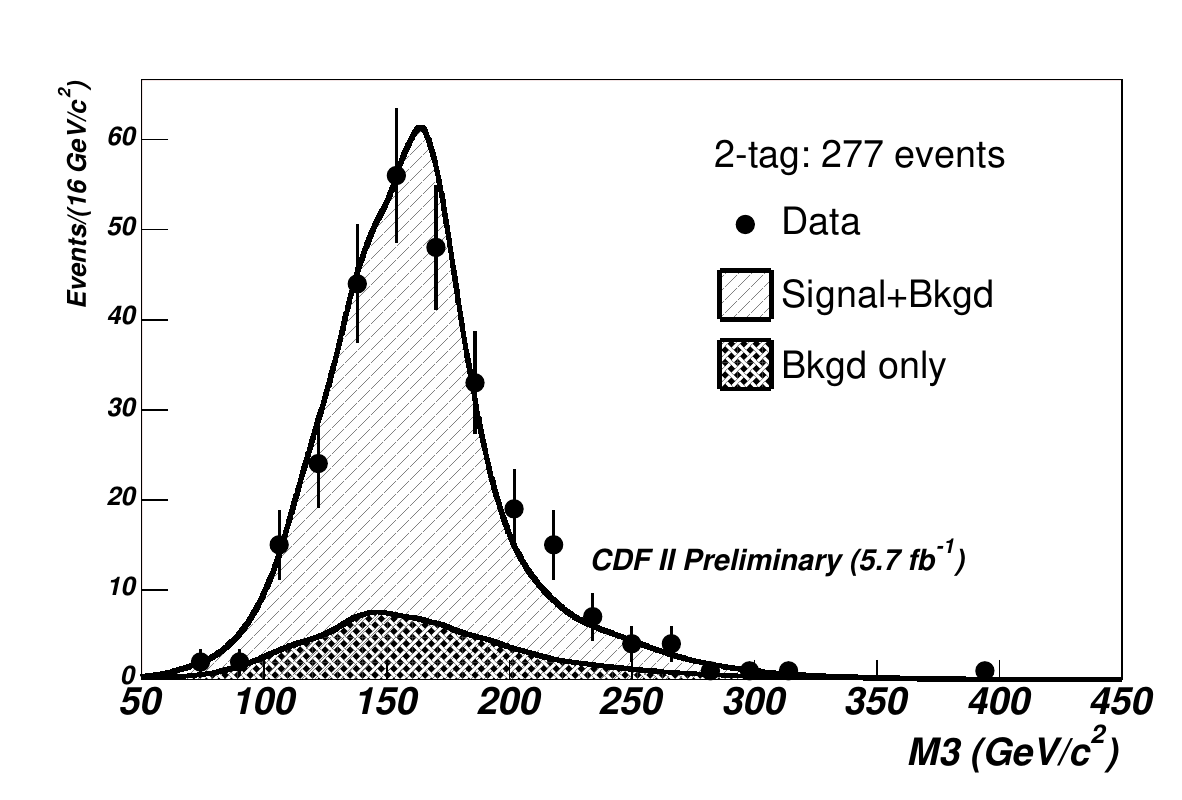}
\caption{Distribution of the reconstructed top quark mass for events 
with at least one $b$-tagged jet. The events are
required to satisfy the neural network event
selection and the goodness-of-fit criterion~\cite{ref:cdf.etmissplusjets}.
}
\label{fig:comb}
\end{figure}
\begin{figure}
\includegraphics[width=8.0cm]{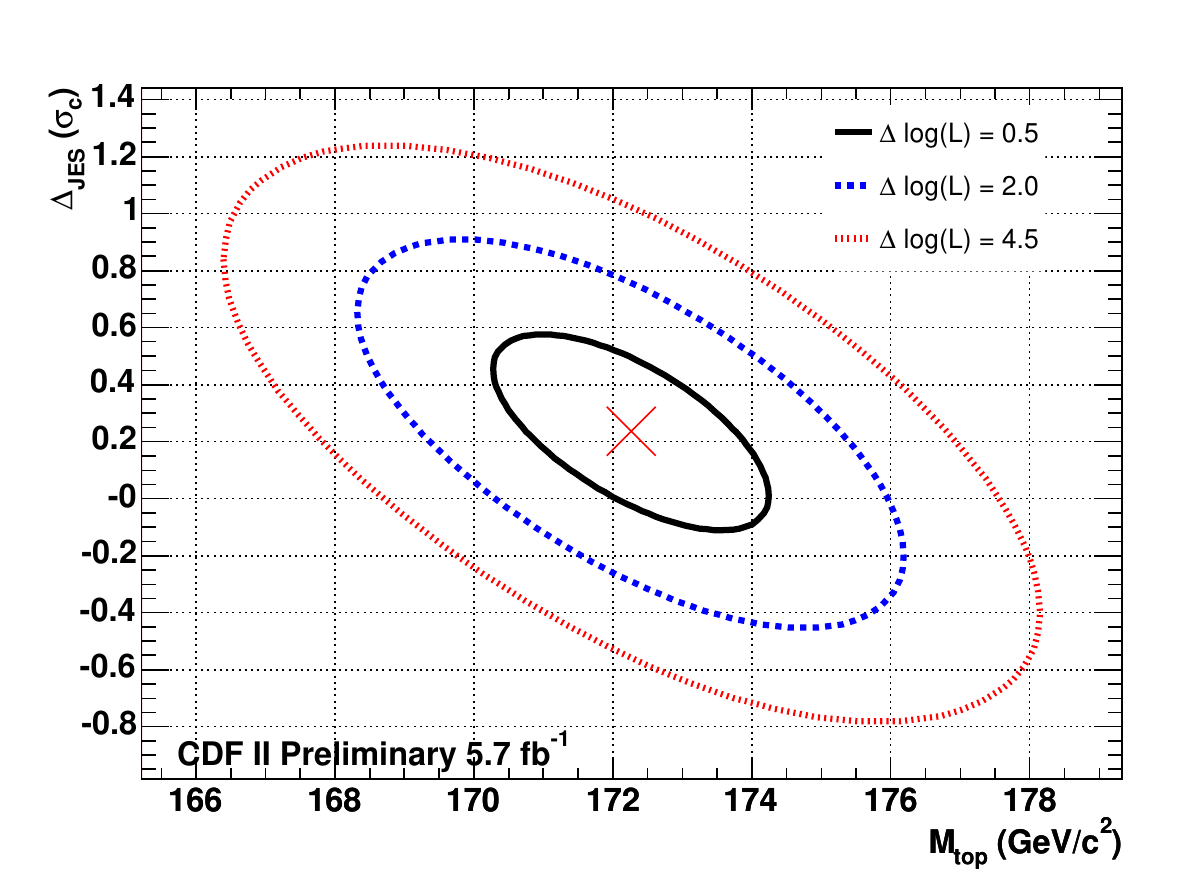}
\caption{Simultaneous measurement of the top quark mass and of the 
jet energy scale in the $\Etmiss$+jets channel.
The $\Delta_{JES}$ parameter is related to the jet
energy scale according to~\eref{eq:jesfromsigma}~\cite{ref:cdf.etmissplusjets}. 
%The uncertainty on the top quark mass is expressed in GeV while the uncertainty in the $\Delta$JES measurement is expressed in number of times the prior systematic of $\sim 3\%$.
}
\label{fig:metjets2D}
\end{figure}

Owing to the large signal acceptance, the good S/B ratio, and the capability to constrain {\em in situ} the jet energy scale uncertainty, this measurement achieves a precision superior to that attainable in the dilepton channel.
The result is $\Mt = 172.3 \pm 1.8 \,(\mbox{stat.}) \pm 1.5 \,(\mbox{JES}) \pm 1.0 \,(\mbox{syst.})$~GeV/$c^2 = 172.3 \pm 2.6 \,(\mbox{total})$~GeV/$c^2$, in agreement with the \Mt\ estimates in 
other decay modes. This result is shown in Figure~\ref{fig:metjets2D},
together with the jet energy scale estimate.

It should be noted that for this channel, just
as for the $\ell$+jets and all-hadronic
signatures, the uncertainty on the JES shift is much lower than the
typical 3\% uncertainty of the prior calibration. In other words, the
collective light quark JES measurements obtained
with about 5,000 $t\bar{t}$ events
in the $\ell$+jets, all-hadronic and $\Etmiss$+jets final states,
validated the data-to-MC correspondence of the overall JES at better
than one percent level.
% , with an approximate 1\% uncertainty. 
The Tevatron
experiments did not use this value in data analyses other than the \Mt\ 
measurements where they were derived, conservatively assuming that
different jet multiplicities could alter the jet shapes and result
in a systematic uncertainty different from the one quoted here.
Still, the \Mt\ measurements in these three channels utilize events
that range in jet multiplicity from four to eight, with compatible
results. Even though the details of the $p_T$ and $\eta$ dependence of
the light quark jet response remain to be understood at this level
of precision, the Tevatron
results suggest that, with the larger dataset available at the LHC, it
should be possible to calibrate light quark jets using the hadronically
decaying $W$s from $t \bar t$ events with uncertainty of 1\% or better.

\subsection{Dilepton measurements}
\label{sec:dilep}

Due to the presence of two charged leptons and two
neutrinos in the final state, the event samples used
to measure \Mt\ in the dilepton channel have excellent
signal purity. Compared to other $t\bar{t}$ final states, physics effects
beyond the SM affecting the \Mt\ estimate are easier to detect.
However, a relatively low branching fraction ($\approx 5$\%, not
counting events with the $\tau$ lepton)
and the absence of the hadronic $W$ decay which provides a natural
reference point for jet energy calibration
in other channels
limit achievable statistical and systematic precision
of the \Mt\ estimate. The dilepton measurements thus complement
more precise results obtained in other channels.

At the time of this writing, the dilepton \Mt\ measurement with
the smallest uncertainty is performed with 5.4~fb\supers{-1} of
data collected by the \D0 \ detector in Tevatron 
Run~II~\cite{ref:d0.dilep.run2bmatrel}. The events are
naturally split into subsamples with $ee$, $e\mu$, or $\mu\mu$
in the final state, with different expected backgrounds.
For the $ee$ and $\mu\mu$ final states, the events are selected
by a set of single-lepton triggers. For the $e\mu$ channel,
a mix of single and multilepton triggers and lepton+jet triggers
is utilized. The events are required to have 
two oppositely charged isolated leptons with
$p_T > 15$~GeV/$c$, and either $|\eta| < 1.1$ or $1.5 < |\eta| < 2.5$ for
electrons and $|\eta| < 2$ for muons, ensuring high efficiency
and low fake rate for lepton identification.
At least two jets are required with $p_T > 20$~GeV/$c$ and $|\eta| < 2.5$.
Background is further suppressed by the $H_T > 115$~GeV requirement
in the $e\mu$  final state
and by requiring substantial missing transverse
momentum in the $ee$ and $\mu\mu$ states.
479 candidate
events are selected with 73, 266, and 140 events, respectively, in the
$e e$, $e \mu$, and $\mu \mu$ channels, of which about $13 \pm 5$,
$48 \pm 15$, and $56 \pm 15$ events, respectively, are expected to arise
from background.
The sample purity
(S/B $\approx$ 3) is already sufficient
so that explicit $b$ flavor identification is not required for jets.

The \Mt\ estimate is constructed by the matrix element method.
In the calculation of event observation probability,
two contributions are taken into account: 
the $q\bar{q} \rightarrow t\bar{t}$
signal process and $p\bar{p} \rightarrow Z$ + 2 jets production
which is expected to be the dominant source of background.
The jet energy transfer functions are modeled
according to \eref{eq:doublegauss}. Directions
of all jets and charged leptons, as well as electron energies,
are assumed to be perfectly measured.
The uncertainty of the  muon track curvature determination
is modeled by a Gaussian resolution function.
Transfer functions are also
constructed for the fraction of parent energy carried by
electrons and muons produced in the
$\tau \rightarrow \ell \nu_{\ell} \nu_{\tau}$
decays (relevant for the $Z \rightarrow \tau^{+}\tau^{-}$ decay mode).
Under the assumption that the direction of the
charged lepton coincides with the direction of the parent tau
and that the daughter leptons are massless, the appropriate
laboratory spectrum is described in~\cite{ref:smc}.

The phase space integration for the signal hypothesis is performed over
the following variables: transverse momentum of the $t\bar{t}$ system
with a prior derived from the ALPGEN event generator~\cite{ALPGEN},
energies of the $b$ quarks, the lepton-neutrino invariant masses,
the differences between neutrino transverse momenta, 
and the muon track curvature ($p_T^{-1}$).

\begin{figure}
\centerline{
\epsfig{file=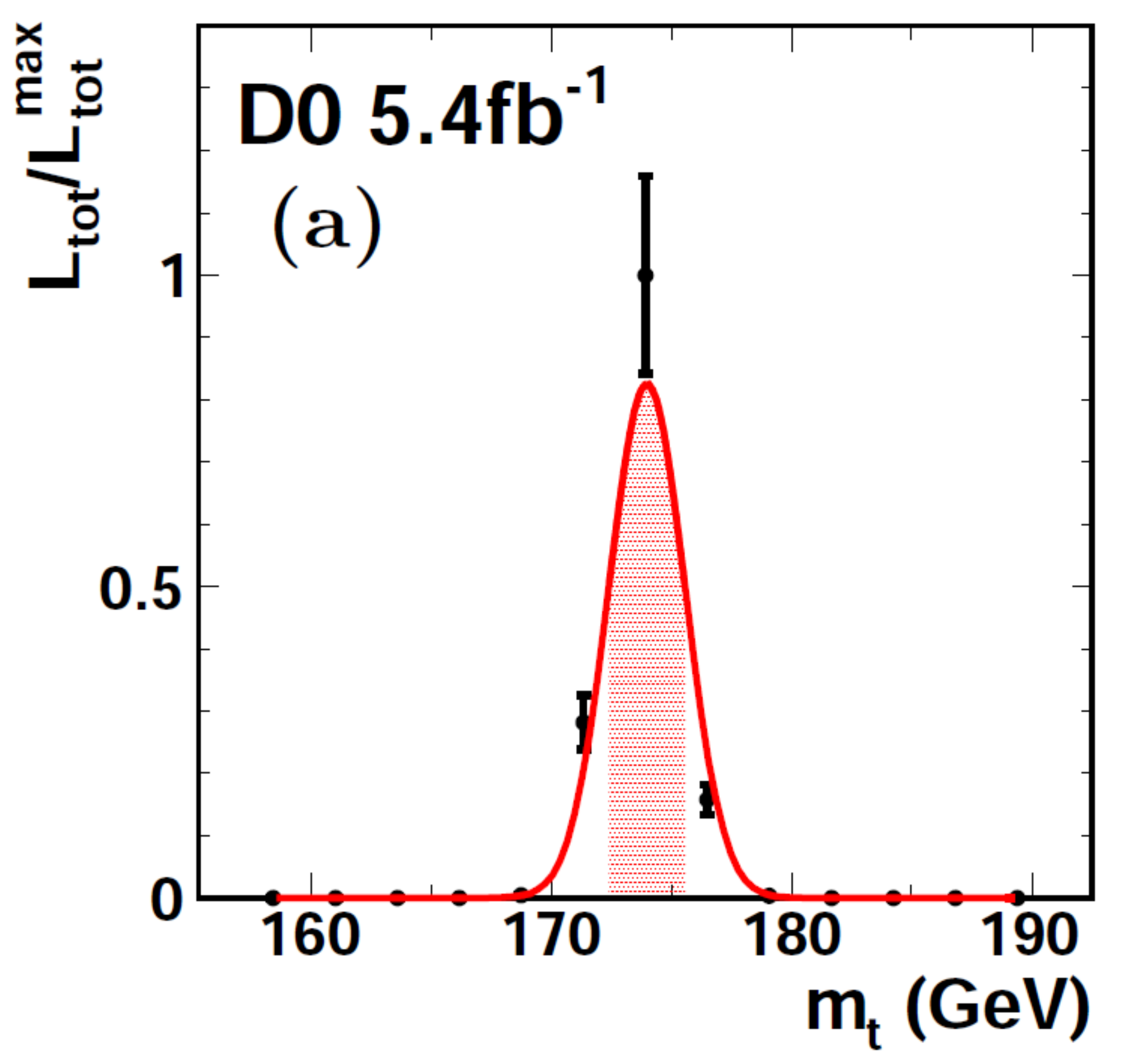,width=.3\textwidth}
}
\caption{Calibrated and normalized
likelihood for data as a function of \Mt\ with best estimate
as well as 68\% confidence level region marked by the
shaded area~\cite{ref:d0.dilep.run2bmatrel}.}
\label{fig:d0_dilepton_likelihood}
\end{figure}

The background matrix element is calculated using VECBOS~\cite{ref:vecbos}.
The background phase space is sampled over the energies of the two partons that produce
the jets and, in the case of $Z \rightarrow \tau^{+}\tau^{-}$ decays, over
the energy fractions of the charged leptons produced in tau decays.
The $p_T$ probability of the $Z$ + 2 jets system is used as the
additional weight, together with the leading order matrix element,
parton distribution functions, and appropriate phase space factors.
Background normalization is adjusted in a manner similar to that
employed in the \D0 \ \Mt\ measurement in the $\ell+$jets channel
(Section~\ref{sec:lepjets}). The likelihood as a function
of \Mt\ is obtained by profiling over the $t\bar{t}$ signal
fraction. The bias and pulls of the final estimate are corrected 
by a mapping function, and the Bayesian credible interval
is extracted. The calibrated likelihood is shown in
Figure~\ref{fig:d0_dilepton_likelihood}.

In this
measurement, the largest component (2.2~GeV/$c^2$)
of the systematic uncertainty is contributed by the calibration
of the jet response. The overall result is 
$\Mt = 174.0 \pm 1.8 \,(\mbox{stat.}) \pm 2.4 \,(\mbox{syst.})$~GeV/$c^2$,
or $\Mt = 174.0 \pm 3.1\,(\mbox{total})$~GeV/$c^2$.

The current most precise CDF \Mt\ measurement in the dilepton channel
is performed by simultaneous application of the template method
in the $\ell$+jets and dilepton final states which allows for
{\it in situ} calibration of the jet energy scale in both
channels~\cite{ref:cdf.template.latest}. 5.6~fb\supers{-1} of data
collected by the CDF detector in Tevatron
Run~II is used, with $\ell$+jets
selection criteria similar to those described in Section~\ref{sec:lepjets}.
The dilepton events are required to have two well-identified
oppositely charged leptons and at least two jets with
$E_T > 15$~GeV and $|\eta| < 2.5$. To further reject backgrounds,
both missing transverse energy and $H_T$ requirements are imposed:
$\Etmiss > 25$~GeV and $H_T > 200$~GeV.

\begin{figure}
  \includegraphics[width=7.5cm]{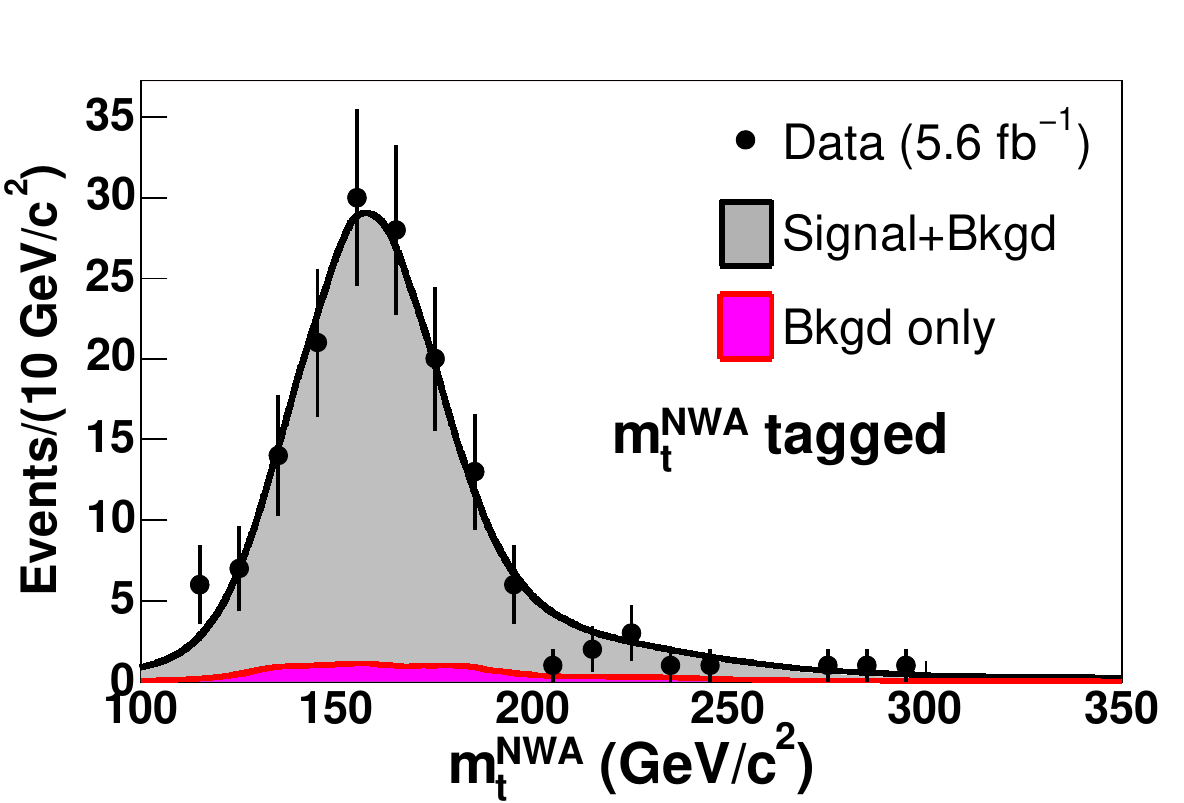}
  \includegraphics[width=7.5cm]{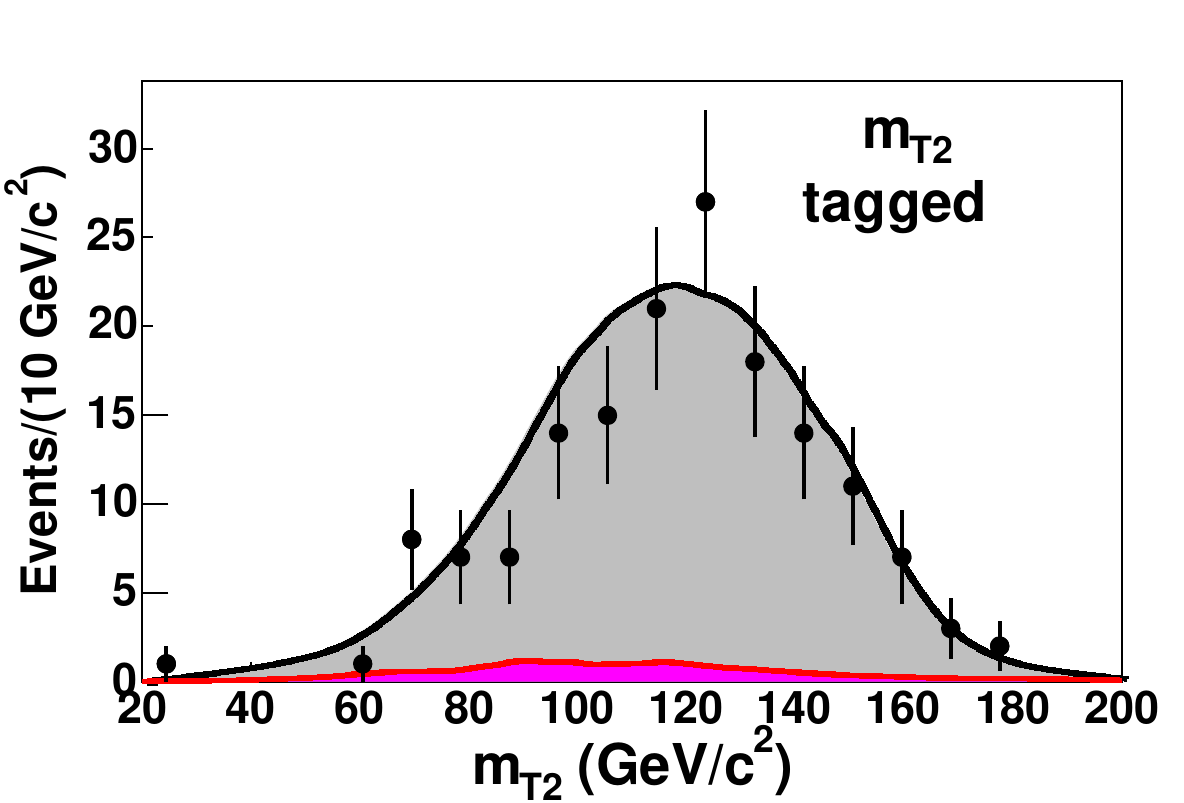}
  \caption{Distributions of $m_t$ values obtained with the $\nu$WT method (left)
           and of the transverse mass $m\sub{T2}$ 
           (right) using CDF $b$-tagged events. The data is overlaid
            with KDE templates~\cite{ref:cdf.template.latest}.}
  \label{fig:dileptondistros}
\end{figure}

Two-dimensional templates are constructed for the \Mt\ determination
by kernel density estimation (KDE).
The variables used are the event-by-event $m_t$ values obtained with
the neutrino weighting algorithm ($\nu$WT, described in
Section~\ref{subsec:kinemfit}) and the transverse mass $m$\sub{T2}.
Distributions of these quantities in the data and in a simulated
sample with $\Mt = 170$~GeV/$c^2$
are shown in Figure~\ref{fig:dileptondistros}.
In the $\ell$+jets channel, $m_t$ values are obtained by kinematic fitting
based on $\chi^2$ minimization (Section~\ref{subsec:kinemfit}).
Twelve jet-to-parton assignments are considered which result
in distinct $\chi^2$.
The $m_t$ values from the two fits with the lowest $\chi^2$ are combined
with $m_{jj}$ (for jets assumed to come
from the hadronic $W$ decay)
from the lowest $\chi^2$ permutation, and three-dimensional
templates are built by KDE.
As both signal and background template shapes
depend on the presence of $b$-tagged jets in the sample,
the templates are made separately
for a number of subsamples: 1-tag and 2-tag
for $\ell$+jets events (at least one secondary vertex
$b$ tag is required),
non-tagged and tagged for dilepton
events. The overall likelihood is obtained
as a function of \Mt\ and $\Delta_{JES}$
by multiplying together the likelihoods from all subsamples.
The $\Delta_{JES}$ nuisance parameter is then eliminated
by profiling. Finally, remaining biases and pulls are corrected
by a mapping function.

Despite cross-calibration from the $\ell$+jets channel,
the systematic uncertainty of this measurement is still dominated
by jet response modeling (of $b$ jets in particular) which
contributes 3.0~GeV/$c^2$. The obtained result is
$\Mt = 170.3 \pm 2.0 \,(\mbox{stat.}) \pm 3.1 \,(\mbox{syst.})$~GeV/$c^2$,
with a total uncertainty of 3.7~GeV/$c^2$. Tevatron \Mt\
measurements in the dilepton channel are summarized
in Table~\ref{table:mtdilep}.
\begin{table*}[hbtp]
\caption{Representative
         Tevatron \Mt\ measurements in the dilepton channel. The ``+'' sign
         between the methods indicates that the overall result is obtained by
         combining multiple techniques. ``$\otimes$'' joins variables used
         in multivariate templates.}
\label{table:mtdilep}
% \begin{indented}
% \item[]\begin{tabular}{@{}lcccccc}
\begin{tabular}{@{}lcccccc}
\br
Experiment & Method & $\int \mathcal{L} dt$ & \Mt\ & $\sigma\sub{stat}$ & $\sigma\sub{syst}$ & Ref. \\
 \ & \ & (fb\supers{-1}) & (GeV/$c^2$) & (GeV/$c^2$) & (GeV/$c^2$) & \ \\
\mr
CDF (Run I)    & $\nu$WT   & 0.11 & 167.4 & 10.3 & 4.8 & \cite{ref:cdf.dilep.run1prl} \\
\D0 \ (Run I)  & ${\mathcal M}$WT + $\nu$WT  & 0.13 & 168.4 & 12.3 & 3.6 & \cite{ref:d0.dilep.run1prl, ref:d0.dilep.run1prd} \\
\D0            & ${\mathcal M}$WT & 0.23 & 155   & $\mbox{}^{+14}_{-13}$ & 7 & \cite{ref:d0.dilep.mdg1} \\

CDF            & MEM  & 0.34 & 165.2 & 6.1 & 3.4 & \cite{ref:cdf.dilep.firstmem, ref:cdf.dilep.matrel.prd} \\

CDF            & DLM  & 0.34 & 166.6 & $\mbox{}^{+7.3}_{-6.7}$ & 3.2 & \cite{ref:cdf.dilep.dlm} \\

CDF            & $\nu$WT + KIN + PHI & 0.36 & 170.1 & 6.0 & 4.1 & \cite{ref:cdf.dilep.3templates} \\

% \D0            & ${\mathcal M}$WT & 0.37 & 177   & 11 & 4 & \cite{ref:d0.dilep.mdg2} \\

\D0            & ${\mathcal M}$WT + $\nu$WT  & 0.37 & 178.1 & 6.7  & 4.8 & \cite{ref:d0.dilepton.matrixweighting2} \\

% \D0            & ${\mathcal M}$WT & 1.0 & 175.2   & 6.1 & 3.4 & \cite{ref:d0.dilep.matrixwr2a} \\
% \D0            & $\nu$WT & 1.0 & 172.5  & 5.8 & 3.5 & \cite{ref:d0.dilep.nuwr2a} \\

\D0            & ${\mathcal M}$WT + $\nu$WT  & 1.0  & 174.7 & 4.4  & 2.0 & \cite{ref:d0.dilep.run2a} \\

CDF            & MEM      & 1.0  & 164.5 & 3.9 & 3.9 & \cite{ref:cdf.dilep.matrel.isrtf} \\

CDF            & KIN + cross section & 1.2  & 170.7 & $\mbox{}^{+4.2}_{-3.9}$ & 3.5 & \cite{ref:cdf.dilep.xsecconstrained} \\

CDF            & $\nu$WT  & 1.9  & 172.0 & $\mbox{}^{+3.6}_{-3.4}$ & 3.8 & \cite{ref:cdf.template.pub1} \\

CDF            & MEM      & 2.0  & 171.2 & 2.7 & 2.9 & \cite{ref:cdf.dilep.neuroevolution} \\

\D0            & MEM      & 2.8  & 172.9 & 3.6  & 2.3 & \cite{ref:d0.dilep.2.8fb} \\

CDF            & PHI      & 2.9  & 165.5 & $\mbox{}^{+3.4}_{-3.3}$ & 3.1 & \cite{ref:cdf.dilep.nuphiweight} \\

CDF            & $\nu$WT $\otimes$ $m\sub{T2}$ & 3.4 & 169.3 & 2.7 & 3.2 & \cite{ref:cdf.dilep.mt2} \\

\D0            & MEM      & 3.6  & 174.8 & 3.3  & 2.6 & \cite{ref:d0.dilep.3.6fb} \\

\D0            & $\nu$WT  & 5.3  & 173.3 & 2.4  & 2.1 & \cite{ref:d0.latest.nuwt} \\
\D0            & MEM      & 5.4  & 174.0 & 1.8  & 2.4 & \cite{ref:d0.dilep.run2bmatrel} \\
CDF            & $\nu$WT $\otimes$ $m\sub{T2}$ & 5.6  & 170.3 & 2.0  & 3.1 & \cite{ref:cdf.template.latest} \\
\br
\end{tabular}
% \end{indented}
\end{table*}

\subsection{Measurements which do not use jets}
\label{sec:other_ms}

A number of unconventional top quark mass measurement ideas
were proposed in the recent literature and tested on the Run~II
Tevatron data. These proposals aim to reduce the systematic
uncertainty (especially its part related
to the jet energy response calibration),
often at the cost of a substantial degradation
in the statistical precision. Even though
at this time such methods are not competitive with the ``traditional''
techniques described earlier in this section,
their special features could make them attractive
for the analysis of massive $t\bar{t}$ event samples
which will be accumulated by the LHC experiments. A few of these ideas
are described below.

\Mt\ measurements based on the tracking information alone were
explored by CDF in~\cite{ref:cdf.ljets.lxyandlpt}. Two variables 
were chosen
to build the templates in the $\ell$+jets channel: L\sub{xy} which is the
transverse distance between the primary and secondary vertices
projected onto the jet direction and $p_T$ of the lepton produced in the
$W \rightarrow \ell \nu_{\ell}$ decay.
Standalone lepton $p_T$ templates were later used by CDF to estimate \Mt\
in the $\ell$+jets~\cite{ref:cdf.ljets.lpt}
and dilepton~\cite{ref:cdf.dilep.lpt} channels.

Both L\sub{xy} and lepton $p_T$ variables 
are sensitive to \Mt\ because $W$ bosons and $b$
quarks produced in the $t \rightarrow W b$ decays receive higher
transverse boosts from heavier top quarks, as illustrated
in Figure~\ref{tracking_templates}. While the sensitivity
of such variables to the calorimeter hadronic jet energy scale calibration
is indeed minimal, reliance upon manifestly non-Lorentz invariant
quantities has its own disadvantages. Both variables are highly
sensitive to the mismodeling of parton distribution functions which
affect typical center-of-mass
energies and therefore average boosts of parent top
quarks. The sensitivity to the uncertainties in background shape and
composition is also increased. In addition, the L\sub{xy} distributions in
Monte Carlo are strongly affected by the $b$ jet fragmentation model, while
the lepton $p_T$ measurements suffer from imprecise energy scale
calibration of the electromagnetic calorimeter.
On the other hand, the L\sub{xy}-based estimate
is essentially complementary to all other methods of top mass determination,
and constitutes an independent source of information about \Mt\ (the same
can not be said about lepton $p_T$ which
is used either explicitly or implicitly by kinematic fitting
and matrix element techniques). 

\begin{figure}
\centerline{
\epsfig{file=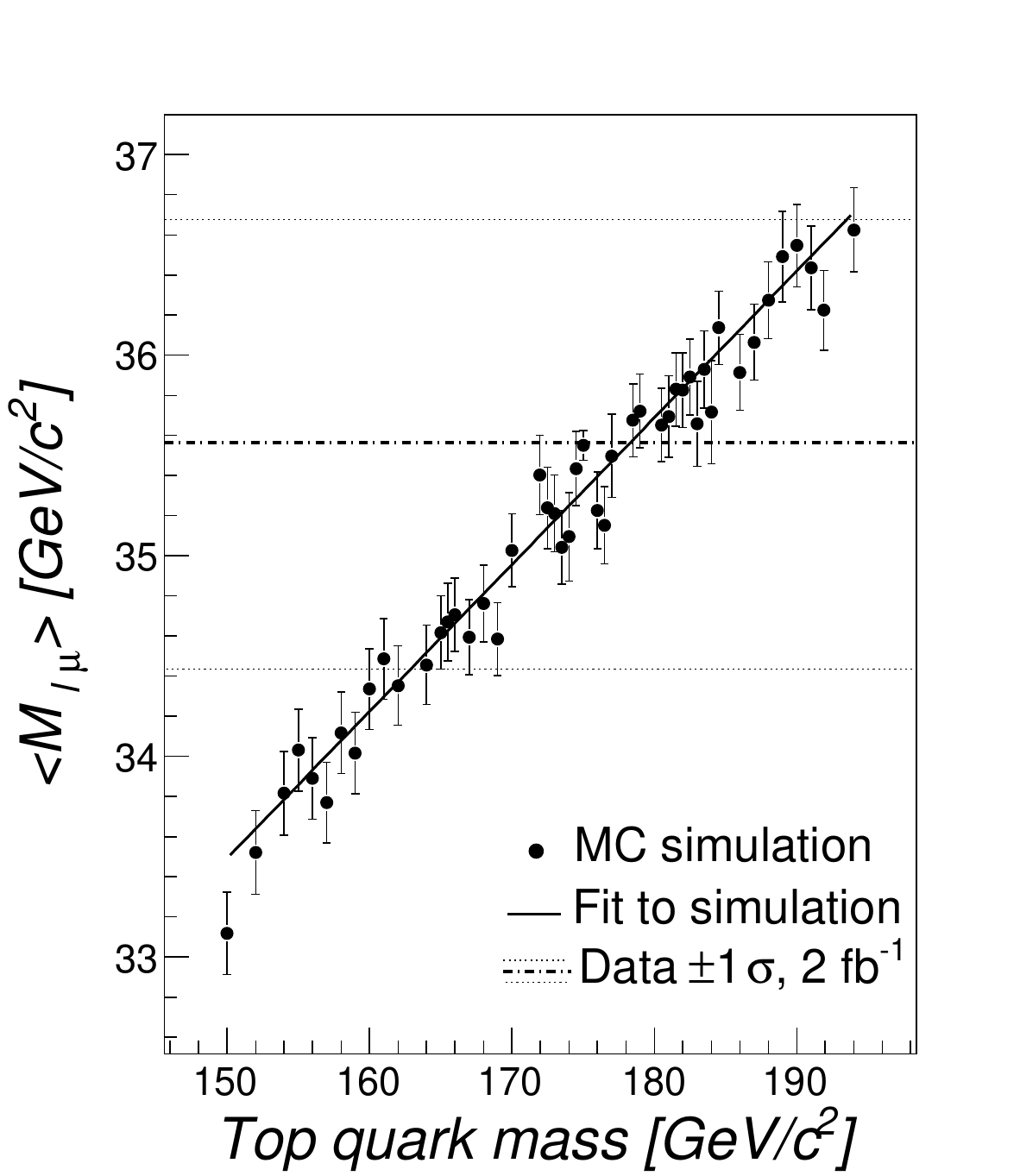,width=.35\textwidth}
}
\caption{
The correlation between the mean value of the $M_{\ell \mu}$
histograms from simulated $t\bar{t}$ and background samples and the
input \Mt. The continuous line shows a linear fit
to the points~\cite{ref:cdf.ljets.linvmass}.}
\label{fig:mlnu}
\end{figure}

The invariant mass, $M_{\ell \mu}$, of the charged lepton from
the $W$ decay and the muon from a semileptonic decay
of the $b$ quark (originated from the same parent top as the leptonically
decaying $W$) was used as template variable in the CDF
study~\cite{ref:cdf.ljets.linvmass}. The efficiency
to ``tag'' a $t\bar{t}$ event passing typical $\ell$+jets event
selection criteria by finding a soft muon produced in a decay 
of a $b$ quark is about 14\%~\cite{ref:cdf.softlepton.xsec}.
This leads to a sizeable event sample: 248 events were selected in 
the 2.0~fb\supers{-1} of CDF Run~II data.
The selection included events with 3 jets which resulted in
a somewhat higher efficiency but lower than usual S/B of about 2.
The average $M_{\ell \mu}$ of the selected sample depends
approximately linearly on \Mt, as illustrated in Figure~\ref{fig:mlnu}.
In this analysis, the dominant components of the systematic
uncertainty are due to imprecise knowledge of
the background shape and fraction
in the $M_{\ell \mu}$ template ($\sigma\sub{bg} = 1.9$~GeV/$c^2$,
statistically limited by the size of the available
event sample) and due to uncertainties in
signal modeling, especially $b$ jet fragmentation, conservatively
estimated by comparing \Mt\ estimates obtained for event
samples generated by HERWIG~\cite{ref:herwig}
and PYTHIA~\cite{PYTHIA} ($\sigma\sub{s} = 2.1$~GeV/$c^2$).

The \Mt\ estimates obtained by methods
which rely mainly on tracking information are collected
in Table~\ref{table:trackingres}.
\begin{table*}[hbtp]
\caption{\Mt\ estimates which rely on tracking information (all by CDF).}
\label{table:trackingres}
% \footnotesize
% \begin{indented}
% \item[]\begin{tabular}{@{}ccccccc}
\begin{tabular}{@{}ccccccc}
\br
Template & Channel & $\int \mathcal{L} dt$ & \Mt\ & $\sigma\sub{stat}$ & $\sigma\sub{syst}$ & Ref. \\
quantity & \ & (fb\supers{-1}) & (GeV/$c^2$) & (GeV/$c^2$) & (GeV/$c^2$) & \ \\
\mr
L\sub{xy} & $\ell$+jets & 0.7 & 180.7 & $\mbox{}_{-13.4}^{+15.5}$ & 8.6 & \cite{ref:cdf.ljets.lxy} \\
Lepton $p_T$ & dilepton & 1.8 & 156 & 20 & 4.6 & \cite{ref:cdf.dilep.oldlpt} \\
L\sub{xy} & $\ell$+jets & 1.9 & 166.9 & $\mbox{}_{-8.5}^{+9.5}$ & 2.9 & \cite{ref:cdf.ljets.lxyandlpt} \\
Lepton $p_T$ & $\ell$+jets & 1.9 & 173.5 & $\mbox{}_{-8.9}^{+8.8}$ & 3.8 & \cite{ref:cdf.ljets.lxyandlpt} \\
$M_{\ell \mu}$ & $\ell$+jets & 2.0 & 180.5 & 12.0 & 3.6 & \cite{ref:cdf.ljets.linvmass} \\
Lepton $p_T$ & $\ell$+jets & 2.7 & 176.9 & 8.0 & 2.7 & \cite{ref:cdf.ljets.lpt} \\
Lepton $p_T$ & dilepton & 2.8 & 154.6 & 13.3 & 2.3 & \cite{ref:cdf.dilep.lpt} \\
\br
\end{tabular}
% \end{indented}
\end{table*}
Potential relevance
of these methods for top mass measurements
with LHC event samples will strongly depend on the improvements
in the modeling of parton distribution functions and $b$ jet fragmentation,
as well as on the availability of precise detector calibrations for
lepton $p_T$.

% The \D0 \ collaboration extracted top mass estimates using the $t\bar{t}$
% cross section measurements in the $\ell$+jets~\cite{ref:d0.ljets.fromxsec}
% and dilepton~\cite{ref:d0.dilep.mtfromxsec} channels.

% With large $t\bar{t}$ samples, a variety of other trade-offs
% between statistical and systematic uncertainty can be considered. For example,
% background-related systematics can be reduced by tightening the selection
% criteria and increasing the sample purity, jet-related systematics can
% be controlled by using only events with energetic, well-separated jets in
% the central detector region where efficiency and energy calibration is
% usually superior, {\it etc.} An even better approach consists in
% evaluating the systematic uncertainty contribution
% for any given source {\it on event-by-event basis}, by constructing
% the sample likelihood as a function of the corresponding nuisance
% parameter. Upon marginalization of the parameter with an appropriate prior,
% the events will be automatically combined
% with optimal weights which take this source of uncertainty into account.

\subsection{Top-antitop mass difference}
\label{sec:mtdelta}

\begin{figure}
% \centerline{
% \epsfig{file=fig2fromd0mdelta.pdf,width=.5\textwidth}
% }
\epsfig{file=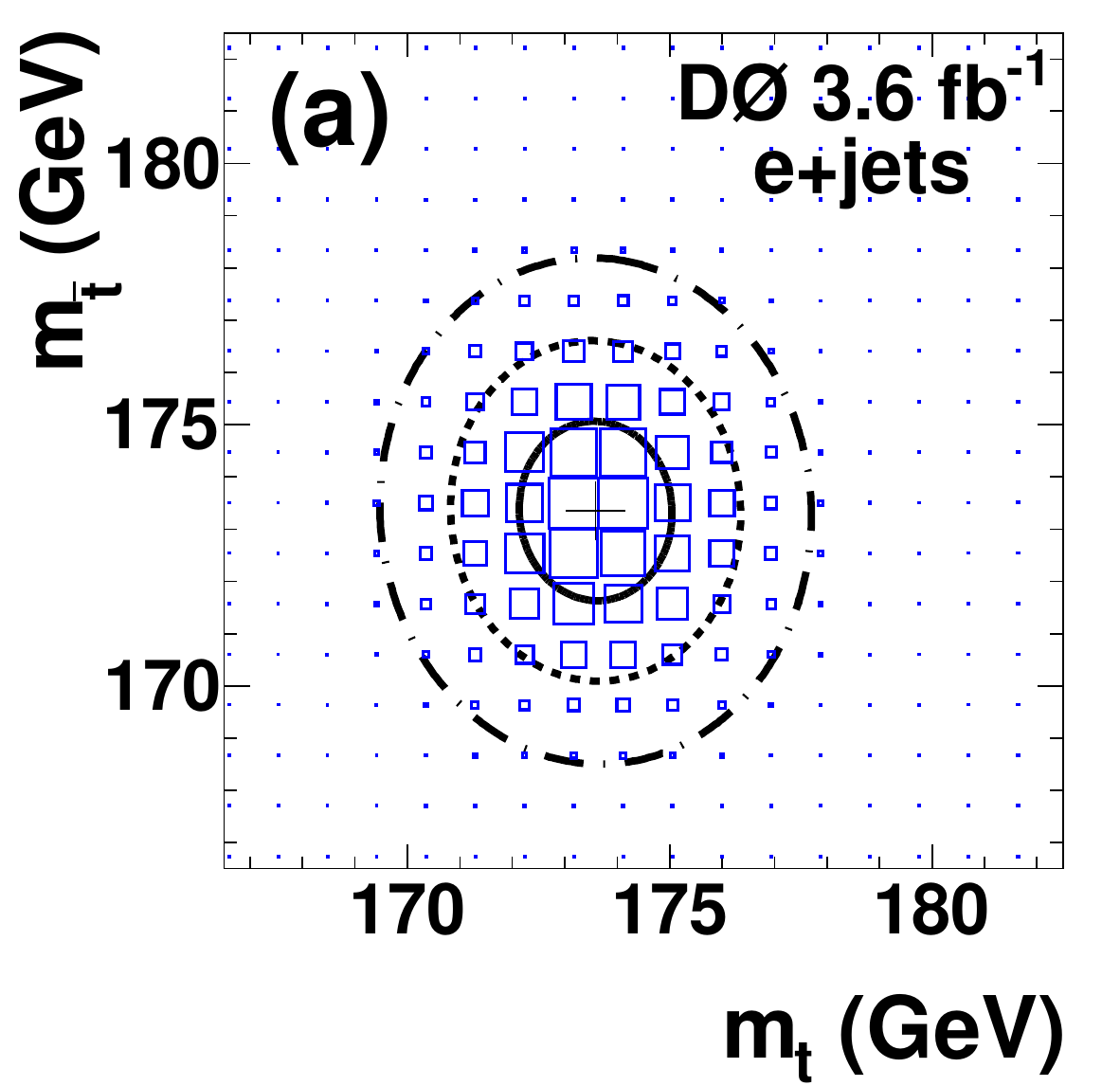,width=.25\textwidth}
\epsfig{file=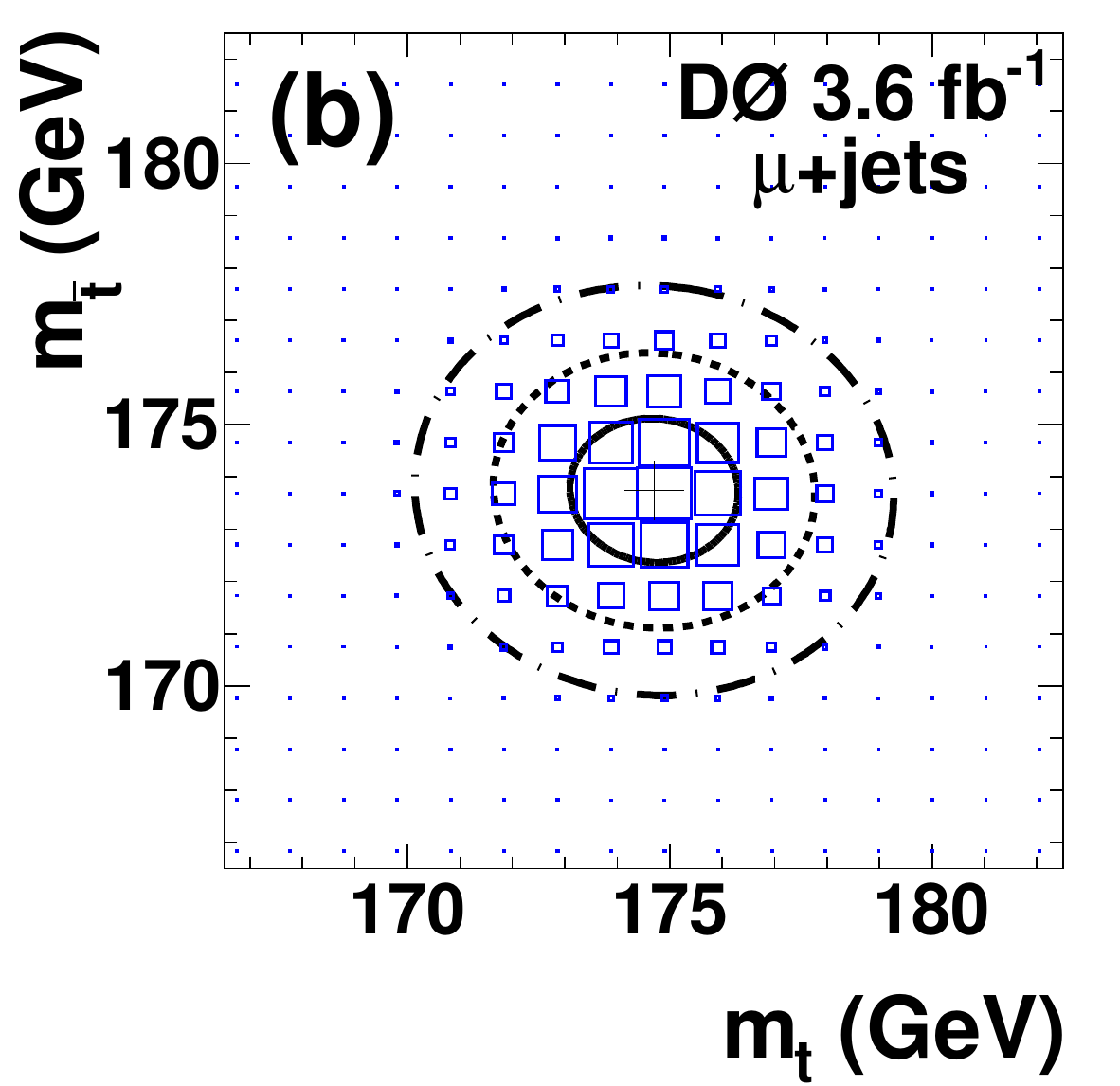,width=.25\textwidth}
\caption{
Sample likelihoods as a function of \Mt\ and \Mtbar\ for
(a) $e$+jets and (b) $\mu$+jets \D0 \ event
samples. The boxes, representing the
bins in the two-dimensional histograms of the likelihoods, have
areas proportional to the bin contents. The solid, dashed,
and dash-dotted lines represent 1, 2, and 3 standard deviation
contours of two-dimensional Gaussian fits before pull
corrections are applied~\cite{ref:d0.mtdelta.latest}.}
\label{fig:fig2fromd0mdelta}
\end{figure}

The \D0 \ collaboration was the first to search for
a violation of the CPT-invariance in the $t\bar{t}$ production and decay
processes by measuring the difference between the masses of the $t$ and
$\bar{t}$ quarks~\cite{ref:d0.mtdelta}.
%
% Both \D0 ~\cite{ref:d0.mtdelta, ref:d0.mtdelta.latest}
% and CDF~\cite{ref:cdf.mtdelta} collaborations have searched for
% a violation of the CPT-invariance in the $t\bar{t}$ production and decay
% processes by measuring the difference between masses of $t$ and
% $\bar{t}$ quarks. Within the confines of the Standard Model in which
% CPT-invariance is strictly enforced,
% these measurements can be viewed as sensitive tests of data
% analysis methods and detector response simulations.
%
The latest \D0 \ measurement~\cite{ref:d0.mtdelta.latest}
of $\Delta \equiv \Mt - \Mtbar$
is performed with an
event sample which corresponds to about 3.6~fb\supers{-1}
of Tevatron Run~II data.
The $\ell$+jets final state is selected
in which positively (negatively) charged leptons are
used to tag the $t$ ($\bar{t}\,$) quarks in each event.
The matrix element technique is utilized to construct the overall
likelihood as a function of $M\sub{sum} = (\Mt + \Mtbar)/2$,
$\Delta$, $JES$, and the signal fraction,
in a manner similar to the \D0 \ \Mt\ measurement described
in Section~\ref{sec:lepjets}. The signal fraction
nuisance parameter is eliminated by profiling, while $JES$
is fixed to a constant value consistent with the $W$ mass constraint.
The resulting likelihoods, represented as functions of
\Mt\ and \Mtbar, are shown in Figure~\ref{fig:fig2fromd0mdelta}
separately for the $e$+jets and  $\mu$+jets channels.
The final likelihood for $\Delta$ is constructed by marginalizing
$M\sub{sum}$ with a flat prior and subsequently combining
$e$+jets and $\mu$+jets channels.

Compared to an \Mt\ estimate, importance of various
sources of systematic uncertainties is substantially different 
in the $\Delta$ measurement.
For example, the jet energy scale-related
uncertainty of a
few percent becomes relatively insignificant, while effects which
can have different impact upon reconstruction
of $t$ vs. $\bar{t}$ decay products
become of major concern.
Such effects include the mismeasurement of the lepton charge
and uncertainties from modeling differences in the response
of the calorimeter to $b$ and $\bar{b}$ jets (most notably,
a different content of $K^{+}/K^{-}$ mesons which have
different interaction cross sections with the calorimeter
material). Taking these
considerations into account, the \D0 \ collaboration obtains
an estimate
$\Delta = 0.8 \pm 1.8 \,(\mbox{stat.}) \pm 0.5 \,(\mbox{syst.})$~GeV/$c^2$,
consistent with $\Mt = \Mtbar$.

The CDF measurement of $\Delta$
utilized 5.6~fb\supers{-1} of Run~II data~\cite{ref:cdf.mtdelta}.
This measurement is also performed in the $\ell$+jets channel,
with the sample selection criteria similar to those used
for the CDF \Mt\ measurement described in Section~\ref{sec:lepjets}.
Events without $b$ tags are admitted as well, with an additional
requirement $H_T > 250$~GeV. The template approach is utilized.
The event-by-event estimate of $\Delta$,
$\Delta m_t$, is obtained from
a kinematic fit. $\chi^2$ is evaluated according to
expression \eref{eq:chisq} in which
the two terms that include $m_t$ are replaced by
$(M_{b \ell \nu} - (M\sub{sum} - dm_{reco}/2))^2 / \Gamma_t^2 + (M_{b j j} - (M\sub{sum} + dm_{reco}/2))^2 / \Gamma_t^2$.
$M\sub{sum}$ is fixed at 172.5~GeV/$c^2$.
$\Delta m_t$ is subsequently
defined by $\Delta m_t = -Q_{\ell} \cdot dm_{reco}^{min}$, where
$Q_{\ell}$ is the sign of the lepton charge which distinguishes
$t$ and $\bar{t}$, and $dm_{reco}^{min}$ is the value of
$dm_{reco}$ that minimizes the $\chi^2$.
Two-dimensional templates are constructed using the two
$\Delta m_t$ values which
correspond to the best and second best (over distinct
jet-to-parton assignments and kinematic solutions)
values of $\chi^2$,
with input $\Delta$ varied between $-20$ and 20~GeV/$c^2$.
To improve the template modeling, the event sample is
split into six subsamples with zero, one, or two $b$ tags
and different values of $Q_{\ell}$.
Templates are subsequently fitted to the distributions of
best and second best $\Delta m_t$ observed in the data.
To determine $\Delta$, log-likelihoods from all event subsamples are added and
the statistical uncertainty is scaled by 1.04 to ensure proper
frequentist coverage (the bias is found to be consistent with zero).

As in the \D0 \ measurement, the dominant part of the systematic
uncertainty is due to signal mismodeling. CDF obtains
$\Delta = -3.3 \pm 1.4 \,(\mbox{stat.}) \pm 1.0 \,(\mbox{syst.})$~GeV/$c^2$,
which is consistent with $\Mt = \Mtbar$ at the about $2 \sigma$ level.

\subsection{Mass from cross section}

\begin{figure}[htbp]
\centerline{
\epsfig{file=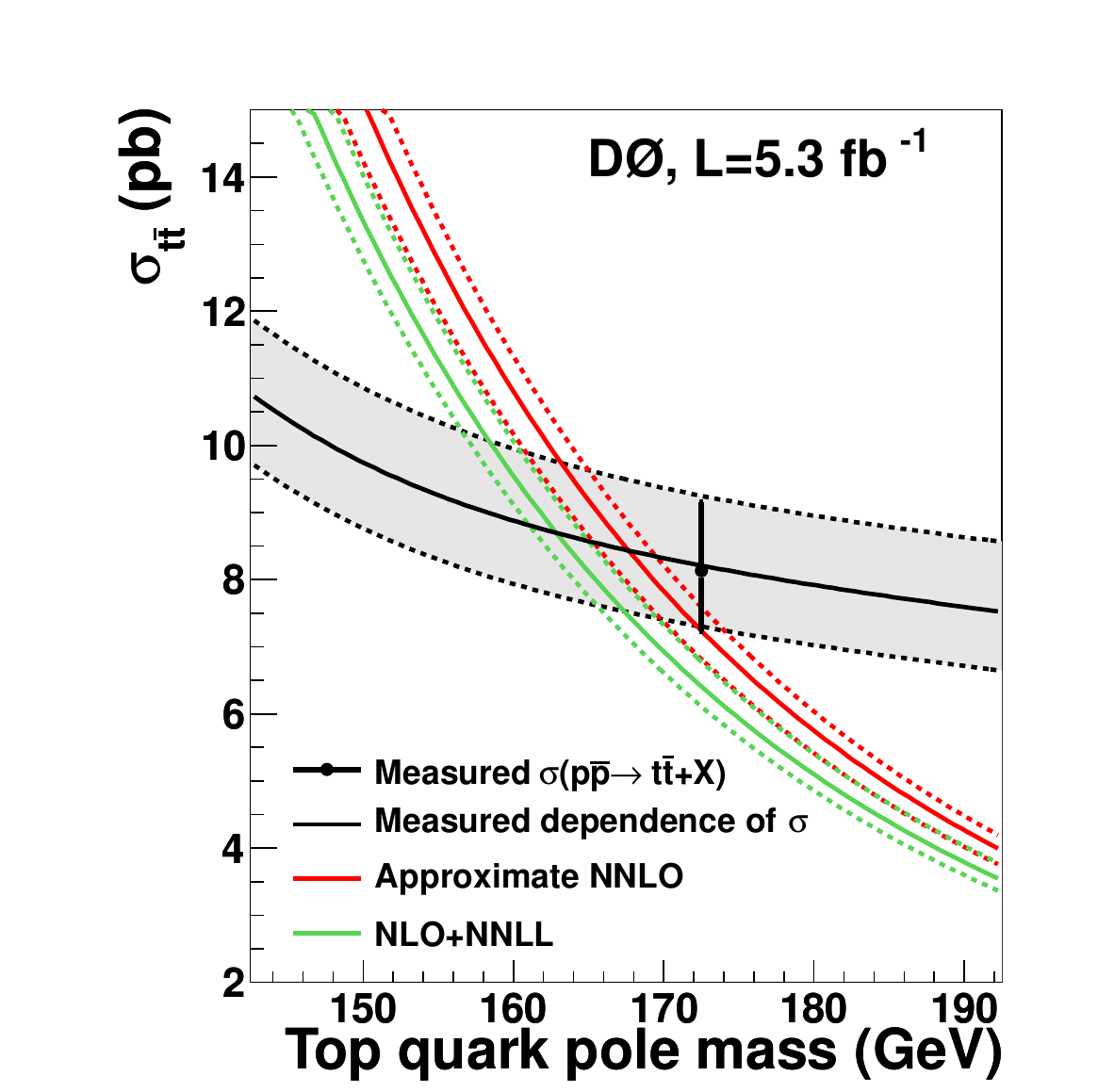,width=.4\textwidth}
}
\caption{Top cross section vs. top quark mass as calculated by different 
         authors~\cite{langenfeld,Ahrens-NNLL}. The data point is the
         \D0 ~ measurement~\cite{d0-sigma-MS} mentioned in the text.  }
\label{sig-polemass-d0}
\end{figure}

The theoretical $t\bar{t}$ cross section prediction
depends on the top mass, as shown in 
Figure~\ref{sig-mass} in Section~\ref{sec:top-sig}. Therefore, given a 
theory curve and a cross section measurement, it is possible to extract
a top mass estimate.
This has been done by the \D0 ~collaboration using several cross section
measurements: in the dilepton channel~\cite{D0-tmass-dilep}, 
using a combination of different channels~\cite{D0-tmass-all}, and, most recently, 
in the $\ell$+jets channel~\cite{d0-sigma-MS}.
The latter measurement, compared with two theoretical
cross section calculations,
is shown in Figure~\ref{sig-polemass-d0}.

The measured value of the cross section is positioned
at $\mt$ = 172.5 \gevocs, 
the mass value used to calculate the acceptance for the measurement.
The gray band depicts the total experimental uncertainty 
plotted as a function of the top mass.
The appoximate NNLO calculations are from Ref~\cite{langenfeld}, the NLO+NNLL 
calculations are from Ref~\cite{Ahrens-NNLL}. The uncertainties in the 
theoretical calculations are shown by the dotted lines. 
% From this plot it is possible
% to extract an estimate of the top mass using the two cross section
% approximations. 
The analysis extracting the top mass estimate from these
cross section approximations
makes the assumption that $M\sub{MC}$ = $M\sub{pole}$ (see 
Section~\ref{sec:tmass-def}). The results are:

 \begin {center}        
        $M\sub{pole} = 163.0 ^{+5.4}_{-4.0}$ \gevocs \ for NLO+NNLL~~~

        $M\sub{pole} = 167.5 ^{+5.4}_{-4.9}$ \gevocs \ for approx. NNLO    
\end{center}

\noindent
The value obtained with the approximate NNLO 
calculation is closer to the  top mass  measured  directly but
the uncertainties are large in both cases. 
% The same authors have calculated the $t\bar{t}$ production cross 
% section in the ${\overline{MS}}$ renormalization scheme.
\D0 ~has also extracted a value 
for the top quark mass in the $\overline{MS}$ renormalization
scheme, as mentioned in 
Section~\ref{sec:tmass-def}.

\section{Combination}
\label{sec:combo}

% The CDF and \D0 \ collaborations measured the \Mt\ parameter using the several techniques described in Section~\ref{sec:meas_techs} in the different decay modes described in Section~\ref{sec:tev_runii}.
While all individual \Mt\ measurements provide important information 
{\it per se}, a more precise estimate can still be obtained by
combining the results.
% 
% it is highly desirable to quote to the high energy
% physics community one single number, rather than leaving the choice to experimentalists and theoreticians of which estimate to utilize
% for their own studies. Also, while some measurements are more precise than others, they are not completely correlated as either the data samples are independent, or different systematics affect the measurements, or both. It is thus important to communicate one single number, and it is advantageous this single number to be the combination of all the available measurements. 
%
Such a combination is regularly performed by the Tevatron Electroweak Working Group (TEVEWWG)~\cite{TEVEWWG}.
Normally, the current most precise measurement in each decay mode per experiment is selected to enter into the Tevatron combination. Utilizing measurements which use overlapping datasets requires the estimation of the statistical correlations among the measurements themselves. 
Many systematic uncertainties are correlated among channels 
as their sources affect either the modeling of the signal and/or
background events or the calibration of the detector response to the final state objects. These correlations have to be understood in detail in order
to combine the measurements properly. This necessitates a collaborative
effort by the authors of the \Mt\ measurements that are being combined.

% For the above reason, it has been of crucial importance that the authors of the several measurements work closely together to understand these correlations in detail.

\subsection{Method and general issues}
\label{sec:uncertainty}

The CDF and \D0 \ collaborations use the best linear unbiased estimator (BLUE) 
method~\cite{Lyons:1988, Valassi:2003} to combine their measurements. 
% An increase in accuracy in the observable measurement is expected as the several analyses are designed to study orthogonal datasets. 
%The method is also granted to provide increased accuracy whenever the systematic uncertainties are at least partially correlated.
% As stressed above, it is of particular importance to understand the correlation - or absence of it - in the impact the systematic sources have on the measurements.  
% The BLUE method utilized here allows a straightforward breakdown of the systematic sources. 
The collaborations have been working jointly for a number of years 
with the purpose of understanding the sources of the systematic 
uncertainty and their interdependence. Once an independent 
uncertainty source is identified, its impact on the \Mt\ measurements
is evaluated by assigning a correlation
matrix which has an entry for each pair of results.
% classifying them into a number of independent categories.
% and breaking the list down to their subcomponents, in such a way as to minimize the overlap between different systematic categories. For simplicity, all systematic sources are considered to be completely independent from each other.  
% The problem of understanding the correlation among measurements thus reduces to the problem of understand how correlated is the impact of each systematic uncertainty between the several measurements, i.e. to define a correlation matrix for each systematic source. 
Before the final combination is performed, systematic sources are collected into groups that share similar physical origins and whose impact on the measurements is described by the same correlation matrix. This grouping is described below: 
{\flushleft{\bf Statistics}}. The statistical uncertainty associated with the  \Mt\ determination.
{\flushleft{\bf iJES}}: Part of the JES uncertainty which originates from
   {\em in situ} calibration procedures that use the $W$ boson
   mass in decay modes with at least
   one hadronically decaying $W$.
{\flushleft{\bf aJES}}: Part of the JES uncertainty which originates from
    the difference in calorimeter 
    electromagnetic over hadronic ($e/h$) response 
    for $b$ jets and light-quark jets. 
  {\flushleft{\bf bJES}}: Part of the JES uncertainty which originates from
    generator modeling of $b$ jets.
    For both CDF and \D0 , it includes the uncertainties 
    arising from variations in the semileptonic branching fractions,
    $b$ fragmentation functions, and differences in the color flow between 
    $b$ jets and light-quark jets.
  {\flushleft{\bf cJES}}: Part of the JES uncertainty which originates from
    the modeling of light-quark 
    fragmentation and out-of-cone corrections. For \D0 \ Run~II measurements,
    it is included in the dJES category.
  {\flushleft{\bf dJES}}: Part of the JES uncertainty which originates from
   the finite size of the data samples used to calibrate the
   jet energy response of the detectors.   This includes uncertainties associated
   with the $\eta$-dependent JES corrections which are estimated
   using dijet data events. For \D0 \ this also includes the uncertainties in 
   the light jet response and uncertainties
   that arise due to the sample dependence of 
   jet corrections derived with the $\gamma$+jets data.
  {\flushleft{\bf rJES}}: The remaining part of the JES uncertainty. 
    For CDF, this is dominated by uncertainties in the calorimeter response to light-quark jets, and also includes small 
    uncertainties associated with the multiple interaction and underlying 
    event corrections. For \D0 \ Run~II measurements,
    uncertainty sources of this kind belong to
    the dJES category.
  {\flushleft{\bf LepPt}}: The systematic uncertainty related to the calibration
    of lepton transverse momentum measurements.
  {\flushleft{\bf Signal}}: The systematic uncertainty arising from uncertainties
    in the $t \bar t$ modeling. This is obtained summing in quadrature several different sources. It 
    includes uncertainties in the ISR and FSR descriptions; the uncertainty on the knowledge of the PDFs; 
    the systematic uncertainty arising from a variation of the phenomenological description of color reconnection between final state particles. The systematic uncertainty associated with variations of the physics model used to calibrate the data analysis methods, correlated across all measurements. It includes variations observed when ISAJET (Run~I) or HERWIG or ALPGEN+PYTHIA are substituted to PYTHIA in modeling the $t \bar t$ signal. For \D0 \ it also includes the uncertainty that arises from considering or ignoring higher order Feynman diagrams.
  {\flushleft{\bf Detector modeling (DetMod)}}: Uncertainty in the modeling of the detector in the MC simulation, such as the uncertainty on the jet energy resolution and identification efficiency. 
  {\flushleft{\bf Background from MC (BGMC)}}:  Uncertainty in the background modeling, including sample
     composition and shape of relevant distributions, including the choice of the factorization scale used to model $W$+jets production.
{\flushleft{\bf Background from data (BGData)}}: This group includes uncertainties associated with the modeling of the QCD  multijet background in the all-hadronic, $\met$+jets and $\ell$+jets channels, uncertainties associated with the modeling of the Drell-Yan background in the dilepton channel
  {\flushleft{\bf Method}}: The systematic uncertainty arising from any source specific
    to a particular data analysis method, including the finite Monte Carlo statistics 
    available for method calibration. 
  {\flushleft{\bf Uranium Noise and Multiple Interactions (UN/MI)}}: This is specific to \D0 \ and includes the uncertainty
    arising from uranium noise in the \D0 \ calorimeter and from the
    multiple interaction corrections to the JES.  For \D0 \ Run~I these
    uncertainties were sizable, while for Run~II, owing to the shorter
    calorimeter electronics integration time and {\em in situ} JES calibration, these uncertainties
    are negligible.
  {\flushleft{\bf Multiple Hadron Interactions (MHI)}}: The systematic uncertainty arising from the
  mismodeling of the distribution of the number of collisions per Tevatron bunch crossing.

Reasonable variations in the assignment of uncertainty sources to these groups,
in the back-propagation of the bJES
uncertainties to Run~I measurements,
% in the approximations made to symmetrize the uncertainties used in the combination, 
and in the assumed 
magnitude of the correlations have produced
a negligible impact on the combined \Mt\ measurement
and its total uncertainty.

% After having grouped the systematic sources in categories sharing similar origins and identical correlations among the many measurements, the BLUE method needs the explicit correlations matrices. As the correlations depend on the actual measurements, we will discuss them in the next section.
%The correlation matrices $\textbf{C}_i$ for all the sources of uncertainties $i$ whose correlation among the generic measurements $a, b$ from CDF and $a, b$ from \D0 is expressed through the correlation coefficients $c$ in the following form: \D0%\[\begin{array}{c|cccc}
%  & \textrm{CDF a} & \textrm{CDF b} & \textrm{\D0 a} & \textrm{\D0 b}  \\ \hline
%CDF a& c_{CaCa} & c_{ CaCb} & c_{CaDa} & c_{CaDb} \\
%CDF b & c_{CbCa} & c_{CbCb} & c_{CbDa} & c_{CbDb} \\
%\D0 a & c_{DaCa} & c_{DaCb} & c_{DaDa} & c_{DaDb} \\
%\D0 b & c_{DbCa} & c_{DbCb} & c_{DbDa} & c_{DbDb} \\
%\end{array}\] 

\subsection{Latest combination}

The latest Tevatron combination~\cite{tevMtop} utilizes twelve different measurements of \Mt.
%  five published Run~I results, three published Run~II results, and three preliminary Run~II results. 
% They are no longer preliminary.
These are the Run~I and Run~II CDF measurements in the $\ell$+jets (l+j in the summary table)~\cite{ref:cdf.ljets.run1prl,ref:plujan}, all-hadronic 
(all-j)~\cite{Abe:1997rh,LucaNote}, and dilepton (di-l)~\cite{ref:cdf.dilep.run1prl,TMTNote} channels,
the \D0 \ Run~I and Run~II measurements in the $\ell$+jets~\cite{ref:d0.ljets.run1prd,ref:d0.ljets.3.6fb} 
and dilepton~\cite{ref:d0.dilep.run1prl,ref:d0.dilep.3.6fb} channels, one CDF measurement in the $\ell$+jets channel that used tracking information with minimal dependence on the jet response (trk)~\cite{ref:cdf.ljets.lxyandlpt}, and one CDF measurement performed in the $\met$+jets signature~\cite{ref:cdf.etmissplusjets}. All these measurements are summarized in Table~\ref{tab:inputs} together with their uncertainties.

\begin{table*}[hbtp]
\caption[Input measurements]{Summary of the measurements used to determine the
  Tevatron average \Mt.  Integrated luminosity ($\int \mathcal{L}\;dt$) has units in
  fb$^{-1}$, and all other numbers are in GeV$/c^2$.  The uncertainty categories and 
  their correlations are described in the Sec.\,\ref{sec:uncertainty}.  The total systematic uncertainty 
  and the total uncertainty are obtained by adding the relevant contributions 
  in quadrature. ``n/a" stands for ``not applicable'',``n/e" for ``not evaluated".}
\label{tab:inputs}
\begin{center}

\renewcommand{\arraystretch}{1.30}
{\small 
\begin{tabular}{l|ccccc|ccccc|cc|c} 
\hline \hline
       & \multicolumn{5}{c|}{{Run~I} published} 
       & \multicolumn{5}{c|}{{Run~II} published} 
       & \multicolumn{2}{c}{{Run~II} prel.}  \\ 
       %\cline{2-12}
       & \multicolumn{3}{c}{ CDF } 
       & \multicolumn{2}{c}{ D\O\ }
       & \multicolumn{3}{|c}{ CDF }
       & \multicolumn{2}{c|}{ D\O\ }
       & \multicolumn{2}{c|}{ CDF }
       & Tevatron
        \\
%                         CDF R1       CDFR1   CDFR1       D0R1       D0R1        CDFRII   CDFRII        CDFRII        D0RIInp       D0RIInp        CDFRIInp   CDFRIInp
                       &     allh &   l+jt   & di-l     &     l+jt &   di-l  &     di-l  &    Lxy     &    l+jt     &    di-l    &    l+jt     &       allh  &  MEt      &  all  \\
\hline
$\int \mathcal{L}\;dt$ &      0.1 &     0.1  &     0.1  &     0.1  &     0.1 &       5.6 &        1.9 &      5.6    &        5.2 &          3.6 &       5.8  & 5.7       & $\le$ 5.8    \\
\hline                         
Result                 & 186.0    & 176.1    & 167.4    & 180.1    & 168.4   &    170.28 &     166.90 &      173.00 &     173.97 &       174.94 &    172.47  & 172.32    & 173.18     \\
\hline                         
iJES                  &   n/a     &      n/a &      n/a &      n/a &      n/a &     n/a  &        n/a &        0.58 &       n/a  &         0.53 &        0.95& 1.54     & 0.39      \\
aJES                  &   n/a     &      n/a &      n/a &      0.0 &      0.0 &     0.14 &        n/a &        0.13 &       1.57 &         0.0 &         0.03& 0.12     & 0.09      \\
bJES                  &   0.6     &      0.6 &      0.8 &      0.7 &      0.7 &     0.33 &        n/a &        0.23 &       0.40 &         0.07 &       0.15 & 0.26     & 0.15      \\
cJES                  &   3.0     &      2.7 &      2.6 &      2.0 &      2.0 &     2.13 &       0.36 &        0.27 &       n/a  &         n/a  &       0.24 & 0.20     & 0.05       \\
dJES                  &   0.3     &      0.7 &      0.6 &      n/a &      n/a &     0.58 &       0.06 &        0.01 &       1.50 &         0.63 &       0.04 & 0.05     & 0.20       \\
rJES                  &   4.0     &      3.4 &      2.7 &      2.5 &      1.1 &     2.01 &       0.24 &        0.41 &       n/a  &         n/a  &       0.38 & 0.45     & 0.12        \\
LepPt                 &   n/e     &      n/e &      n/e &      n/e &      n/e &     0.27 &      n/a  &        0.14 &       0.49 &         0.18 &       -    &  -        & 0.10        \\
Signal                &   2.0     &      2.6 &      2.9 &      1.1 &      1.8 &     0.73 &      0.90 &        0.56 &       0.74 &         0.77 &       0.62 &  0.74     & 0.51        \\
DetMod                &   0.0     &      0.0 &      0.0 &      0.0 &      0.0 &     0.0  &      0.0  &        0.0  &       0.33 &         0.36 &       0.0  &  0.0      & 0.10        \\
UN/MI                 &      n/a  &      n/a &      n/a &      1.3 &      1.3 &     n/a  &      n/a  &        n/a  &       n/a  &        n/a  &       n/a  &  n/a      &  0.00        \\
BGMC                  &      1.7  &      1.3 &      0.3 &      1.0 &      1.1 &     0.24  &     0.80  &       0.27 &       0.0  &        0.18  &        0.0 & 0.0       & 0.14        \\
BGData                &      0.0  &      0.0 &      0.0 &      0.0 &      0.0 &     0.14  &     0.20  &       0.06 &       0.47 &        0.23  &        0.56& 0.12      & 0.11         \\
Method                &   0.6     &      0.0 &      0.7 &      0.6 &      1.1 &      0.12 &      2.50 &       0.10 &       0.10 &         0.16 &        0.38& 0.14      & 0.09         \\
MHI                   &      n/e  &      n/e &      n/e &      n/e &      n/e &      0.23 &      0.0  &       0.10 &       0.0  &         0.05 &        0.08& 0.16      & 0.08     \\
\hline                         
Syst                  &   5.7    &       5.3 &      4.9 &      3.9 &      3.6 &      3.13 &      2.82 &        1.06 &       2.45 &         1.24 &        1.40& 1.82     & 0.75    \\
Stat                  &  10.0    &       5.1 &     10.3 &      3.6 &     12.3 &      1.95 &      9.00 &        0.65 &       1.83 &         0.83 &        1.43& 1.80     & 0.56     \\
\hline                         
Total                 &  11.5    &       7.3 &     11.4 &       5.3 &    12.8 &      3.69 &      9.43 &        1.23 &       3.06 &         1.50 &        2.00&  2.56    & 0.94     \\ 
\hline
\hline
\end{tabular}
}
\end{center}
\end{table*}

All Run~I measurements
have relatively large statistical uncertainties due to the limited size of the $t\bar t$ samples collected. 
Their systematic uncertainties are dominated by the total jet energy scale (JES) uncertainty. In Run~II, both CDF and \D0 \ take advantage of the larger $t \bar t$ samples available and employ new analysis techniques to reduce both of these uncertainties. 

The statistical uncertainty is uncorrelated among the measurements due to the
non-overlapping event samples used. The only possible exception to this rule is the CDF Run~II \Mt\ measurements in the $\ell$+jets channel named ``l+j'' and ``trk'' in the Table~\ref{tab:inputs} caption. The latter employs a subset of the data sample used by the former. The statistical correlation between the trk analysis and an older Run~II CDF l+j measurement was studied using Monte Carlo signal-plus-background pseudo-experiments which correctly accounted for the sample overlap and was found to be consistent with zero. The statistical part of the JES systematic uncertainty (iJES) is also uncorrelated, for the same reasons. The correlation matrices for the statistical uncertainty and for the iJES category are thus the trivial unit
matrices. Unit matrix is also used to represent correlations for the Method category.

The correlation among the  measurements for the bJES, cJES, and Signal uncertainties is taken to be 100\%.
The uncertainties in the aJES, dJES, LepPt, MHI, DetMod and BGData categories are taken to be 100\% correlated among all Run~I and all Run~II measurements within the same experiment, but uncorrelated between Run~I and Run~II, and uncorrelated between the experiments. The uncertainties in the rJES and UN/MI categories are taken to be 100\% correlated among all measurements within the same experiment but uncorrelated between the experiments. The uncertainty in the BGMC category is taken to be 100\% correlated among all measurements in the same channel and among experiments.

The combined value of the top quark mass is
%
%\begin{eqnarray}
%$M_{top}$=\gevcc{\measStatSyst{173.32}{0.56}{0.89}} \\
%$M_{top}$=173.32 \pm 1.06\,\GeV/c^2 \\
%$M_{top}$=173.32 \pm 0.61\% \\
%$M_{top}$=173.3 \pm 1.1\,\GeV/c^2 
%\end{eqnarray}
%$$M_{top}$=\gevcc{\measStatSyst{173.32}{0.56}{0.89}}$. 
$\Mt = 173.18 \pm 0.56 \,(\mbox{stat.}) \pm 0.75 \,(\mbox{syst.})$~GeV/$c^2$.
The $\chi^2$ value of this result is 8.3 for 
11 degrees of freedom. The corresponding 
confidence level is 68\%, indicating good agreement 
among the input measurements.
Adding the statistical and systematic uncertainties in quadrature yields a total uncertainty of 0.94~GeV/$c^2$, corresponding to a
relative precision of 0.54\% on the top quark mass.
Rounded off to two significant digits in the uncertainty, the combined result is $\Mt = 173.2 \pm 0.9$~GeV/c$^2$.
% The breakdown of the uncertainties is 
% shown in Table~\ref{tab:inputs}. 
%This value supersedes previous combinations~\cite{Mtop1-tevewwg04}.
%, Mtop-tevewwgSum05,  Mtop-tevewwgWin06, Mtop-tevewwgSum06, Mtop-tevewwgWin07, Mtop-tevewwgWin08, Mtop-tevewwgSum08, Mtop-tevewwgWin09}. 

%The pull and weight for each of the inputs are listed in Table~\ref{tab:stat}.

\begin{figure}[htbp]
 \begin{center}
 \includegraphics[width=0.4\textwidth]{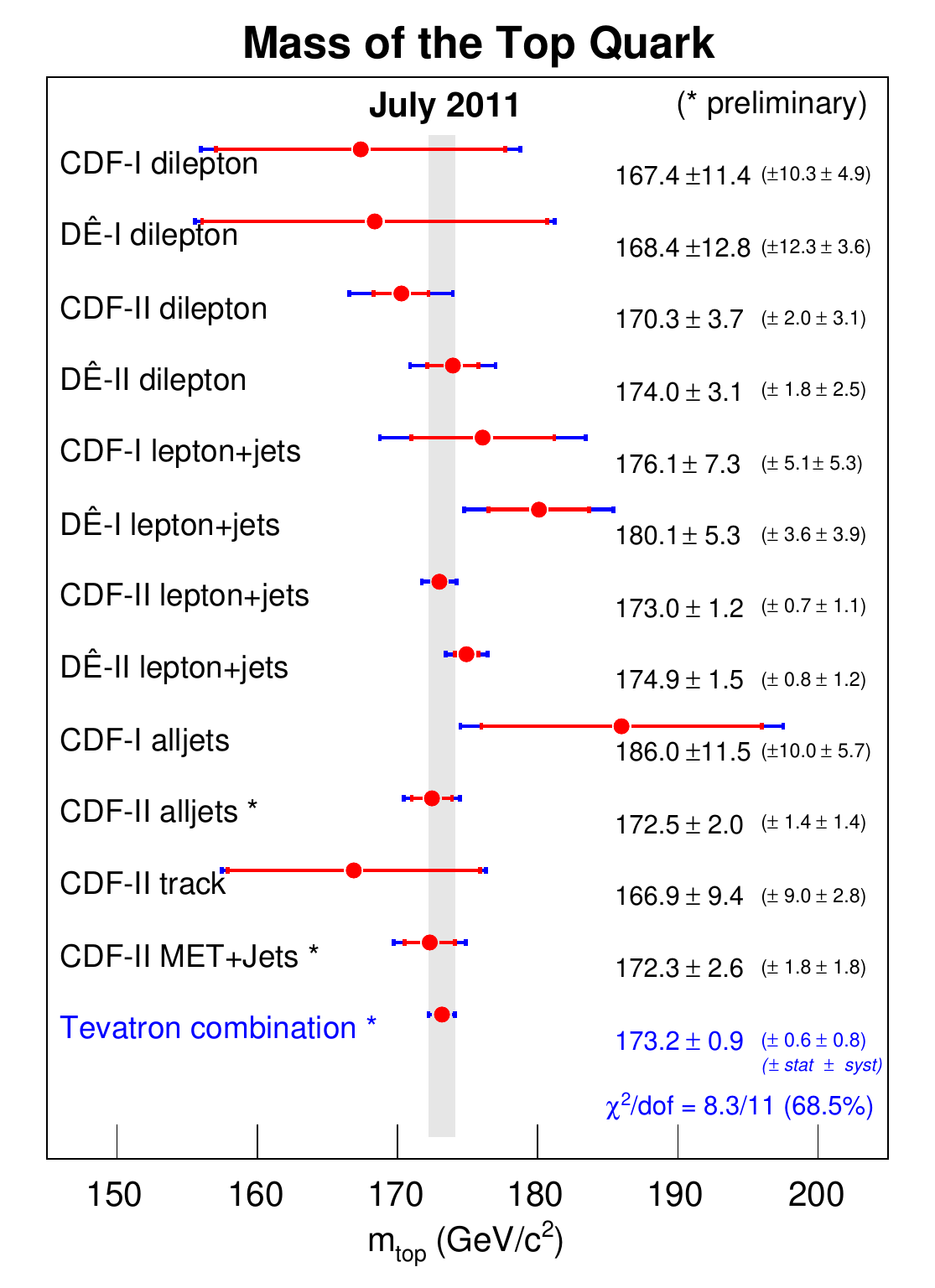}
 \end{center}
 \caption[Summary plot for the Tevatron average top-quark mass]
  {Summary of the input measurements and resulting Tevatron average
  mass of the top quark~\cite{tevMtop}.}
  \label{fig:summary} 
 \end{figure}

The input measurements and the current combination
are displayed in Figure~\ref{fig:summary}.
%\vspace*{0.10in}
%
%\begin{table}[tbh]
%\caption{\label{tab:BLUEuncert} 
%Summary of the Tevatron combined average $M_{top}$ . The uncertainty categories are 
%described in the text. The total systematic uncertainty and the total uncertainty are obtained 
%by adding the relevant contributions in quadrature.}
%\begin{center}
%\begin{tabular}{lc} \hline \hline
%  & Tevatron combined values (GeV/$c^2$) \\ \hline
%$M_{top}$            & 173.32 \\ \hline
% iJES                     &    0.46 \\
% aJES                    &  0.21  \\
%  bJES                   &  0.20 \\  
%  cJES                   &  0.13 \\ 
%  dJES                  &   0.19 \\ 
%  rJES                   &   0.15 \\
%  LepPt                 &  0.10 \\
%  Signal                &   0.19  \\
%  Background     &   0.23 \\
%  Fit                       &   0.11 \\
%  MC                     &   0.40 \\
%  UN/MI                &    0.02 \\
 % Background     &   0.23 \\
 % Fit                       &   0.11 \\
%  CR                     &   0.39 \\
%  MHI                    &   0.08 \\ \hline
%  Systematics      &   0.89 \\ 
%  Statistics            &    0.56 \\ \hline
%  Systematics (w/o iJES)      &   0.76 \\
%  Statistical (w/ iJES) & 0.72 \\ \hline
%  
%  Total                  &   1.06 \\
%   \hline \hline
%
 % \end{tabular}
% \end{center}
 % \end{table}
The most recent CDF and \D0 \ measurements in the $\ell$+jets channel which use the matrix element method enter the combination with the largest weights, followed by the CDF template measurement in the all-hadronic channel.

It can be deduced from Table~\ref{tab:inputs} that the total
JES-related uncertainty is $0.49$~GeV/$c^2$, with $0.39$~GeV/$c^2$ coming 
from the statistical component and $0.30$~GeV/$c^2$ from other sources.
As the most significant fraction of the JES uncertainty 
is still statistical in nature,
the \Mt\ precision can be further improved by simply analyzing larger collision
datasets already collected by CDF and \D0 .
%In fact, as discussed in Section~\ref{sec:tev_runii},
%CDF has already analyzed more data in the all-hadronic channel
%and added an independent measurement in the $\Etmiss$+jets channel,
%while \D0 \ produced new results in both $\ell$+jets and dilepton channels.
%With the new data added, the latest CDF-only \Mt\ combination~\cite{CDFcombo11}
%reaches a precision similar to the Tevatron average just described.

When the complete dataset produced by the Tevatron will be analyzed,
the total uncertainty of the \Mt\ determination will likely drop around $0.8$~GeV/$c^2$,
even without further understanding of the systematic sources.
By the time Tevatron operations will close at the end of 2011, LHC
will have produced much larger $t \bar t$ samples, so the expected
\Mt\ statistical uncertainty will soon be negligible. It will be 
important to establish the correlations among the measurements 
performed at the Tevatron and at the LHC so that
one single estimate of the top quark mass can still be provided
to the particle physics community.

\section{Conclusions}
\label{sec:conclusion}

% Top quark is the heaviest elementary object known. Its Yukawa coupling
% is of order unity in the Standard Model which results in a strong
% dependence of the indirectly estimated Higgs boson mass on \Mt.

From the first direct observation of top quark pair production and
throughout the whole lifetime of the Tevatron collider program,
precision determination of the top quark mass has been the subject of
intensive studies at the CDF and \D0 \ experiments. A number of
sophisticated \Mt\ measurement techniques have been developed.
Important methodological advances include the first practical
application of the matrix element analysis method to hadron collider
data and introduction of the {\it in situ} jet energy scale
calibration with the $W$ boson mass. These particular improvements led
to a substantial reduction in the statistical and systematic
uncertainty, respectively. 
Recent \Mt\ measurements performed
in different final states of the $t\bar{t}$ system, with
multiple analysis methods and with
independent experimental setups are all in close agreement with
each other, in accordance with Standard Model expectations.

The current combined estimate of the top quark mass by CDF and \D0 \ is
\begin{eqnarray*}
   \fl \Mt & = 173.18 \pm 0.56 \,(\mbox{stat.}) \pm 0.75 \,(\mbox{syst.}) \,\,\mbox{GeV}/c^2 \\
   \fl     & = 173.18 \pm 0.94\,(\mbox{total}) \,\,\mbox{GeV}/c^2
\end{eqnarray*}
Its accuracy is limited
by systematic uncertainties resulting from a number of imprecisely
known calibration parameters, both theoretical and experimental.
Although the knowledge of some of these parameters will automatically
improve as more data becomes available, further reduction of the
uncertainty will require more detailed understanding of the underlying
physics processes. Whenever possible, this understanding should
be translated into the calculation of the corresponding systematic
uncertainty on event-by-event basis. Modeling of parton
showering and hadronization, together with 
accurate representation of nonlinearities in the detector jet
response, will remain critical for CDF and \D0 .

In the future, when
much larger $t\bar{t}$ samples are accumulated by the LHC experiments,
\Mt\ measurements which do not use jets could outperform more conventional
techniques. The ultimate precision will be reached by
scanning the $t\bar{t}$ center-of-mass energy production threshold at a high energy
lepton collider.

\section{Acknowledgments}
\label{sec:ack}

We thank Christian Bauer for guidance regarding the relationship
between the experimental and theoretical top quark mass. We are also
indebted to Paul Lujan, Jeremy Lys, Yvonne Peters, and Elizaveta Shabalina
for reading the manuscript and providing useful comments.

\section*{References}

\end{document}